\documentclass[aps,prx,amsmath,amssymb,twocolumn,superscriptaddress,showpacs,longbibliography]{revtex4-2}
\usepackage{amsmath,amssymb}
\usepackage{graphicx}
\usepackage{appendix}
\usepackage[usenames,dvipsnames]{color}
\usepackage[colorlinks=true, citecolor=blue, linkcolor=blue, urlcolor=blue]{hyperref}
\usepackage{bm}
\usepackage{layouts}
\usepackage{booktabs}
\usepackage[normalem]{ulem}

\usepackage[nottoc,numbib]{tocbibind}

\newcommand{\D}{\textrm{d}}
\newcommand{\I}{\textrm{i}}
\newcommand{\ket}[1]{\left|#1\right\rangle}
\newcommand{\bra}[1]{\left\langle#1\right|}
\newcommand{\dmu}{\partial_{\mu}}
\newcommand{\dnu}{\partial_{\nu}}
\newcommand{\Amu}{A_{\mu}}
\newcommand{\Anu}{A_{\nu}}
\newcommand{\hc}{\text{h.c.}}
\newcommand{\fk}{\bm{k}}

\newcommand\TstrutSmall{\rule{0pt}{3.5ex}}
\newcommand\Tstrut{\rule{0pt}{5ex}}         
\newcommand\TstrutLarge{\rule{0pt}{10ex}}
\newcommand\Bstrut{\rule[-3.8ex]{0pt}{0pt}}   
\newcommand\BstrutSmall{\rule[-1.7ex]{0pt}{0pt}}   
\newcommand\BstrutLarge{\rule[-7.0ex]{0pt}{0pt}}

\newlength{\thickarrayrulewidth}
\newcommand{\REV}[1]{#1}

\begin{document}

\title{Light-matter coupling and quantum geometry in moir{\'e} materials}

\author{Gabriel E.~Topp}
\email{gabriel.topp@aalto.fi}
\affiliation 
{Department of Applied Physics, Aalto University, FI-00076 Aalto, Finland}

\author{Christian J.~Eckhardt}
\email{christian.eckhardt@mpsd.mpg.de}
\affiliation 
{Institute for Theory of Statistical Physics, RWTH Aachen University, and JARA Fundamentals of Future Information Technology, 52062 Aachen, Germany}
\affiliation 
{Max Planck Institute for the Structure and Dynamics of Matter, Center for Free Electron Laser Science, Luruper Chaussee 149, 22761 Hamburg, Germany}

\author{Dante M.~Kennes}
\email{dante.kennes@rwth-aachen.de}
\affiliation 
{Institute for Theory of Statistical Physics, RWTH Aachen University, and JARA Fundamentals of Future Information Technology, 52062 Aachen, Germany}
\affiliation 
{Max Planck Institute for the Structure and Dynamics of Matter, Center for Free Electron Laser Science, Luruper Chaussee 149, 22761 Hamburg, Germany}

\author{Michael A.~Sentef}
\email{michael.sentef@mpsd.mpg.de}
\affiliation 
{Max Planck Institute for the Structure and Dynamics of Matter, Center for Free Electron Laser Science, Luruper Chaussee 149, 22761 Hamburg, Germany}

\author{P\"aivi T\"orm\"a}
\email{paivi.torma@aalto.fi}
\affiliation 
{Department of Applied Physics, Aalto University, FI-00076 Aalto, Finland}

\date{\today}

\begin{abstract}
Quantum geometry has been identified as an important ingredient for the physics of quantum materials and especially of flat-band systems, such as moir\'e materials. On the other hand, the coupling between light and matter is of key importance across disciplines and especially for Floquet and cavity engineering of solids.
Here we present fundamental relations between light-matter coupling and quantum geometry of Bloch wave functions, with a particular focus on flat-band and moir\'e materials, in which the quenching of the electronic kinetic energy could allow one to reach the limit of strong light-matter coupling more easily than in highly dispersive systems. We show that, despite the fact that flat bands have vanishing band velocities and curvatures, light couples to them via geometric contributions. Specifically, the intra-band quantum metric allows diamagnetic coupling inside a flat band; the inter-band Berry connection governs dipole matrix elements between flat and dispersive bands. We illustrate these effects in two representative model systems: (i) a sawtooth quantum chain with a single flat band, and (ii) a tight-binding model for twisted bilayer graphene. For (i) we highlight the importance of quantum geometry by demonstrating a nonvanishing diamagnetic light-matter coupling inside the flat band. For (ii) we explore the twist-angle dependence of various light-matter coupling matrix elements. Furthermore, at the magic angle corresponding to almost flat bands, we show a Floquet-topological gap opening under irradiation with circularly polarized light despite the nearly vanishing Fermi velocity. We discuss how these findings provide fundamental design principles and tools for light-matter-coupling-based control of emergent electronic properties in flat-band and moir\'e materials.
\end{abstract}
\maketitle

\section{Introduction}\label{INTRO}
It has become increasingly clear that not only the band structure but also the properties of the Bloch functions are of key importance in understanding and designing the physical properties of periodic structures. Prominent examples are the quantum Hall effect and topological insulators~\cite{Klitzing1980,Thouless1982,Haldane1988,Kane2005,Bernevig2006QuantumWells,Konig2007QuantumWells,Hasan2010,Bernevig2013,Jotzu:2014}, which are governed by quantum geometric concepts such the Chern number (the Berry curvature integrated over the Brillouin zone) or other topological invariants. Recently, the set of quantum geometric quantities known to bear significance to physical observables has broadened further. For instance the quantum metric (Fubini-Study metric), which describes the distance between quantum states in a submanifold of the Hilbert space, has been predicted to influence diverse phenomena ranging from superconductivity~\cite{peotta_superfluidity_2015,julku_geometric_2016,torma_quantum_2018}, orbital magnetic susceptibility~\cite{Gao_FieldIncuded_2014,ogata_orbital_2015,piechon_geometric_2016,rhim_quantum_2020}, exchange constants~\cite{PhysRevB.95.184428} and exciton Lamb shift~\cite{Srivastava:2015} to the nonadiabatic anomalous Hall effect~\cite{Gao_FieldIncuded_2014,Bleu2018}. The quantum metric has recently been measured experimentally~\cite{Gianfrate:2019,Tan2019ExpQubit,Weitenberg:2019,Yu2020expdiamond}. The Berry curvature and the quantum metric are the imaginary and real parts of the quantum geometric tensor, respectively~\cite{Provost80}. Therefore, the distance between quantum states and the topology of the system are intimately connected. This provides, for instance, a fundamental lower bound of the superfluid weight (superfluid density) in terms of the Chern number and Berry curvature of the band~\cite{peotta_superfluidity_2015,liang_band_2017}. 

It is intuitive to expect that the effects from the quantum geometric properties of the Bloch states outsize those of the band structure if the latter is featureless. Indeed, the geometric contribution to the supercurrent of a superconductor is maximized in so-called flat (dispersionless) bands. The importance of quantum geometric concepts is therefore amplified by the recent breakthrough results on flat-band-related superconducting and correlated phases in twisted bilayer graphene (TBG)~\cite{Neto07,Morell2010,Bistritzer12233,Li2010,Cao2018a, Cao2018, Yankowitz2019,Kerelsky2019, Sharpe605, lu2019superconductors, Serlin2019,MacDonald2019,Andrei2020,Balents2020,Kennes2021}. The precise mechanism of superconductivity in TBG is heavily debated \cite{ PhysRevLett.121.257001,PhysRevB.98.241412,PhysRevB.98.220504,PhysRevX.9.041010,PhysRevLett.122.257002,PhysRevB.102.064501,PhysRevB.98.245103,PhysRevX.8.041041,PhysRevB.98.205151,PhysRevB.98.214521,PhysRevB.98.075154,PhysRevLett.121.087001,PhysRevB.98.085436,PhysRevLett.121.217001,PhysRevB.98.195101,PhysRevB.98.241407,PhysRevX.8.031089,You2019,PhysRevB.99.195120,PhysRevB.99.134515,PhysRevLett.122.026801,PhysRevB.100.085136,Claassen2019,PhysRevB.101.155413,PhysRevB.103.L041103,PhysRevB.103.024506,NematicSCExp,NematicSCRaf,TaoNemSC,ScMacDoEL,BlackSchafferSC,Balents2020,Stepanov2020unt}. However, for this nearly flat-band system the superfluid weight has also been proposed to contain a significant geometric contribution~\cite{julku_superfluid_2020,hu:2019} and to be governed by the topological $C_{2z}T$ Wilson loop winding number~\cite{xie_topology-bounded_2020}. 
Although much research has focused on TBG, for which flat bands appear only for certain narrow ranges around specific twist angles (so-called magic angles, the largest of which is $1.1\pm0.1^{o}$ \cite{Bistritzer12233,Cao2018,Yankowitz2019}), the general concept of flat-band engineering is much more broadly applicable \cite{Kennes2021}. For example, recent advances into multilayered and/or electrostatically gated graphitic systems \cite{Liu2019, Cao2019, Shen2019,Chen2019a, Chen2019, chen2019tunable,TutucBi,rubioverdu2020universal,Tritsaris_2020}, twisted bilayer boron nitride \cite{Xian2019BN},  twisted homo- or heterobilayers of semiconducting transition metal dichalcogenides \cite{Wu2018, Wu2019, Naik2018, Ruiz-Tijerina2019,Wu2018, Wu2019, Schrade2019,Wang2020WSe2,XianMoS2,LischnerTMDs}, monochalcogenides  \cite{KennesGeSe} or twisted bilayers of magnetic MnBi$_2$Te$_4$ \cite{Lian20}, have been made. These studies demonstrate that, in the absence of semi-metallic behavior typical for graphene, the bandwidth (kinetic energy scales) can be reduced continuously and smoothly with the twist angle by the moir\'e superpotential induced by the interference pattern between layers, which in turn allows increasing the relative importance of interactions.

\begin{figure}
    \centering
    \includegraphics[width=0.9\columnwidth]{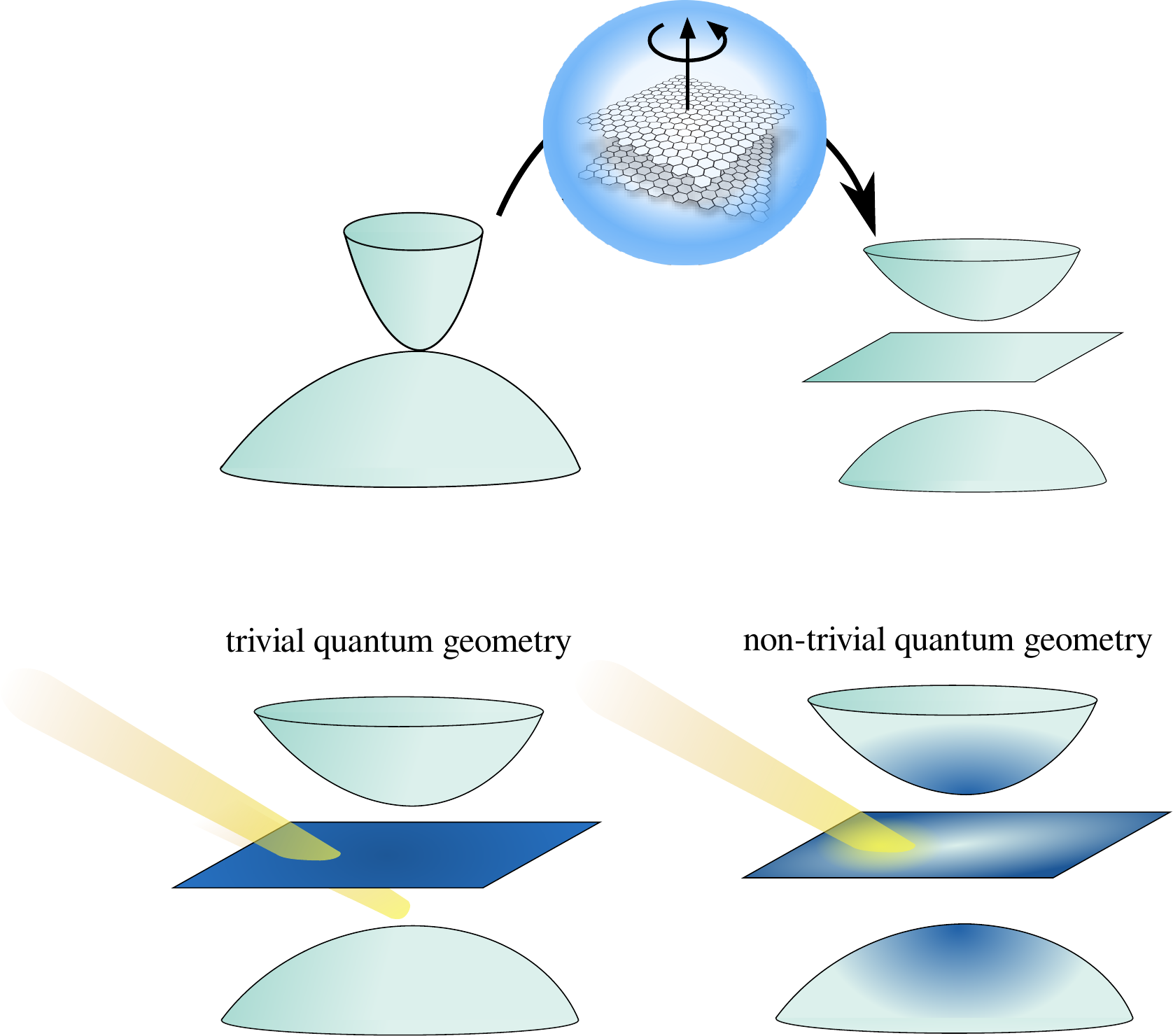}
     \caption{Illustration of the main concept of this paper. Moir\'e engineering has been demonstrated as a flexible route towards two-dimensional flat-band engineering (top row). From the band structure one might na\"ively expect that the coupling to light in flat bands vanishes. However, a geometric contribution (present, e.g., in TBG), can contribute to light-matter coupling and dominates in flat band systems (bottom row).}
    \label{fig:Anchor}
\end{figure}

In this article, we show how quantum geometry affects one more important physical phenomenon, namely the light-matter coupling (LMC) between electromagnetic fields and Bloch electrons. The LMC of quantum materials is of critical importance, for their potential integration into optoelectronic devices \cite{wang_electronics_2012,liu_strong_2015}, for their ability to host polaritonic excitations with strongly hybridized light and matter contributions \cite{carusotto_quantum_2013,torma_strong_2015,flick_atoms_2017,flick_strong_2018,frisk_kockum_ultrastrong_2019,forn-diaz_ultrastrong_2019}, as well as for the fundamental possibility to engineer new states of matter in them through LMC. The light-matter engineering concept is particularly intriguing. In the regime of classical light, Floquet matter \cite{wang_observation_2013, PhysRevX.4.031027} -- matter driven periodically in time -- has brought about key ideas for the so-called Floquet engineering of topology \cite{mahmood_selective_2016,PhysRevX.9.041001,rudner_band_2020} and other emergent properties of quantum materials \cite{oka_floquet_2019}. Going from classical to quantum light, is has been noticed how strong LMC in the quantum-electrodynamical regime is potentially useful to engineer cavity quantum materials \cite{laussy_exciton-polariton_2010,cotlet_superconductivity_2016,kavokin_exciton-polariton_2016,sentef_cavity_2018,schlawin_cavity-mediated_2019,hagenmuller_enhancement_2019,curtis_cavity_2019,wang_cavity_2019,sentef_quantum_2020,thomas_exploring_2019,kiffner_manipulating_2019,mazza_superradiant_2019,andolina_cavity_2019,gao_photoinduced_2020,chakraborty_non-bcs-type_2020,li_manipulating_2020,hubener_engineering_2020,ashida_quantum_2020,latini_ferroelectric_2021}.

In flat-band and moir\'e systems, LMC would be of even greater impact. The quenching of the kinetic energy of flat-band electrons might allow one to reach the regime of strong LMC more easily than in strongly dispersive materials \cite{Kennes2021}. Specifically for TBG and other moir\'e systems, there have been first theoretical explorations of Floquet engineering \cite{topp_topological_2019,vogl_effective_2020,vogl_floquet_2020,katz_optically_2020,rodriguez-vega_floquet_2020,li_floquet-engineered_2020,lu_valley-selective_2020,vogl_floquet_2021,rodriguez-vega_low-frequency_2021} and nonlinear optical responses \cite{ikeda_high-order_2020}. Recently a measured strong mid-infrared photoresponse in bilayer graphene at small twist angles was reported \cite{deng_strong_2020}. Specifically in Ref.~\onlinecite{topp_topological_2019} the band gap that opens at the Dirac points under laser driving with circularly polarized light was found to be significantly larger than the value that is na\"ively expected as one approaches the flat-band regime at the magic angle in TBG. It is these observations that prompt two key questions: Where does the LMC in flat-band systems come from? How is it influenced by the twist angle as a control parameter of band flatness? Both of these questions will be addressed in this paper.

The role of quantum geometry for the LMC in moir\'e materials such as TBG has not been studied, although connections between quantum geometry and coupling of electrons to electromagnetic fields have been found in other contexts. Following earlier work for first-principles computations of magnetic susceptibilities in insulators \cite{mauri_magnetic_1996,pickard_first-principles_2002,marques_response_2012}, it has been discussed how the quantum geometry of Bloch wave functions affects the orbital magnetic susceptibility~\cite{ogata_orbital_2015,piechon_geometric_2016,rhim_quantum_2020}, \REV{linear \cite{chen_probing_2021}} and nonlinear optical responses~\REV{\cite{sipe_second-order_2000,li_detection_2020,ahn_low-frequency_2020,ahn_riemannian_2021}}, and spin susceptibility of spin-orbit coupled superfluids~\cite{Iskin2018}, and enables Higgs spectroscopy in flat-band superconductors \cite{villegas2020anomalous}.

In this article, we demonstrate the key importance of quantum geometry for LMC in flat-band materials (Fig.~\ref{fig:Anchor}). We first derive fundamental relations between the LMC and quantum geometric quantities in a general multi-band system in Sec.~\ref{Sec:LMC_ToyModel}. We consider separately the linear and quadratic LMC (the para- and diamagnetic terms), and the intra- and inter-band contributions in each. In Sec.~\ref{Sec:Saw_tooth} we illustrate these general results within a simple model, namely the sawtooth quantum chain. Sec.~\ref{Sec.I} explores the LMC and its geometric properties in a tight-binding model system that captures the salient features of TBG and similar moir\'e materials. Finally, we conclude in Sec.~\ref{DISC}.

\begin{table*}
    \centering
    \vspace{5mm}
    \begin{tabular}{@{} l l l @{}}
        \toprule
        & \hspace{2mm} \textbf{Linear} ($\Amu$) & \hspace{2mm} \textbf{Quadratic} ($\Amu \Anu$) \BstrutSmall \TstrutSmall\\ \midrule
        \textbf{Intra-band} ($n$) \Tstrut \Bstrut & \hspace{2mm} $\dmu \varepsilon_n$ & \hspace{2mm}  $\dmu \dnu \varepsilon_n -\sum_{n \neq n'} (\varepsilon_n {-} \varepsilon_{n'})  \left(\langle \dmu n \ket{n'} \langle n' \ket{\dnu n} + \hc \right)$ \\[-0.6cm]
        \textbf{Inter-band} ($n$, $m$) \TstrutLarge \BstrutLarge & \hspace{2mm} $(\varepsilon_n - \varepsilon_m) \langle m \ket{\dmu n}$ & \hspace{2mm} $\begin{aligned}  & \left[ (\dmu \varepsilon_n - \dmu \varepsilon_m) \langle m \ket{ \dnu n} + \frac{1}{2} \varepsilon_m \langle \dmu \dnu m \ket{n} \right.  \end{aligned} $ \\[-0.6cm]
         &  & \hspace{2mm} $\begin{aligned}  & \left. + \frac{1}{2} \varepsilon_n \langle m \ket{\dmu \dnu n} + \sum_{n'} \varepsilon_{n'} \left( \langle \dmu m \ket{n'}\langle n' \ket{\dnu n}\right) \right] + (\mu \leftrightarrow \nu) \end{aligned} $ \\[0.4cm]
        \bottomrule
    \end{tabular}
     \caption{Linear and quadratic intra- and inter-band light-matter couplings. The  dependence on the quasi-momentum $\fk$ is not marked explicitly, it should be kept in mind that all the couplings are $\fk$-local. The light-matter couplings are determined not only by the bandstructure but also by the quantum geometric properties of the wavefunctions.} 
    \label{tab:LMCsSummarized}
\end{table*}

\section{Light-matter coupling in multi-band systems} \label{Sec:LMC_ToyModel}
To set the stage, we briefly recall LMC of free and band electrons in single-band settings. In a free electron gas, the coupling to an electromagnetic vector potential $\mathbf{A}(\mathbf{r},t)$ is given by the gauge-invariant minimal coupling prescription $\mathbf{p} \rightarrow \mathbf{p}-e\mathbf{A}$, leading to the kinetic energy $(\mathbf{p}-e\mathbf{A})^2/{2m}$. This implies the usual linear paramagnetic ($\mathbf{j} \cdot \mathbf{A}$) LMC proportional to the electronic current density $\mathbf{j}$, and the quadratic diamagnetic ($\frac{n e^2}{2m} \mathbf{A}^2$) LMC proportional to the electronic density $n$ and inversely proportional to the mass $m$ of the electrons. For band electrons with energy dispersion $\epsilon(\fk)$, the paramagnetic light-matter coupling is then dictated by the band velocity $\bf{v}(\fk) = \partial\epsilon(\fk)/\partial\fk$, whereas the diamagnetic coupling features the effective mass related to the inverse of the band curvature $\partial^2\epsilon(\fk)/\partial\fk^2$. Specifically, the na\"ive expectation for the LMC in flat-band systems is that LMC should vanish since both the band velocity and the band curvature are zero in a strictly flat band. Obviously this is the case in single flat bands corresponding to the atomic limit, and may happen also in multi-band systems. However, we will show in the following that in multi-band systems with specific geometric properties the LMC actually does not vanish. 

We now proceed to calculate the light-matter couplings (LMCs) for generic multi-orbital tight-binding models with the
Hamiltonian 
\begin{equation}
    H_0 = \sum_{i, j} \sum_{a, b} t_{a, b}(i, j) c^{\dagger}_{i, a} c_{j, b}\;.
    \label{eq:H0}
\end{equation}
Here $i$, $j$ are sites on a Bravais-lattice and $a$, $b$ are orbital indices. 
Furthermore, $c^{(\dagger)}$ are annihilation (creation) operators of electrons and $t$ denotes the hopping integral.
Throughout the paper we omit the spin of the electrons.
We couple this system to light via the Peierls substitution adding a phase to the hopping integral
\begin{equation}
    H = \sum_{i, j} \sum_{a, b} t_{a, b}(i, j)e^{i \int_{\mathbf{R}_{j, \REV{b}}}^{\mathbf{R}_{i, \REV{a}}} d r_{\mu}' A_{\mu}(\mathbf{r}',t)}  c^{\dagger}_{i, a} c_{j, b} ,
    \label{eq:HPeierls1}
\end{equation}
where $A_{\mu}(\mathbf{r}, t)$ is the component in $\mu$ direction of the electromagnetic vector potential in the Coulomb gauge.
We use natural units setting ${e=\hbar=c=1}$ and throughout the paper employ Einstein's summation convention. 
One can expand the exponential in Eq.~(\ref{eq:HPeierls1}) which yields
\begin{equation}
    \begin{aligned}
    H = H_0 + &\sum_{i, j} \sum_{a, b} t_{a, b}(i, j)\\ 
    &\left[ L^{A}_{\mu} A_{\mu} + \frac{1}{2} L^{AA}_{\mu \nu} A_{\mu} A_{\nu}  + \dots \right]  c^{\dagger}_{i, a} c_{j, b}\;,
    \end{aligned}
    \label{eq:HPeierls2}
\end{equation}
where we have defined the light-matter couplings as
\begin{equation}
    \begin{aligned}
    &L_{\mu}^{A} = \left( \partial_{A_{\mu}(\mathbf{r}, t)} H \right)|_{A = 0},\\
    &L_{\mu \nu}^{AA} = \left( \partial_{A_{\mu}(\mathbf{r},t)} \partial_{A_{\nu}(\mathbf{r}', t')} H \right)|_{A = 0}.
    \end{aligned}
    \label{eq:defLMCs}
\end{equation}
We denote these terms as linear and quadratic LMC, respectively. \REV{The LMC tensors are gauge dependent, but their absolute values are not, and it is the latter that are observable and influence various physical phenomena that depend on LMC \cite{morimoto_topological_2016}.}

In this work we are mainly interested in the long-wavelength limit and thus set $A_{\mu}(\mathbf{r}, t) = A_{\mu}(t)$ in what follows.
In this part we also suppress the time-dependence of the vector potential denoting it as $A_{\mu}$ as it is irrelevant to the derivations. 

We now diagonalize the Hamiltonian $H_0$ to
\begin{equation}
    H_0 = \int dk \, \psi^{\dagger}_{\fk} \, h(\fk) \, \psi_{\fk}\;,
\end{equation}
where $\psi^{(\dagger)}_{k}$ is an annihilation (creation) operator in the orbital basis and $h(\fk)$ is a matrix in orbital space that depends continuously on the quasi-momentum parameter $\fk$.
Performing a Fourier transform of the light-matter coupled Peierls Hamiltonian $H$ and assuming a spatially constant vector potential one obtains 
\begin{equation}
    H = \int dk \, \psi^{\dagger}_{\fk} \, h(\fk + \mathbf{A}) \, \psi_{\fk}\;.
\end{equation}
In this manner we may interpret the Hamiltonian and the LMCs as matrices that continuously depend on the quasi-momentum $\fk$ as a parameter.
We can thus calculate the matrix elements of the LMCs in a convenient way as
\begin{equation}
    \begin{aligned}
    L_{\mu, a b}^{A}(\fk) &= \bra{a} \left( \partial_{A_{\mu}} H(\fk) \right)|_{A = 0} \ket{b}\\
    &= \bra{a} \left( \partial_{k_{\mu}} H_{0}(\fk) \right) \ket{b}, \\
    L_{\mu \nu, a b}^{AA}(\fk) &= \bra{a} \left( \partial_{A_{\mu}} \partial_{A_{\nu}} H(\fk) \right)|_{A = 0} \ket{b} \\
    &= \bra{a} \left( \partial_{k_{\mu}} \partial_{k_{\nu}} H_{0}(\fk) \right) \ket{b} ,
    \end{aligned}
    \label{eq:LMCderivativeKOrbital}
\end{equation}
where $\ket{a}$ and $\ket{b}$ are again vectors in the orbital basis.
Through a standard basis transform the above equation can be written in any basis and not only the orbital one.
However, in order for Eq.~(\ref{eq:LMCderivativeKOrbital}) to hold one must only differentiate the k-dependence of the LMC with respect to the quasi-momentum $\fk$ and \emph{not} a possible k-dependence of the basis vectors.

In what follows we will be particularly interested in the matrix elements of the LMCs in the band basis -- i.e.~the eigenbasis of $h(\fk)$ of the non-light-matter coupled system.
We thus define
\begin{equation}
    \begin{aligned}
    &L_{\mu, n m}^{A}(\fk) = \bra{n(\fk)} \left( \partial_{k_{\mu}} H_0(\fk) \right)\ket{m(\fk)} \\
    &L_{\mu \nu, n m}^{AA}(\fk) = \bra{n(\fk)} \left( \partial_{k_{\mu}} \partial_{k_{\nu}} H_0(\fk) \right)\ket{m(\fk)}, 
    \end{aligned}
    \label{eq:defLMCsBands}
\end{equation}
where $\ket{n(\fk)}$ and $\ket{m(\fk)}$ are vectors in the band basis. In general, they will naturally depend on $\fk$. 
For $n = m$ we call the coupling \emph{intra}-band \REV{and keep only one index i.e.~writing $L_{\mu, n = m}^{A} \equiv L_{\mu, n}^{A}$ and similar for quadratic couplings.} For $n \neq m$ we call it \emph{inter}-band. Next, we calculate the general expressions for the linear and quadratic intra-band and inter-band couplings.
We abbreviate derivatives with respect to the quasi-momentum $k_{\mu}$ as $\partial_{k_{\mu}} \rightarrow \partial_{\mu}$. \REV{We note that in the case of a degenerate $k$-point $\fk^*$ the basis vectors and through them also the LMCs, as defined in Eq.~\ref{eq:defLMCs}, are not unique. In this case, one may evaluate the LMC by taking the limit $\fk \to \fk^*$.}

\subsection{Linear intra-band coupling}
\label{sec:LinIntra}
The linear intra-band coupling
\begin{equation}
    L_{n, \mu}^A(\fk) = \bra{n(\fk)} \left( \dmu H_0(\fk) \right) \ket{n(\fk)}
\end{equation}
can be calculated by utilizing the Schr\"odinger equation
\begin{equation}
    H_0(\fk) \ket{n(\fk)} = \varepsilon_n(\fk) \ket{n(\fk)} ,
    \label{eq:Schroedinger}
\end{equation}
where $\varepsilon_n(\fk)$ is the eigenvalue of $H_0(\fk)$ corresponding to the eigenvector $\ket{n(\fk)}$ at a specific k-point, i.e., the k-dependent dispersion of band $n$.
By performing a derivative of the equation with respect to $k_{\mu}$ and acting from the left with $\bra{n(\fk)}$ one obtains
\begin{equation}
    L_{n, \mu}^A(\fk) = \partial_{\mu} \varepsilon_n(\fk)
\end{equation}
Thus the linear intra-band LMC of band $n$ is simply given by its band velocity -- which is a well known result.
\subsection{Linear inter-band coupling}
\label{sec:LinInter}
In an analogous way we derive the linear inter-band coupling
\begin{equation}
    L_{m n, \mu}^A(\fk)_{m \neq n} = \bra{m(\fk)} \left( \dmu H_0(\fk) \right) \ket{n(\fk)}.
\end{equation}
We use the Schr\"odinger equation, Eq.~(\ref{eq:Schroedinger}), and perform a derivative with respect to $k_{\mu}$, but this time act with $\bra{m(\fk)} \neq \bra{n(\fk)}$ from the left.
This yields
\begin{equation}
    L^A_{m n, \mu}(\fk)|_{m \neq n} = (\varepsilon_n(\fk) - \varepsilon_m(\fk)) \langle m(\fk) \ket{\partial_{\mu} n(\fk)}
    \label{eq:LinearInter}
\end{equation}
The mathematical form of this result is known from calculations of other physical quantities: The superfluid weight of a multi-band system can be calculated by performing the Peierls substitution and expanding the vector potential to second order (including the paramagnetic and diamagnetic terms), and then evaluating the current-current response in its static and long-wavelength limit~\cite{liang_band_2017}. In such calculation, Eq.~(\ref{eq:LinearInter}) appears as the inter-band part of the paramagnetic current. However, the superfluid weight is distinct from this as it involves the current-current commutator, and also the diamagnetic term, as will be discussed below.  
\subsection{Quadratic intra-band coupling}
\label{sec:QuadraticIntra}
We continue with the quadratic intra-band coupling
\begin{equation}
    L_{n, \mu \nu}^{AA}(\fk) = \bra{n(\fk)} \left( \dmu \dnu H_0(\fk) \right) \ket{n(\fk)}. 
\end{equation}
Following analogous steps to the derivation of the linear case we find
\begin{equation}
\begin{aligned}
    &L_{n, \mu \nu}^{AA}(\fk) = \partial_{\mu} \partial_{\nu} \varepsilon_{n}(\fk) - \sum_{n', n' \neq n} (\varepsilon_{n}(\fk) {-} \varepsilon_{n'}(\fk)) \times\\
    & \times (\langle \partial_{\mu} n(\fk) \ket{n'(\fk)}\langle n'(\fk) \ket{\partial_{\nu} n(\fk)} + \hc)
    \label{eq:QuadraticIntra}
\end{aligned}
\end{equation}
Details of the derivation are given in App.~\ref{sec:calcQuadraticIntra}.

The quadratic inter-band coupling in band $n$ is thus given by two terms. The first is the curvature of the band as in case of a single band. The second term is a manifestly multi-band contribution and, interestingly, \REV{is similar to the quantum metric}
\begin{equation}
\begin{aligned}
    g^{n}_{\mu \nu}(\fk) &= \bra{\dmu n (\fk)} (1 - \ket{n(\fk)}\bra{n(\fk)}) \ket{\dnu n(\fk)} + \hc \\
    & = \sum_{n', n' \neq n} \langle \dmu n(\fk) \ket{n'(\fk)} \langle n'(\fk) \ket{\dnu n(\fk)} + \hc.
\end{aligned}
\end{equation}
Thus finite quantum metric may enable finite LMC even in a flat band where the effective mass is infinite (i.e.~the first term is zero). The quantum metric generates a "geometric" effective mass that can be finite. This has been previously noticed in different physical contexts. First, the two-body problem of two attractively interacting fermions in a flat band gives quantum-metric-dependent pair effective mass in certain limits~\cite{torma_quantum_2018}. In~\cite{Iskin19,Iskin2020_Goldstone}, an effective band mass was also calculated and the result is mathematically equivalent to Eq.~(\ref{eq:QuadraticIntra}) although the physical context of the result is different.

It is of interest to compare the result (\ref{eq:QuadraticIntra}) to the geometric contribution of multi-band superfluid weight $D^s_{\text{geom}, \mu \nu}$~\cite{liang_band_2017}:
\begin{equation}
    \begin{aligned}
    D^s_{\text{geom}, \mu \nu} &= \sum_{k, m, n} \left[ \frac{\tanh(\beta E_m {/} 2)}{E_m} - 
    \frac{\tanh(\beta E_n {/} 2)}{E_n}\right] \times \\
    & \times \frac{\Delta^2 (\varepsilon_n - \varepsilon_m)}{\varepsilon_n + \varepsilon_m - 2 \mu}%
    \left( \langle \partial_{\mu} m \ket{n} \langle n \ket{\partial_{\nu} m} + \text{h.c.} \right) .
    \end{aligned}
    \label{eq:GeometricalSuperfluidWeight}
\end{equation}
Here $E_n$ and $\Delta$ are the Bogoliubov eigenenergies and the order parameter of a superconducting system, respectively, $\beta$ the inverse temperature, and $\mu$ the chemical potential. Terms from the quantum metric multiplied by the band energy difference appear here too, but otherwise the form differs from Eq.~(\ref{eq:QuadraticIntra}). Physically, the two results describe distinct processes. The LMC in Eq.~(\ref{eq:QuadraticIntra}) corresponds solely to the diamagnetic term, while Eq.~(\ref{eq:GeometricalSuperfluidWeight}) contains the diamagnetic term and the product of two paramagnetic terms (the current-current commutator); for a band-integrated quantity, it is possible to combine these utilizing integration by parts. Furthermore, the superfluid weight depends on the properties of the superconducting ground state such as the order parameter $\Delta$.  

\subsubsection{Quadratic intra-band coupling in special cases: two-band systems, and lowest/highest flat bands}
For a two band system the second term in Eq.~(\ref{eq:QuadraticIntra}) becomes directly proportional to the metric. 
To see this let $E_g({\fk}) = \varepsilon_1(\fk) - \varepsilon_2(\fk)$ be the band-gap between the two bands.
Then the quadratic LMC into band $1$ is (the coupling into band $2$ follows in the same manner)
\begin{equation}
    L_{1, \mu \nu}^{AA}(\fk) = \dmu \dnu \varepsilon_1(\fk) - E_g({\fk}) g^{1}_{\mu \nu}(\fk) .
\end{equation}
Another interesting case is the intra-band coupling into an exactly flat band.
In that case the linear intra-band coupling vanishes identically as seen in Sec.~\ref{sec:LinIntra}.
The curvature term in Eq.~(\ref{eq:QuadraticIntra}) does not contribute either.
Thus, to quadratic order in $A$ the LMC (superscript $A,AA$ denoting that linear and quadratic order is included) will be given as
\begin{equation}
    \begin{aligned}
    &L_{n, \mu \nu}^{A, AA}(\fk) = - \sum_{n', n' \neq n} (\varepsilon_n(\fk) {-} \varepsilon_{n'}(\fk)) \times \\
    & \times \left(\langle \dmu n(\fk) \ket{n'(\fk)} \langle n'(\fk) \ket{\dnu n(\fk)} + \hc \right) \\
    \end{aligned}
\end{equation}
If the flat band is the one with the lowest or highest energy we can define a lower bound to the magnitude of the light-matter coupling:
\begin{equation}
   | L_{1/N, \mu \mu}^{A, AA} | \geq \tilde{E}_g(\fk) g^{1/N}_{\mu \mu}(\fk) . \label{lowerbound}
\end{equation}
Here $\tilde{E}_g(\fk)$ denotes the separation between the considered flat band one ($N$) and the next higher (lower) lying band at each $k$-point.
$g^{1(N)}_{\mu \mu}(\fk)$ denotes the diagonal elements of the quantum metric in band one (N), which are always positive. Although finite quantum metric suggests the possibility of finite LMC also in the off-diagonal case, a relation similar to Eq.~(\ref{lowerbound}) cannot be derived. The bound Eq.~(\ref{lowerbound}) is meaningful nevertheless, since often, for instance in our TBG LMC study below, the diagonal light-matter couplings are the relevant ones. 

\subsection{Quadratic inter-band coupling}
\label{sec:QuadraticInter}
Last, we calculate the quadratic inter-band coupling
\begin{equation}
    L_{m n, \mu \nu}^{AA}|_{m \neq n}(\fk) = \bra{m(\fk)} \left( \dmu \dnu H_0(\fk) \right) \ket{n(\fk)}
\end{equation}
By a similar approach as above, we find
\begin{equation}
\begin{aligned}
    &L_{m n, \mu \nu}^{AA}|_{m \neq n}(\fk) =\\
    &\left(\dmu \varepsilon_n(\fk) - \dmu \varepsilon_m(\fk)\right) \langle m(\fk) \ket{\partial_\nu n(\fk)}\\
    & + \frac{1}{2} \varepsilon_m(\fk) \langle \dmu \dnu m(\fk) \ket{n(\fk)} + \frac{1}{2} \varepsilon_n(\fk) \langle m(\fk) \ket{\dmu \dnu n(\fk)}\\
    & + \sum_{n'} \varepsilon_{n'(\fk)} \left( \langle \dmu m(\fk) \ket{n'(\fk)}\langle n'(\fk) \ket{\dnu n(\fk)} \right)\\
    & + \mu \leftrightarrow \nu.
\end{aligned}
\end{equation}
Here the first term contains the inter-band Berry connection, defined as $\textbf{A}_{mn} (\fk) = i \langle m(\fk) \ket{\partial_{\fk} n(\fk)}$.
Products of components of the inter- and intra-band Berry connection also appear under the sum in the last term.
The Berry connection is not a gauge-independent quantity, which implies that the quadratic inter-band LMC depends on gauge, too, which in fact is the case.

We summarize our findings on the LMC in  Tab.~\ref{tab:LMCsSummarized}.

\section{Saw-tooth chain}\label{Sec:Saw_tooth}
\begin{figure*}
    \centering
    \includegraphics{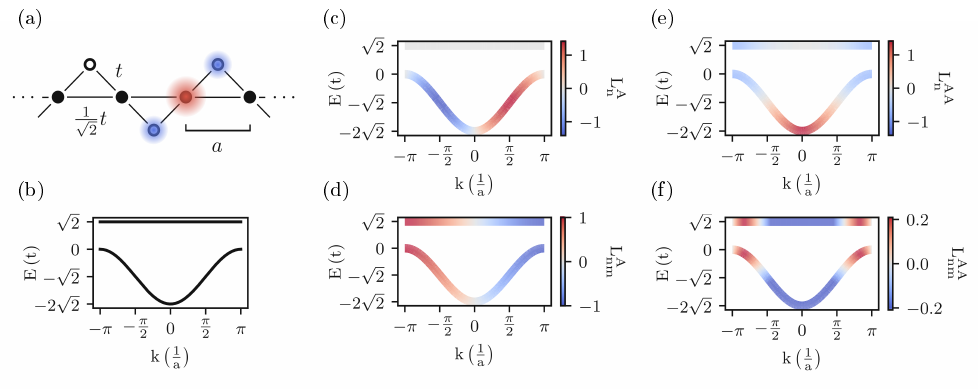}
    \caption{The sawtooth chain exhibiting a flat and a dispersive band, and its light-matter couplings.
    (a) Sketch of the model. We illustrate the destructive interference that causes the flat band by plotting its Bloch wavefunction on three adjacent sites. The amplitude is denoted by the radius of the colored circle (The amplitude on the full black dot is $\sqrt{2}$-times that of the amplitude on the empty circle). The sign is determined by the color: red for positive and blue for negative. (b) The bare bandstructure exhibiting a completely flat and a dispersive band separated by a gap. (c) Linear intra-band coupling. This quantity vanishes identically in the completely flat band. (d) Linear inter-band coupling. Note that this is a gauge dependent quantity. We have chosen the same gauge for the plot as in the text, Eq.~(\ref{eq:sawtoothLinearCouplings3}). (e) Quadratic intra-band coupling. Importantly, this is non-zero even in the completely flat band. (f) Quadratic inter-band coupling. We have again chosen the same gauge as in the text Eq.~(\ref{eq:sawtoothQuadraticCouplings3}).}
    \label{fig:sawtoothChain}
\end{figure*}
Particle localization in a periodic system due to strong confinement, so-called atomic limit, leads to a trivial flat band. Non-trivial flat bands can emerge in multi-orbital lattices, without strong confinement, due to interference effects~\cite{Leykam:2018}. Suitable unit cells containing several sites can be provided, for instance, by geometry as in the case of the famous Lieb~\cite{Lieb:1989} and kagome lattices. Another possibility is to use magnetic fields or artificial gauge potentials which create unit cells determined by the magnetic flux; an obvious example are Landau levels. Quasi-one-dimensional ladder systems exhibit flat bands as well. The essence of all such systems is that it is possible to have eigenstates which are located at some sites of the lattice with suitable wavefunction phases so that hopping to neighboring sites is prevented by destructive interference. 

One of the simplest models offering a flat band is the sawtooth ladder or quantum chain~\cite{Huber2010_sawtooth, Zhang2015_sawtooth}, which we use here to illustrate our findings on LMC. 
The Hamiltonian of the sawtooth chain, depicted in Fig.~\ref{fig:sawtoothChain}(a), is
\begin{equation}
    \begin{aligned}
    H_0 = & - \sum_i t \, c_{i, B}^{\dagger} c_{i, A} + \hc - \sum_i t \, c_{i + 1, A}^{\dagger} c_{i, B} + \hc\\
    & - \sum_i \frac{t}{\sqrt{2}} c_{i + 1, A}^{\dagger} c_{i, A} + \hc   .
    \end{aligned}
\end{equation}
The dispersion of the resulting two bands reads
\begin{equation}
    \begin{aligned}
        & \varepsilon_1(k) = -\sqrt{2}t (\cos(k a) + 1)\\
        & \varepsilon_2(k) = \sqrt{2}t ,
    \end{aligned}
    \label{eq:dispersionsSawtooth}
\end{equation}
This bandstructure is shown in Fig.~\ref{fig:sawtoothChain}(b).
The bands are separated by the $k$-dependent gap
\begin{equation}
    E_g(k) = \sqrt{2}t (\cos(k a) + 2)
    \label{eq:sawtoothGap}
\end{equation}
We also explicitly state the metric for later comparison to the LMCs. It is equal for both bands and reads
\begin{equation}
    g^{1/2}(k) = \frac{\sin(\frac{k \, a}{2})^2}{2(\cos(k a) + 2)^2} .
    \label{eq:sawtoothMetric}
\end{equation}
We next calculate the linear intra- and inter-band LMCs.
\REV{Since the inter-band couplings are only determined up to a phase (similar to all transition matrix elements) we explicitly state the eigenfunctions $v_1$ and $v_2$ effectively fixing the gauge}
\REV{\begin{equation}
    \begin{aligned}
       v_1 = \frac{1}{\sqrt{2 \cos(\frac{1}{2}k)^2 + 1}} %
    \begin{pmatrix}%
    \sqrt{2} \cos(k/2) \\%
    1%
    \end{pmatrix}\\
       v_2 = \frac{1}{\sqrt{2 \cos(\frac{1}{2}k)^2 + 1}} %
    \begin{pmatrix}%
    -1 \\%
    \sqrt{2} \cos(k/2)%
    \end{pmatrix} .
    \end{aligned}
    \label{eq:sawtoothWaveFunctions}
\end{equation}}
\REV{With these the linear light-matter couplings read}
\begin{align}
    & L_{1}^{A}(k) = \sqrt{2}t \sin(k a), \label{eq:sawtoothLinearCouplings1}\\
    & L_{2}^{A}(k) = 0, \label{eq:sawtoothLinearCouplings2}\\
    & L_{12}^{A}(k) = -t \sin\left(\frac{k a}{2}\right) \label{eq:sawtoothLinearCouplings3} .
\end{align}
As expected, the linear LMC inside the second band, Eq.~(\ref{eq:sawtoothLinearCouplings1}), vanishes due to the vanishing band velocity.
At the same time, the linear LMC inside the dispersive band, Eq.~(\ref{eq:sawtoothLinearCouplings2}), can easily be read off as the derivative of the band dispersion, Eq.~(\ref{eq:dispersionsSawtooth}).
Note that the linear inter-band coupling is not gauge-invariant; we have fixed the gauge for the shown result Eq.~(\ref{eq:sawtoothLinearCouplings3}).
The linear intra-band and inter-band LMCs are plotted in Fig.~\ref{fig:sawtoothChain} (c) and (d), respectively.

We continue by calculating the quadratic LMCs and obtain
\begin{align}
     L_{1}^{AA}(k) &= \sqrt{2}t \cos(k a) + \sqrt{2}t (\cos(k a) + 2) \times \\
    & \hspace{15mm} \times \frac{\sin(\frac{k a}{2})^2}{2 (\cos(k a) + 2)^2}, \label{eq:sawtoothQuadraticCouplings1}\\
    & = \sqrt{2}t \cos(k a) + E_g(k) g^{1/2}(k) \\
     L_{2}^{AA}(k) &= - \sqrt{2}t (\cos(k a) + 2) \times \frac{\sin(\frac{k a}{2})^2}{2 (\cos(k a) + 2)^2}, \label{eq:sawtoothQuadraticCouplings2}\\
    & = - E_g(k) g^{1/2}(k) \\
    L_{12}^{AA}(k) &= \frac{1}{2}t \cos \left(\frac{k a}{2} \right) \frac{5 - 2 \cos(\frac{k \, a}{2})}{1 + 2 \cos(\frac{k \, a}{2})} . \label{eq:sawtoothQuadraticCouplings3}
\end{align}
Indeed the quadratic LMC inside the flat band, Eq.~(\ref{eq:sawtoothQuadraticCouplings2}), is given by the negative energy gap Eq.~(\ref{eq:sawtoothGap}) multiplied by the metric of the band Eq.~(\ref{eq:sawtoothMetric}).
The quadratic LMC inside the dispersive band, on the other hand, has the same contribution with a flipped sign.
In addition, the band curvature contributes to the quadratic LMC Eq.~(\ref{eq:sawtoothQuadraticCouplings2}).
Again, the inter-band coupling, Eq.~(\ref{eq:sawtoothQuadraticCouplings3}), is not gauge-invariant and we fixed the gauge for the form shown here.
We illustrate the quadratic intra- and inter-band LMCs in Fig.~\ref{fig:sawtoothChain} (e) and (f), respectively.

\section{Twisted bilayer graphene}\label{Sec.I}
We analyze LMCs in magic-angle ($\Theta_\text{M}=1.05^\circ$) twisted bilayer graphene (MATBG) as a prototypical flat band moir{\'e} material, utilizing our analytical results from Sec.~\ref{Sec:LMC_ToyModel}. In the first part, after introducing the full-unit cell tight-binding Hamiltonian, we explore the static light-matter coupled Hamiltonian. In the second part we employ Floquet theory to investigate dynamical LMC effects in the high-frequency driving regime. 

\subsection{Full unit-cell tight-binding model}\label{SSec.A}
The TBG tight-binding Hamiltonian has the general form
\begin{equation}
    H_0 = \sum_{i, j} \sum_{a, b} t_{a, b}(i, j) c^{\dagger}_{i, a} c_{j, b},
\end{equation}
where the indices $i$, $j$ denote the Bravais lattice and the indices $a$, $b$ refer to the $p_\text{z}$-orbitals of the carbon atoms within both layers. We assume Slater-Koster hopping matrix elements of the general form:
\begin{align}
    t_{a,b}(i,j) =& t_0 \exp{\left[-\beta \frac{r-b}{b}\right]}\frac{x^2+y^2}{r^2}\notag\\&+t_1 \exp{\left[-\beta \frac{r-c_0}{b}\right]}\frac{z^2}{r^2}, \label{Eq:Hopping_TBG}
\end{align}
where $r=||(x,y,z)^T||$ denotes the distance between sites $\bm{r}_{ia}$ and $\bm{r}_{jb}$. We choose our model parameters according to \cite{julku_superfluid_2020}. The parameters $b=a_0/\sqrt{3}$, $a_0=2.46$ $\mathrm{\AA}$, and $c_0=3.35$ $\mathrm{\AA}$ denote the nearest-neighbour distance of the carbon atoms, the monolayer lattice constant, and the interlayer distance, respectively. For our model calculations we assume an intralayer hopping $t_0=-2.7$ eV, an interlayer hopping $t_1=0.297$ eV, and $\beta=7.2$ as fitting parameter for the exponential decay of the hopping matrix elements.

\begin{figure}
    \centering
    \includegraphics[scale=0.49]{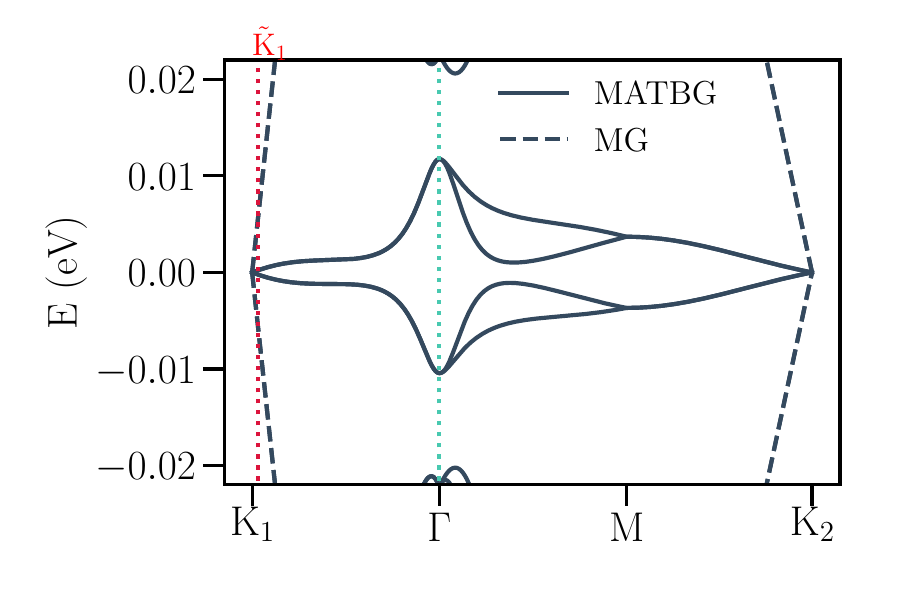}
    \caption{Low-energy electronic bandstructure of the magic-angle twisted bilayer graphene (MATBG) model ($\Theta_\text{M}=1.05^\circ$) along the two Dirac points of the mini Brilluoin zone. Black dashed lines show the monolayer Graphene (MG) Dirac bands. The red and turquoise dotted lines indicate the two different momenta considered in Sec.~\ref{SSec.B}.} 
    \label{fig:EQ_BANDS_TBG}
\end{figure}

Fig.~\ref{fig:EQ_BANDS_TBG} shows the electronic bandstructure of the TBG model Hamiltonian at the first magic angle ${\Theta_\text{M}=1.05^\circ}$. The reference monolayer graphene Dirac bands (black dashed lines) illustrate the strong band-flattening effect that originates from the interlayer-coupling-induced back-folding to the mini Brilluoin zone corresponding to a moir{\'e} super cell of $N=11908$ lattice sites. The red (turquoise) dotted line indicates a quasi-momentum cut in close proximity to the Dirac point (at $\Gamma$) for which we investigate the LMC in the following section. 

\subsection{Light-matter coupling}\label{SSec.B}
We explore electronic LMC of the nearly-flat band manifold in MATBG. In accordance with our analytic model investigations of Secs.~\ref{Sec:LMC_ToyModel} and \ref{Sec:Saw_tooth}, we couple a spatially homogeneous external vector potential $\bm{A}$ via Peierls substitution. This introduces gauge phase factors ${t_{a,b}(i,j)\rightarrow t_{a,b}(i,j)\exp{\left[\I g \bm{A}(\bm{r_{i,a}}-\bm{r_{j,b}})\right]}}$ to the hopping elements of Eq.~(\ref{Eq:Hopping_TBG}). Expanding the field-dependent hopping elements up to second order, we define the LMC elements of the Fourier transformed Hamiltonian in the original band basis ${H_0(\fk)\ket{n(\fk)} = \varepsilon_n (\fk)\ket{n(\fk)}}$ according to Eqns.~(\ref{eq:defLMCsBands}) as
\begin{equation}
    \begin{aligned}
    &L_{m n, \mu}^{A}(\fk) = \bra{m(\fk)} \left( \partial_{\mu} H_0(\fk) \right)\ket{n(\fk)}, \\
    &L_{m n, \mu \nu}^{AA}(\fk) = \bra{m(\fk)} \left( \partial_{\mu} \partial_{\nu} H_0(\fk)\right)\ket{n(\fk)}. 
    \end{aligned}
    \label{EQ:LMC_TBG}
\end{equation}
We characterize the LMC of the nearly flat electronic bands by introducing the three quantities 
\begin{equation}
    \begin{aligned}
        \mathrm{LMC}^{m=n}_{A}(\fk) &=& \frac{1}{4} \sum \limits_{m} \sqrt{\sum_\mu|L_{\mu, m m}^{A}(\fk)|^2}, \\
        \mathrm{LMC}^{m \neq n}_{A}(\fk) &=& \frac{1}{4} \sum \limits_{mn,m \neq n} \sqrt{\sum_\mu|L_{\mu, m n}^{A}(\fk)|^2}, \\
        \mathrm{LMC}^{m=n}_{AA}(\fk) &=& \frac{1}{4} \sum \limits_{m} \sqrt{\frac{1}{2}\sum_{\mu\nu}|L_{\mu\nu, m m}^{A}(\fk)|^2}. \\
    \end{aligned}
\end{equation}    
The first expression $\mathrm{LMC}^{m=n}_{A}$ defines a quantitative measure of the linear-field intra-band coupling. The index $m \in \{-2,-1,0,+1\}$ runs over the four flat bands (we choose a counting where the first conduction band has band index $m=0$). As we only consider fields within the x-y-plane we have two spatial indices $\mu,\nu \in \{x,y\}$. The second quantity $\mathrm{LMC}^{m\neq n}_{A}$ provides a measure for the linear-field inter-band coupling. The indices $m \neq n \in \{-4,-3,-2,-1,0,+1,+2,+3\}$ run over the four flat bands and the two higher and lower lying dispersive bands. For computational simplicity, we neglect the coupling to higher and lower lying bands at this point; the limited number of bands however is sufficient for making the important qualitative points, discussed below. The last expression, $\mathrm{LMC}^{m=n}_{AA}$, quantifies the quadratic-field intra-band coupling. Again, the band index $m \in \{-2,-1,0,+1\}$ runs over the four flat bands. 
\REV{We do not present the quadratic inter-band coupling as our numerical analysis has shown that, for the chosen driving regime, it has no impact on our results presented in Sec.~\ref{Sec.:C}.}

\begin{figure*}
    \centering
    \includegraphics[width=\textwidth]{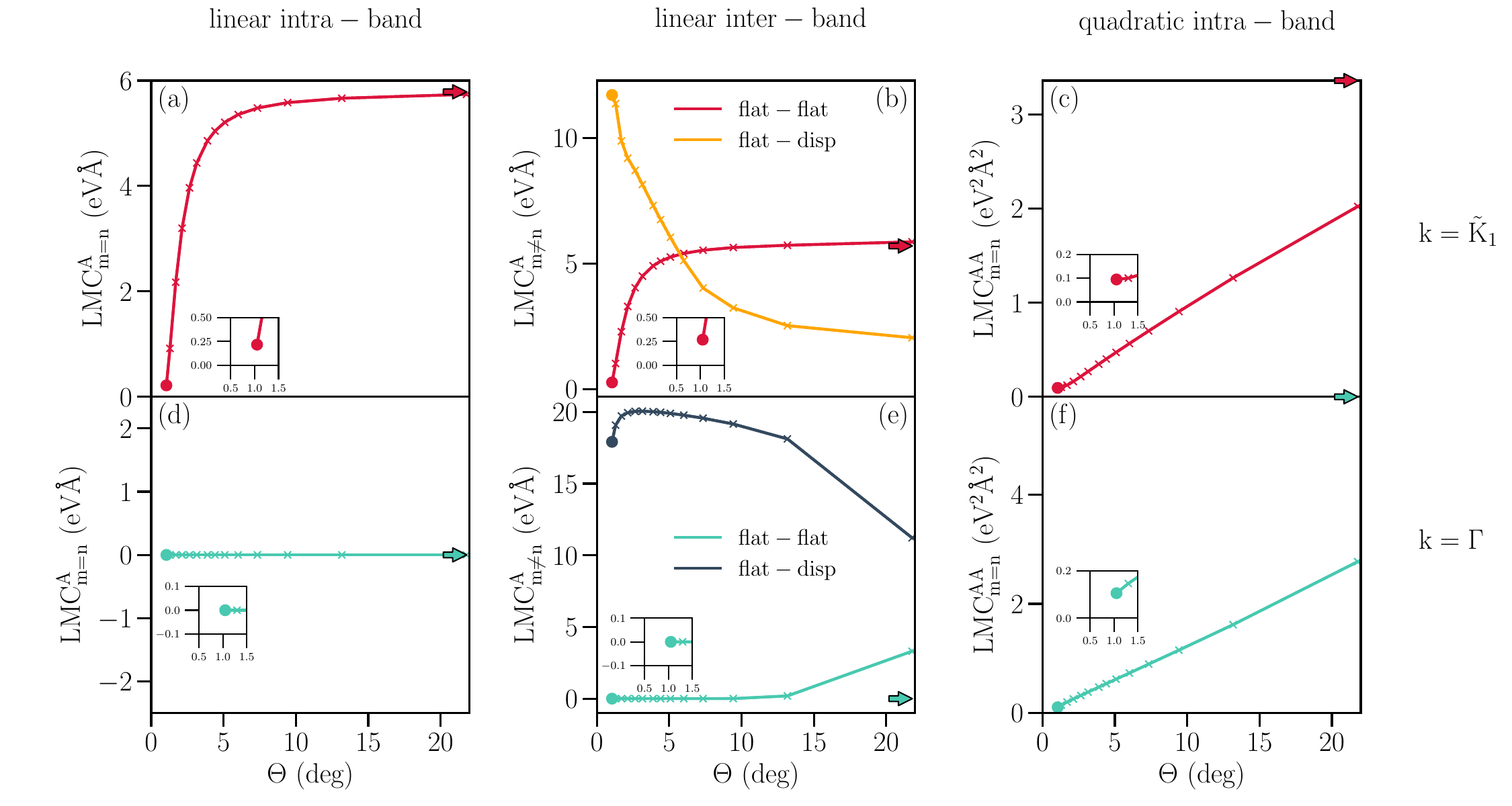}
    \caption{LMCs of the Peierls-substituted MATBG Hamiltonian. (a)-(c) show the LMCs at $\mathrm{\tilde{K_1}}$ in the linear regime of the first Dirac cone around $\mathrm{K_1}$. (d)-(f) show the LMCs at the $\mathrm{\Gamma}$-point. Coloured dots indicate the magic-angle values that are presented on a finer grid in the insets. Coloured arrows indicate the monolayer graphene reference values. (a),(d) provides a measure of the linear-field intra-band coupling (band velocity) summed over the four flat bands. (b),(e) show the linear-field inter-band couplings. The red and turquoise lines indicate the coupling between the four flat bands. The orange and black line provide a measure of the coupling of the four flat bands to the two higher and two lower lying dispersive bands. (c),(f) measure of the quadaratic-field intra-band coupling.} 
    \label{fig:STATIC_COUPLING_TBG}
\end{figure*}

The numerical results for these quantities for the field-coupled Hamiltonian are presented in Fig.~\ref{fig:STATIC_COUPLING_TBG}. The upper row, Fig.~\ref{fig:STATIC_COUPLING_TBG}(a-c), shows the LMCs at a momentum $\tilde{K}_1$ within the linear region close to the first Dirac point. We avoid the exact Dirac point as the band degeneracy imposes a vanishing inter-band coupling between the flat bands (see Tab.~\ref{tab:LMCsSummarized}). Moreover, the second k-derivatives vanish at the Dirac point (see App.~\ref{Sec:LMC_DIRAC}). The lower row illustrates the same couplings at the Brilluoin zone center $\Gamma$. Colored arrows indicate reference values of two uncoupled graphene sheets, which can easily be accessed by switching off the interlayer coupling in our model. Within the Dirac cone, the decrease of the band velocity (interlayer coupling) with twist angle reflects the well-known quenching of the bandwidth when approaching the magic angle from above. For increasing twist angles the band velocity approaches that of monolayer graphene (Fig.~\ref{fig:STATIC_COUPLING_TBG}(a)). At the $\mathrm{\Gamma}$-point, the band velocity vanishes for all twist angles as it does for graphene (Fig.~\ref{fig:STATIC_COUPLING_TBG}(d)). 

To investigate the linear-field inter-band coupling, we consider the coupling within the flat bands and the coupling between the flat bands and the two higher and the two lower lying dispersive bands separately. At the Dirac cone, the coupling within the flat bands shows the same angle-dependency as the intra-band coupling, which is a direct consequence of the form of the Dirac Hamiltonian (see App.~\ref{Sec:LMC_DIRAC}). However, while the inter-band coupling within the flat bands becomes very small at the magic angle, we find an increasingly pronounced coupling to the higher and lower lying dispersive bands for small twist angles (Fig.~\ref{fig:STATIC_COUPLING_TBG}(b)). This significant inter-band coupling between the four flat bands and the other bands is the central result of this section as it provides, at first glance counter-intuitively, a strong light-matter engineering channel at the Dirac points despite vanishingly small Fermi velocities. Importantly, the true inter-band coupling could be even higher, as we restrict the summation to the four nearest lying dispersive bands. \REV{The significance of the inter-band coupling manifests in two ways. First, one can directly compare its absolute LMC amplitude at the magic angle to either the inter-band coupling between the flat bands or to the linear-field intraband coupling (Fig.~\ref{fig:STATIC_COUPLING_TBG}(a)). These are quantities that dominate the LMC in monolayer graphene (see red arrows in Figs.~\ref{fig:STATIC_COUPLING_TBG} (a), (b)). Moreover, the interband coupling is significant indirectly, as we will demonstrate in Sec.~\ref{Sec.:C}: the experimentally observable light-induced Floquet gap at the Dirac points is much larger than expected from naive Dirac band physics and strongly depends on this interband-coupling to higher and lower lying dispersive bands.}

The qualitative behaviour of the linear-field inter-band coupling at the Brillouin zone center is similar (Fig.~\ref{fig:STATIC_COUPLING_TBG}(e)). Again the coupling between the flat bands is very small while the coupling to the dispersive bands becomes very strong towards approaching the magic angle. Interestingly, the coupling within the flat bands seems to reach a local maximum for an intermediate twist angle at the $\Gamma$-point. These results highlight the limited applicability of the four-band models often used for describing TBG. This general coupling mechanism between flat bands could play an important role within measured strong midinfrared light-matter responses in small angle TBG \cite{deng_strong_2020}.

The quadratic intra-band coupling increases monotonically with the twist angle both close to the Dirac point and at the Brillouin zone center. Note that since we look not exactly at the Dirac point, the second derivatives in k can have finite values in contrast to the Dirac Hamiltonian itself (see App.~\ref{Sec:LMC_DIRAC}). At the magic angle the quadratic intra-band coupling is finite but small.  

\subsection{Floquet engineering}\label{Sec.:C}
Exploiting the significant LMC of the quasi-flat MATBG bands shown in the previous section, we now investigate the effect of a time-periodic external field on the MATBG electronic bandstructure via the Floquet formalism. In particular, we explore two different effects of a circularly polarized laser field. The first effect is the opening of a topological band gap $\Delta_K$ at the Dirac points, that is a direct consequence of broken time-reversal symmetry \cite{oka_photovoltaic_2009,PhysRevB.91.155422, topp_topological_2019, mciver_light-induced_2020, sentef_theory_2015, sato_microscopic_2019, nuske_floquet_2020}. The second effect is a bandwidth renormalization of the low-energy manifold at the Brilluoin zone center $\Delta_\Gamma$, usually referred to as dynamical localization. We quantify these effects as a function of the field strength and trace back their origin to different parts of the LMC Hamiltonian.      
By coupling a time-periodic laser field, the Hamiltonian itself becomes time dependent ${H(t)=H(t+T)}$ with time period $T$. Expanding up to 2nd order in the driving field, the matrix elements of the time-dependent Hamiltonian in the original band basis take the general form
\begin{equation}
    \begin{aligned}
   \bra{m} H(t) \ket{n} = \varepsilon_n\delta_{mn} 
   &+ \sum\limits_\mu L^{A}_{\mu, mn}  A_{\mu}(t) \\
   &+ \frac{1}{2}\sum\limits_{\mu\nu} L^{AA}_{\mu \nu, mn} A_{\mu}(t)A_{\nu}(t).
    \end{aligned}
    \label{eq:HPeierls3}
\end{equation}
Note that for simplicity we drop the explicit k-dependency from here on. By mapping the periodic time-evolution to a quasi-static eigenvalue problem via a discrete Fourier transform in time, the eigenspectrum of the driven system can be described by a Floquet matrix
\begin{eqnarray}
    \mathcal{H}^{\alpha\beta} = \frac{1}{T} \int \limits_0^T \D \REV{t} H(t) e^{\I (\alpha-\beta)\Omega t} + \delta_{\alpha\beta}\alpha\Omega,
    \label{Eq:FLOQUET_MAT}
\end{eqnarray}
where the indices $\alpha$ and $\beta$ span a multi-photon Hilbert space with single-photon energy $\Omega=\frac{2\pi}{T}$. For the \REV{experimentally relevant} frequency regime ($\Omega=\REV{3}$ eV) and moderate field strengths, we find converged Floquet results by keeping contributions up to the linear photon order (see Fig.~\ref{fig:FLOQ_GAP_CON_M} in App.~\ref{sec:Floquet convergence}). However, even in the single-photon limit the huge size of the full moir{\'e} supercell yields a Floquet matrix of leading order $3\times 11908$, which is utterly challenging to treat numerically. To effectively reduce the dimension of the problem, we introduce a band cutoff $N_\text{c}$ to the full time-dependent Hamiltonian (\ref{eq:HPeierls3}) in the original band basis. Instead of all 11908 bands, we consider only a number of $N_\text{c}$ bands with band indices $m, n$ running from $\left[ -N_\text{c}/2,N_\text{c}/2-1\right]$. As shown in Fig.~\ref{fig:FLOQ_GAP_CON} of App.~\ref{sec:Floquet convergence}, a band cutoff $N_\text{c}=512$ yields converged results.

In order to disentangle distinct LMC contributions to the light-induced Floquet gap at $K$ and the band renormalization at $\Gamma$, we employ an effective downfolded Floquet Hamiltonian \cite{bukov_universal_2015} via a high-frequency expansion in $\Omega$. Up to order $\Omega^{-1}$ this yields
\begin{eqnarray}
    H_\textrm{eff} = \mathcal{H}^{00} + \frac{1}{\Omega}\sum \limits_{l=1}^{\infty} \frac{1}{l}\left[\mathcal{H}^{0-l},\mathcal{H}^{0+l} \right]. \label{Eq:Heff}
\end{eqnarray}
As a great advantage of the above expression, the zero-photon sector of the Floquet Hamiltonian now decouples from the higher photon orders, recovering the original Hilbert space of the unperturbed Hamiltonian. This allows to straightforwardly track down these parts of the effective Hamiltonian that dominantly contribute to light-induced effects exhibited by the Floquet bandstructure. We keep terms up to $l\le2$ and assume circular field polarization within the x-y plane of the form ${\mathbf{A}(t) = A_0(\sin{\Omega t}, \cos{\Omega t}, 0)}$. 

In the following, we investigate different contributions to the effective Floquet Hamiltonian Eq.~(\ref{Eq:Heff}) from distinct LMC orders up to second order. Again, ${H_0 \ket{m} = \epsilon_m \ket{m}}$ indicates the eigenbasis of the field-free Hamiltonian.

\paragraph{Zeroth order}
The zeroth order contribution corresponds to the field-free Hamiltonian
\begin{eqnarray}
    \mathcal{H}^{00}_0 &=& H_0 \label{Eq:H000}.
\end{eqnarray}
 
\paragraph{Linear coupling $L_{\mu, n m}^{A}$} Taking only linear-field LMC into account the expressions appearing in Eq.~(\ref{Eq:Heff}) can be read off as
\begin{equation}
    \begin{aligned}
    \mathcal{H}^{00}_{A} &= 0, \\ 
    \mathcal{H}^{0-1}_{A} &= \frac{A_0}{2}\left( \partial_y H_0 + \I \partial_x H_0 \right),  \\
    \mathcal{H}^{0+1}_{A} &= \frac{A_0}{2}\left(  \partial_y H_0 - \I \partial_x H_0 \right),  \\
    \mathcal{H}^{0-2}_{A} &= \mathcal{H}^{0+2}_{A} = 0. \label{Eq:H0+2_A} 
    \end{aligned}
\end{equation}    
Plugging Eqns.~(\ref{Eq:H0+2_A}) into Eq.~(\ref{Eq:Heff}) and employing the definitions of Eqns.~(\ref{EQ:LMC_TBG}) yields  
\begin{eqnarray}
   \bra{m} H^{A}_\textrm{eff} \ket{n} = \frac{\I A_0^2}{2\Omega} \left[ \sum_l L^{A}_{x,ml}L^{A}_{y,ln} - L^{A}_{y,ml}L^{A}_{x,ln} \right] \label{Eq:MaEl1_A}
\end{eqnarray}
for the matrix elements of the effective Floquet Hamiltonian. 

\paragraph{Quadratic coupling $L_{\mu\nu, n m}^{AA}$ } Considering solely quadratic-field LMC yields
\begin{equation}
    \begin{aligned}
   \mathcal{H}^{00}_{AA} &= \frac{A_0^2}{4}\left[ \partial^2_x H_0 + \partial^2_y H_0  \right], \\
   \mathcal{H}^{0-1}_{AA} &=  \mathcal{H}^{0+1}_{AA} = 0, \\
   \mathcal{H}^{0-2}_{AA} &= \frac{A_0^2}{8}\left[ -\partial^2_x H_0 + 2\I\partial_x\partial_y H_0 +\partial^2_y H_0 \right], \\
   \mathcal{H}^{0+2}_{AA} &= \frac{A_0^2}{8}\left[ -\partial^2_x H_0 - 2\I\partial_x\partial_y H_0 +\partial^2_y H_0 \right]. \label{Eq:H0+2_AA}
    \end{aligned}
\end{equation}      
 Inserting Eqns.~(\ref{Eq:H0+2_AA}) into Eq.~(\ref{Eq:Heff}) we obtain
\begin{eqnarray}
   \bra{m} H^{AA}_\textrm{eff} \ket{n} =& \frac{A_0^2}{4}\left[ L_{xx, mn}^{AA} +  L_{yy, mn}^{AA} \right]
   &+\mathcal{O}(A_0^4)
   \label{Eq:LMC_AA_HEFF}
\end{eqnarray}
for the quadratic-field contribution to the matrix elements of the effective Floquet Hamiltonian.

\begin{figure*}
    \centering
    \includegraphics[scale=0.60]{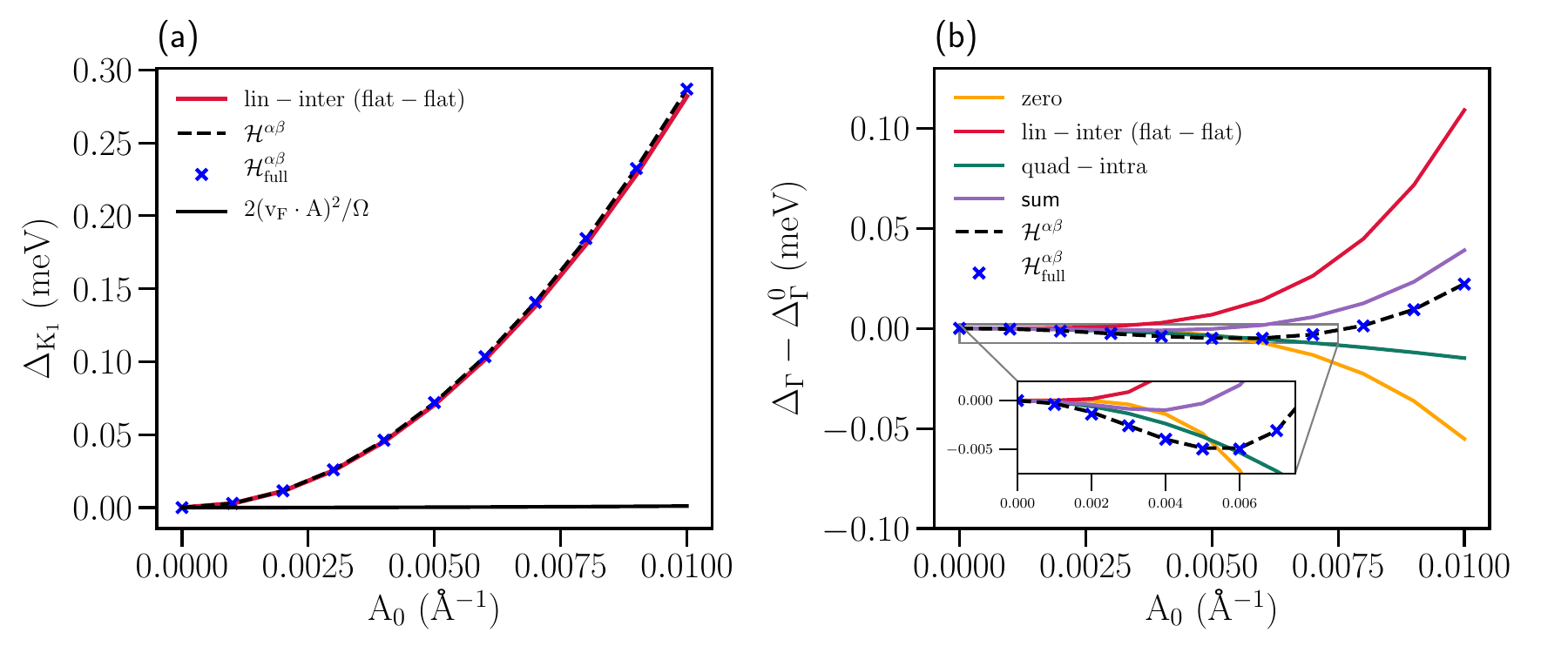}
    \caption{Floquet band effects as function of the driving amplitude $A_0$ at a driving frequency of $\Omega=\REV{3}$ eV. The coloured curves illustrate the amplitude dependency of the dominant contribution to the gap renormalization obtained from the effective HFA Floquet Hamiltonian, $H_\text{eff}$. \REV{The dashed black lines indicate the values obtained from the band-truncated Floquet matrix $\mathcal{H}^{\alpha\beta}$ up to quadratic field order. Blue crosses indicate reference  values obtained from the full Floquet matrix without band truncation with full exponential LMC.} (a) shows a quadratic gap opening at the Dirac point $K_1$ as function of the driving amplitude. The dominant contribution clearly stems from the linear-field inter-band matrix elements between the flat bands of the effective Floquet Hamiltonian (lin-inter, see Eq.~(\ref{Eq:MaEl1_A}) with $m\neq n \in \{-2,1,0,1\}$). The black solid line indicates the gap value obtained from a na\"ive 2-band approximation with the Moire renormalized Fermi velocity of $v_F=0.23$ $\mathrm{eV}/\mathrm{\AA}$. \REV{We find excellent agreement between the HFA and the full Floquet calculations.} (b) shows the bandwidth renormalization of the flat band manifold at the $\Gamma$-point. For small driving amplitudes $A_0\le \REV{0.004}$ $\mathrm{\AA}^{-1}$ the bandwidth is dominated by the quadratic intra-band coupling (quad-intra, see Eq.~(\ref{Eq:LMC_AA_HEFF}) with $m=n \in \{-2,1,0,1\}$), which leads to a reduction of the bandwidth. For higher driving amplitudes the bandwidth is dominated by the linear-field inter-band coupling between the flat bands (lin-inter, see Eq.~(\ref{Eq:MaEl1_A}) with $m\neq n \in \{-2,1,0,1\}$), which induces a trend towards an increasing bandwidth. The yellow curve shows the bandwidth contribution from the unperturbed Hamiltonian (zero, see Eq.~(\ref{Eq:H000})). As reference, we show the sum of all three dominant contributions from $H_\text{eff}$ (purple curve). \REV{Despite a small overestimation of the renormalized bandwidth, the HFA (the purple curve) qualitatively reproduces the trajectories of full Floquet calculations (blue crosses).}} 
    \label{fig:FLOQUET_GAPS}
\end{figure*}

In Fig.~\ref{fig:FLOQUET_GAPS} we explore the Floquet bandstructure effects as a function of the driving amplitude $A_0$. As  a reference, we first calculate the values of the Floquet gap at $K_1$ and the electronic low-energy bandwidth at $\Gamma$ via Eq.~(\ref{Eq:FLOQUET_MAT}) with a truncated band basis according to the band cutoff (see black dashed lines in Fig.~\ref{fig:FLOQUET_GAPS}(a),(b)). In a next step, we use the effective Floquet Hamiltonian, Eq.~(\ref{Eq:Heff}) to calculate the different contributions of distinct LMC orders to the gap and bandwidth renormalization, respectively. As in Sec.~\ref{SSec.B} we consider intra-band and inter-band contributions separately. Moreover, we distinguish the inter-band coupling between flat bands and the coupling between flat and dispersive bands. We do this by first calculating the Floquet eigenbasis of the full effective Floquet Hamiltonian. Afterwards we set all matrix elements that are not of current interest to zero and transform the resulting matrix to the Floquet eigenbasis. We then calculate the quantity of interest from the diagonal elements, which for the full matrix correspond to the Floquet eigenvalues of the effective Hamiltonian. We only plot the dominant contribution. \REV{Importantly, we quantify the impact of higher-order field effects by extracting reference values from a full Floquet Hamiltonian, Eq.~(\ref{Eq:FLOQUET_MAT}), where $H(t)$ contains all bands and the full exponential LMC. The excellent agreement between the full (blues crosses) and the truncated results (black dashed line) renders higher order LMC effects (beyond second order) insignificant within the chosen driving regime.} 

For the light-induced band gap at the Dirac point $K_1$, Fig.~\ref{fig:FLOQUET_GAPS}(a), the linear-field inter-band coupling between the flat bands (see Eq.~(\ref{Eq:MaEl1_A}), $m \neq n \in \{-2,-1,0,+1\}$) is by far the dominant contribution. This agrees with the very small Fermi velocity and vanishing second k-derivatives $\partial^2_\mu H_0=0$ at $K_1$. The quadratic plot trajectory agrees with the quadratic field-dependency. At first glance, this result is counter-intuitive, since the dipole coupling matrix elements between the flat bands become very small at the magic angle (see Fig.~\ref{fig:STATIC_COUPLING_TBG}(b)). However, as Eq.~(\ref{Eq:MaEl1_A}) illustrates, the dynamical transition matrix elements of the effective Floquet Hamiltonian depend via the index $l$ on the dipole matrix elements to all higher and lower lying bands, which we have shown to be significant (note that $l$ now runs over all bands within the chosen band cutoff.) Due to its importance, we want to rephrase this key result: The MATBG Floquet Dirac gap is predominantly opened via Floquet-induced coupling between the flat bands that is \textit{mediated via strong dipole coupling of the flat bands to the higher and lower lying dispersive bands}. This result is in stark contrast to monolayer graphene, where the band gap in the high-frequency regime is solely governed by the Fermi velocity $\Delta_K^{MG} \approx \frac{2(v_F A_0)^2}{\Omega}$ \cite{oka_photovoltaic_2009} as demonstrated experimentally for the surface Dirac cone in Bi2Se3 \cite{mahmood_selective_2016}. For reference we plot the na\"ive expectation for the gap using the renormalized Fermi velocity of MATBG (black line) to highlight the difference between both gap opening mechanisms. In agreement with the weak intra-band coupling ($v_F \approx 0.23$ $\mathrm{eV}$/$\mathrm{\AA}$) (see Fig.~\ref{fig:STATIC_COUPLING_TBG}(a)), it is vanishingly small at the magic angle. The importance of the other bands beyond the four flat bands, via mediated couplings, resembles at an overall level the finding that the higher bands are important for the superfluid weight even when superfluid pairing mainly takes place in the four flat bands~\cite{julku_superfluid_2020} --- also due to virtual processes to the higher bands.

The band renormalization at $\Gamma$ (Fig.~\ref{fig:FLOQUET_GAPS}(b)) shows a more complex behaviour as a function of the driving amplitude $A_0$ than at the Dirac point. For small driving amplitudes $A_0\le \REV{0.004}$ $\mathrm{\AA}^{-1}$ the significant contribution to the renormalization stems from the quadratic intra-band coupling (green curve, see Eq.~(\ref{Eq:LMC_AA_HEFF})). As the valence (conduction) bands show a pronounced positive (negative) band curvature at the Brilluoin zone center, we suppose that a considerable part of the renormalization might have its origin in the finite band mass (see the first term on the upper right-hand side in table \ref{tab:LMCsSummarized}). However, quantifying the geometric contribution (second term on the upper right-hand side in table \ref{tab:LMCsSummarized}) to the renormalization effect and exploring possible ways to enhance it are interesting questions to tackle. As the sign of the geometric part is not fixed, this term could potentially either reinforce or counteract the curvature induced renormalization, depending on the mass sign of the surrounding dispersive bands. For stronger driving ($A_0>\REV{0.004}$ $\mathrm{\AA}^{-1}$) the linear-field inter-band coupling between the flat bands (red curve, see Eq.~(\ref{Eq:MaEl1_A})) dominates the Floquet bandwidth leading to an overall bandwidth increase at $\Gamma$ towards higher driving amplitudes. This increase is partly counteracted by zero-field contribution from the unperturbed Hamiltonian Eq.~(\ref{Eq:H000}). \REV{Interestingly, the high-frequency approximation (HFA) shows very good agreement with the full Floquet calculations at the Dirac point while it overestimates the bandwidth renormalization at higher driving amplitudes. However, it should be emphasized that in the context of this work the HFA is rather employed as means to obtain a profound understanding of the underlying physical mechanism than to give exact quantitative results as, e.g., provided in \cite{vogl_effective_2020, vogl_floquet_2020} for Floquet driven MATBG. Qualitatively, the HFA has overall good agreement with the full Floquet results.} Moreover, notice that we found the gap opening mechanism at the Dirac point to be qualitatively stable as function of the driving frequency whereas it has a strong impact on the bandwidth at $\Gamma.$ 

\section{DISCUSSION}\label{DISC}

Reducing the kinetic energy scale is a promising route for stabilizing quantum many-body effects that emerge from interactions and/or coupling to external perturbations. The extreme example of making the interaction or coupling energy scales dominant are so-called flat bands where the energy dispersion is constant. For instance, enhancement of ferromagnetism~\cite{Tasaki:1992,Mielke:1991}, superconductivity~\cite{kopnin:2011}, and light-matter coupling (LMC)~\cite{Kennes2021} have been predicted for (nearly) flat band systems. However, many important physical observables and responses, especially when related to displacement of the particles, depend on the effective mass and the Fermi velocity, which in a flat band diverge and vanish, respectively. This apparent problem has recently been pointed out to be refutable in multi-band systems of suitable quantum geometry; for instance, in the context of superconductivity the quantum metric, Berry curvature and Chern number of the band guarantee stable supercurrent~\cite{peotta_superfluidity_2015,liang_band_2017}. Here, we have shown that large light-matter couplings are possible in flat bands of suitable quantum geometry, and that this has remarkable ramifications in the case of a multi-band model describing twisted bilayer graphene (TBG).  

We calculated the LMC to second order in the field amplitude in a generic multi-band system and distinguished the intra- and inter-band contributions. For the linear in amplitude case (the paramagnetic term), the intra-band coupling is given by the group velocity as is well known, while the inter-band coupling depends on the properties of the Bloch functions. The quadratic (diamagnetic) intra-band term contains the inverse effective mass, which would be present also in a single-band case, but also a multi-band term that relates to the quantum metric of the band. In two limits, namely the two-band case and a system with exactly flat band either as the highest or the lowest one, we showed that the LMC is bounded from below by the quantum metric of the band. This highlights that the quadratic intra-band coupling, usually given by the inverse effective mass, can be finite also in a flat band of diverging effective mass, provided that it has a finite quantum metric. As the quantum metric is bounded from below by the Berry curvature and the Chern number of the band, topologically non-trivial (nearly) flat bands are thus favorable for large quadratic LMCs. Finally, we showed that the quadratic inter-band coupling depends on the inter-band Berry connection and can be non-zero also for flat bands. To illustrate these findings within a simple flat-band model, we presented the LMCs dependence on quantum geometry and how it varies over the Brillouin zone in the case of a sawtooth quantum chain.  

Recent advances in moir\'e materials, where the band curvature can be tuned by a twist angle between atomic layers, motivated us to explore the LMC within a generic tight-binding model that describes the basic features of twisted bilayer graphene. We performed the analysis both for static and dynamic (Floquet engineering) LMC effects, which is challenging due to the extremely large size of the moir\'e unit cell at the magic angle. In the static case, we found a strong dependence of the LMC on the twist angle. Obviously the band flattening tuned by the twist angle modifies the Fermi velocity and the effective mass. We found that the intra-band linear LMC simply reflects these quantities, as would be expected by a na\"ive approach. In contrast, the linear inter-band LMC shows intriguing behavior: it becomes very large at the magic angle. We found this to be due to the couplings between the flat and the dispersive bands while the terms involving two different flat bands were negligible. The quadratic (diamagnetic) LMC inside the flat bands was found to decrease but remain finite as the magic angle in TBG is approached.   

Understanding the dynamic case was strongly motivated by the potential of Floquet engineering in moir\'e materials. For instance, topological gaps have been predicted to open at the K point of the MATBG dispersion, but since the gap was much larger than expected by na\"ive rescaling of a Dirac-fermion-only model, its origin remained a puzzle~\cite{topp_topological_2019}. Here we showed that the possibility of large Floquet gaps at the K-point arises from inter-band coupling between the four flat bands. The mechanism was revealed to be quite subtle and intriguing. In the static case, the matrix elements between the flat bands would be vanishing; however, in the dynamic one they actually contain virtual transitions to the higher (and lower) dispersive bands, which facilitates a sizeable effective coupling between two flat bands. We further considered the $\Gamma$-point and showed that band-flattening can be engineered there via quadratic intra-band terms.

Intuitive physical understanding of the connection between quantum geometry and various responses such as supercurrent and LMC is provided by the connection between Wannier function overlaps and topology. In a band with non-zero Chern number, Wannier functions cannot be exponentially localized~\cite{Brouder:2007}. The significant overlap of Wannier functions of nearby lattice sites facilitates particle displacement and transport driven by inter-particle interactions or an external perturbation --- this effect becomes dominant in a flat band where non-interacting, uncoupled particles localize due to destructive interference of the wavefunctions~\cite{Tovmasyan:2016}. 

Our results provide general guidelines for designing strong light-matter interaction phenomena in multi-band systems. Quenching the kinetic energy is indeed a feasible strategy, since inter-band LMCs as well as the quadratic intra-band one can remain large provided the bands have non-trivial quantum geometric properties, in particular finite quantum metric and/or inter-band Berry connection. The quantum metric can be non-zero in a topologically trivial system, however, it is bounded from below by the Chern number, thus topologically non-trivial systems with nearly flat bands are excellent candidates to host large LMCs. In case of TBG and other moir\'{e} materials, this points to the potential importance of the so-called fragile topology associated with these systems \cite{zou_band_2018,po_origin_2018,ahn_failure_2019}. Our findings on TBG emphasize that inter-band processes involving the higher and lower dispersive bands, either directly (static) or indirectly (Floquet), generate large LMC effects even when the naïve theory based on the energy dispersions would indicate otherwise. This is definitely good news for Floquet or cavity engineering and LMC in moir\'e materials. It emphasizes, however, that models that take into account only a few (in case of TBG, four) flat bands can drastically miss important effects. The need to consider more than four bands, due to geometric contributions, was pointed out also in the context of superconductivity in TBG~\cite{julku_superfluid_2020}.     

An obvious future task is to explore how including more bands than in this work influences LMCs in TBG: what will be the maximum value achievable? To overcome limitations related to the large size of the unit cell, renormalized models~\cite{su:2018,gonzalez-arraga:2017} could be applied, but there one needs to carefully investigate whether the renormalization still works in the context of LMC, and dynamic processes in particular. To provide direct predictions for Floquet engineering experiments, simulations with smaller frequencies need to be performed. Our results should also inspire and guide the way for searching large LMCs and efficient Floquet engineering in other moir\'e materials beyond TBG, where vast possibilities for different parameter regimes and quantum geometric properties can be found.

\section*{ACKNOWLEDGEMENT} 
We acknowledge discussions with J.~W.~McIver. G.T.~and P.T.~acknowledge support by the Academy of Finland under project numbers 303351, 307419 and 327293. M.A.S.~acknowledges financial support through the DFG Emmy Noether program (SE 2558/2).  D.M.K.~acknowledges the support of the Deutsche  Forschungsgemeinschaft (DFG, German Research Foundation) through RTG 1995, within the Priority Program SPP 2244 ``2DMP'' and under Germany’s Excellence Strategy-Cluster of Excellence Matter and Light for Quantum Computing (ML4Q) EXC2004/1 - 390534769. We acknowledge support from the Max Planck-New York City Center for Non-Equilibrium Quantum Phenomena.

\bibliography{zotero_refs,ref_DMK.bib,TBG_COUPLING,FB_LMC,bib_tbg.bib,bec_paper2019.bib,Refs_BGV_3.bib,Refs.bib,References.bib,hexarray-theory.bib,biblio_S.bib,refsChris.bib}

\begin{thebibliography}{173}%
\makeatletter
\providecommand \@ifxundefined [1]{%
 \@ifx{#1\undefined}
}%
\providecommand \@ifnum [1]{%
 \ifnum #1\expandafter \@firstoftwo
 \else \expandafter \@secondoftwo
 \fi
}%
\providecommand \@ifx [1]{%
 \ifx #1\expandafter \@firstoftwo
 \else \expandafter \@secondoftwo
 \fi
}%
\providecommand \natexlab [1]{#1}%
\providecommand \enquote  [1]{``#1''}%
\providecommand \bibnamefont  [1]{#1}%
\providecommand \bibfnamefont [1]{#1}%
\providecommand \citenamefont [1]{#1}%
\providecommand \href@noop [0]{\@secondoftwo}%
\providecommand \href [0]{\begingroup \@sanitize@url \@href}%
\providecommand \@href[1]{\@@startlink{#1}\@@href}%
\providecommand \@@href[1]{\endgroup#1\@@endlink}%
\providecommand \@sanitize@url [0]{\catcode `\\12\catcode `\$12\catcode
  `\&12\catcode `\#12\catcode `\^12\catcode `\_12\catcode `\%12\relax}%
\providecommand \@@startlink[1]{}%
\providecommand \@@endlink[0]{}%
\providecommand \url  [0]{\begingroup\@sanitize@url \@url }%
\providecommand \@url [1]{\endgroup\@href {#1}{\urlprefix }}%
\providecommand \urlprefix  [0]{URL }%
\providecommand \Eprint [0]{\href }%
\providecommand \doibase [0]{https://doi.org/}%
\providecommand \selectlanguage [0]{\@gobble}%
\providecommand \bibinfo  [0]{\@secondoftwo}%
\providecommand \bibfield  [0]{\@secondoftwo}%
\providecommand \translation [1]{[#1]}%
\providecommand \BibitemOpen [0]{}%
\providecommand \bibitemStop [0]{}%
\providecommand \bibitemNoStop [0]{.\EOS\space}%
\providecommand \EOS [0]{\spacefactor3000\relax}%
\providecommand \BibitemShut  [1]{\csname bibitem#1\endcsname}%
\let\auto@bib@innerbib\@empty
\bibitem [{\citenamefont {Klitzing}\ \emph {et~al.}(1980)\citenamefont
  {Klitzing}, \citenamefont {Dorda},\ and\ \citenamefont
  {Pepper}}]{Klitzing1980}%
  \BibitemOpen
  \bibfield  {author} {\bibinfo {author} {\bibfnamefont {K.~v.}\ \bibnamefont
  {Klitzing}}, \bibinfo {author} {\bibfnamefont {G.}~\bibnamefont {Dorda}},\
  and\ \bibinfo {author} {\bibfnamefont {M.}~\bibnamefont {Pepper}},\
  }\bibfield  {title} {\bibinfo {title} {{New Method for High-Accuracy
  Determination of the Fine-Structure Constant Based on Quantized Hall
  Resistance}},\ }\href {https://doi.org/10.1103/PhysRevLett.45.494} {\bibfield
   {journal} {\bibinfo  {journal} {Phys. Rev. Lett.}\ }\textbf {\bibinfo
  {volume} {45}},\ \bibinfo {pages} {494} (\bibinfo {year} {1980})}\BibitemShut
  {NoStop}%
\bibitem [{\citenamefont {Thouless}\ \emph {et~al.}(1982)\citenamefont
  {Thouless}, \citenamefont {Kohmoto}, \citenamefont {Nightingale},\ and\
  \citenamefont {Den~Nijs}}]{Thouless1982}%
  \BibitemOpen
  \bibfield  {author} {\bibinfo {author} {\bibfnamefont {D.~J.}\ \bibnamefont
  {Thouless}}, \bibinfo {author} {\bibfnamefont {M.}~\bibnamefont {Kohmoto}},
  \bibinfo {author} {\bibfnamefont {M.~P.}\ \bibnamefont {Nightingale}},\ and\
  \bibinfo {author} {\bibfnamefont {M.}~\bibnamefont {Den~Nijs}},\ }\bibfield
  {title} {\bibinfo {title} {{Quantized Hall Conductance in a Two-Dimensional
  Periodic Potential}},\ }\href {https://doi.org/10.1103/PhysRevLett.49.405}
  {\bibfield  {journal} {\bibinfo  {journal} {Phys. Rev. Lett.}\ }\textbf
  {\bibinfo {volume} {49}},\ \bibinfo {pages} {405} (\bibinfo {year}
  {1982})}\BibitemShut {NoStop}%
\bibitem [{\citenamefont {Haldane}(1988)}]{Haldane1988}%
  \BibitemOpen
  \bibfield  {author} {\bibinfo {author} {\bibfnamefont {F.~D.~M.}\
  \bibnamefont {Haldane}},\ }\bibfield  {title} {\bibinfo {title} {{Model for a
  Quantum Hall Effect without Landau Levels: Condensed-Matter Realization of
  the "Parity Anomaly"}},\ }\href {https://doi.org/10.1103/PhysRevLett.61.2015}
  {\bibfield  {journal} {\bibinfo  {journal} {Phys. Rev. Lett.}\ }\textbf
  {\bibinfo {volume} {61}},\ \bibinfo {pages} {2015} (\bibinfo {year}
  {1988})}\BibitemShut {NoStop}%
\bibitem [{\citenamefont {Kane}\ and\ \citenamefont {Mele}(2005)}]{Kane2005}%
  \BibitemOpen
  \bibfield  {author} {\bibinfo {author} {\bibfnamefont {C.~L.}\ \bibnamefont
  {Kane}}\ and\ \bibinfo {author} {\bibfnamefont {E.~J.}\ \bibnamefont
  {Mele}},\ }\bibfield  {title} {\bibinfo {title} {{Quantum Spin Hall Effect in
  Graphene}},\ }\href {https://doi.org/10.1103/PhysRevLett.95.226801}
  {\bibfield  {journal} {\bibinfo  {journal} {Phys. Rev. Lett.}\ }\textbf
  {\bibinfo {volume} {95}},\ \bibinfo {pages} {226801} (\bibinfo {year}
  {2005})}\BibitemShut {NoStop}%
\bibitem [{\citenamefont {Bernevig}\ \emph {et~al.}(2006)\citenamefont
  {Bernevig}, \citenamefont {Hughes},\ and\ \citenamefont
  {Zhang}}]{Bernevig2006QuantumWells}%
  \BibitemOpen
  \bibfield  {author} {\bibinfo {author} {\bibfnamefont {B.~A.}\ \bibnamefont
  {Bernevig}}, \bibinfo {author} {\bibfnamefont {T.~L.}\ \bibnamefont
  {Hughes}},\ and\ \bibinfo {author} {\bibfnamefont {S.-C.}\ \bibnamefont
  {Zhang}},\ }\bibfield  {title} {\bibinfo {title} {{Quantum Spin Hall Effect
  and Topological Phase Transition in HgTe Quantum Wells}},\ }\href
  {https://doi.org/10.1126/science.1133734} {\bibfield  {journal} {\bibinfo
  {journal} {Science}\ }\textbf {\bibinfo {volume} {314}},\ \bibinfo {pages}
  {1757} (\bibinfo {year} {2006})}\BibitemShut {NoStop}%
\bibitem [{\citenamefont {K{\"{o}}nig}\ \emph {et~al.}(2007)\citenamefont
  {K{\"{o}}nig}, \citenamefont {Wiedmann}, \citenamefont {Br{\"{u}}ne},
  \citenamefont {Roth}, \citenamefont {Buhmann}, \citenamefont {Molenkamp},
  \citenamefont {Qi},\ and\ \citenamefont {Zhang}}]{Konig2007QuantumWells}%
  \BibitemOpen
  \bibfield  {author} {\bibinfo {author} {\bibfnamefont {M.}~\bibnamefont
  {K{\"{o}}nig}}, \bibinfo {author} {\bibfnamefont {S.}~\bibnamefont
  {Wiedmann}}, \bibinfo {author} {\bibfnamefont {C.}~\bibnamefont
  {Br{\"{u}}ne}}, \bibinfo {author} {\bibfnamefont {A.}~\bibnamefont {Roth}},
  \bibinfo {author} {\bibfnamefont {H.}~\bibnamefont {Buhmann}}, \bibinfo
  {author} {\bibfnamefont {L.~W.}\ \bibnamefont {Molenkamp}}, \bibinfo {author}
  {\bibfnamefont {X.~L.}\ \bibnamefont {Qi}},\ and\ \bibinfo {author}
  {\bibfnamefont {S.~C.}\ \bibnamefont {Zhang}},\ }\bibfield  {title} {\bibinfo
  {title} {{Quantum Spin Hall Insulator State in HgTe Quantum Wells}},\ }\href
  {https://doi.org/10.1126/science.1148047} {\bibfield  {journal} {\bibinfo
  {journal} {Science}\ }\textbf {\bibinfo {volume} {318}},\ \bibinfo {pages}
  {766} (\bibinfo {year} {2007})}\BibitemShut {NoStop}%
\bibitem [{\citenamefont {Hasan}\ and\ \citenamefont {Kane}(2010)}]{Hasan2010}%
  \BibitemOpen
  \bibfield  {author} {\bibinfo {author} {\bibfnamefont {M.~Z.}\ \bibnamefont
  {Hasan}}\ and\ \bibinfo {author} {\bibfnamefont {C.~L.}\ \bibnamefont
  {Kane}},\ }\bibfield  {title} {\bibinfo {title} {{Colloquium : Topological
  Insulators}},\ }\href {https://doi.org/10.1103/RevModPhys.82.3045} {\bibfield
   {journal} {\bibinfo  {journal} {Rev. Mod. Phys.}\ }\textbf {\bibinfo
  {volume} {82}},\ \bibinfo {pages} {3045} (\bibinfo {year}
  {2010})}\BibitemShut {NoStop}%
\bibitem [{\citenamefont {Bernevig}\ and\ \citenamefont
  {Hughes}(2013)}]{Bernevig2013}%
  \BibitemOpen
  \bibfield  {author} {\bibinfo {author} {\bibfnamefont {B.~A.}\ \bibnamefont
  {Bernevig}}\ and\ \bibinfo {author} {\bibfnamefont {T.~L.}\ \bibnamefont
  {Hughes}},\ }\href@noop {} {\emph {\bibinfo {title} {{"Topological Insulators
  and Topological Superconductors"}}}}\ (\bibinfo  {publisher} {Princeton
  University Press},\ \bibinfo {year} {2013})\ p.\ \bibinfo {pages}
  {247}\BibitemShut {NoStop}%
\bibitem [{\citenamefont {Jotzu}\ \emph {et~al.}(2014)\citenamefont {Jotzu},
  \citenamefont {Messer}, \citenamefont {Desbuquois}, \citenamefont {Lebrat},
  \citenamefont {Uehlinger}, \citenamefont {Greif},\ and\ \citenamefont
  {Esslinger}}]{Jotzu:2014}%
  \BibitemOpen
  \bibfield  {author} {\bibinfo {author} {\bibfnamefont {G.}~\bibnamefont
  {Jotzu}}, \bibinfo {author} {\bibfnamefont {M.}~\bibnamefont {Messer}},
  \bibinfo {author} {\bibfnamefont {R.}~\bibnamefont {Desbuquois}}, \bibinfo
  {author} {\bibfnamefont {M.}~\bibnamefont {Lebrat}}, \bibinfo {author}
  {\bibfnamefont {T.}~\bibnamefont {Uehlinger}}, \bibinfo {author}
  {\bibfnamefont {D.}~\bibnamefont {Greif}},\ and\ \bibinfo {author}
  {\bibfnamefont {T.}~\bibnamefont {Esslinger}},\ }\bibfield  {title} {\bibinfo
  {title} {{Experimental Realization of the Topological {H}aldane Model With
  Ultracold Fermions}},\ }\href@noop {} {\bibfield  {journal} {\bibinfo
  {journal} {Nature}\ }\textbf {\bibinfo {volume} {515}},\ \bibinfo {pages}
  {237} (\bibinfo {year} {2014})}\BibitemShut {NoStop}%
\bibitem [{\citenamefont {Peotta}\ and\ \citenamefont
  {T\"orm\"a}(2015)}]{peotta_superfluidity_2015}%
  \BibitemOpen
  \bibfield  {author} {\bibinfo {author} {\bibfnamefont {S.}~\bibnamefont
  {Peotta}}\ and\ \bibinfo {author} {\bibfnamefont {P.}~\bibnamefont
  {T\"orm\"a}},\ }\bibfield  {title} {\bibinfo {title} {Superfluidity in
  topologically nontrivial flat bands},\ }\href
  {https://doi.org/10.1038/ncomms9944} {\bibfield  {journal} {\bibinfo
  {journal} {Nature Communications}\ }\textbf {\bibinfo {volume} {6}},\
  \bibinfo {pages} {8944} (\bibinfo {year} {2015})}\BibitemShut {NoStop}%
\bibitem [{\citenamefont {Julku}\ \emph {et~al.}(2016)\citenamefont {Julku},
  \citenamefont {Peotta}, \citenamefont {Vanhala}, \citenamefont {Kim},\ and\
  \citenamefont {T\"orm\"a}}]{julku_geometric_2016}%
  \BibitemOpen
  \bibfield  {author} {\bibinfo {author} {\bibfnamefont {A.}~\bibnamefont
  {Julku}}, \bibinfo {author} {\bibfnamefont {S.}~\bibnamefont {Peotta}},
  \bibinfo {author} {\bibfnamefont {T.~I.}\ \bibnamefont {Vanhala}}, \bibinfo
  {author} {\bibfnamefont {D.-H.}\ \bibnamefont {Kim}},\ and\ \bibinfo {author}
  {\bibfnamefont {P.}~\bibnamefont {T\"orm\"a}},\ }\bibfield  {title} {\bibinfo
  {title} {Geometric {Origin} of {Superfluidity} in the {Lieb}-{Lattice} {Flat}
  {Band}},\ }\href {https://doi.org/10.1103/PhysRevLett.117.045303} {\bibfield
  {journal} {\bibinfo  {journal} {Physical Review Letters}\ }\textbf {\bibinfo
  {volume} {117}},\ \bibinfo {pages} {045303} (\bibinfo {year}
  {2016})}\BibitemShut {NoStop}%
\bibitem [{\citenamefont {T\"orm\"a}\ \emph {et~al.}(2018)\citenamefont
  {T\"orm\"a}, \citenamefont {Liang},\ and\ \citenamefont
  {Peotta}}]{torma_quantum_2018}%
  \BibitemOpen
  \bibfield  {author} {\bibinfo {author} {\bibfnamefont {P.}~\bibnamefont
  {T\"orm\"a}}, \bibinfo {author} {\bibfnamefont {L.}~\bibnamefont {Liang}},\
  and\ \bibinfo {author} {\bibfnamefont {S.}~\bibnamefont {Peotta}},\
  }\bibfield  {title} {\bibinfo {title} {Quantum metric and effective mass of a
  two-body bound state in a flat band},\ }\href@noop {} {\bibfield  {journal}
  {\bibinfo  {journal} {Physical Review B}\ }\textbf {\bibinfo {volume} {98}},\
  \bibinfo {pages} {220511} (\bibinfo {year} {2018})}\BibitemShut {NoStop}%
\bibitem [{\citenamefont {Gao}\ \emph {et~al.}(2014)\citenamefont {Gao},
  \citenamefont {Yang},\ and\ \citenamefont {Niu}}]{Gao_FieldIncuded_2014}%
  \BibitemOpen
  \bibfield  {author} {\bibinfo {author} {\bibfnamefont {Y.}~\bibnamefont
  {Gao}}, \bibinfo {author} {\bibfnamefont {S.~A.}\ \bibnamefont {Yang}},\ and\
  \bibinfo {author} {\bibfnamefont {Q.}~\bibnamefont {Niu}},\ }\bibfield
  {title} {\bibinfo {title} {Field induced positional shift of {B}loch
  electrons and its dynamical implications},\ }\href
  {https://doi.org/10.1103/PhysRevLett.112.166601} {\bibfield  {journal}
  {\bibinfo  {journal} {Phys. Rev. Lett.}\ }\textbf {\bibinfo {volume} {112}},\
  \bibinfo {pages} {166601} (\bibinfo {year} {2014})}\BibitemShut {NoStop}%
\bibitem [{\citenamefont {Ogata}\ and\ \citenamefont
  {Fukuyama}(2015)}]{ogata_orbital_2015}%
  \BibitemOpen
  \bibfield  {author} {\bibinfo {author} {\bibfnamefont {M.}~\bibnamefont
  {Ogata}}\ and\ \bibinfo {author} {\bibfnamefont {H.}~\bibnamefont
  {Fukuyama}},\ }\bibfield  {title} {\bibinfo {title} {Orbital {Magnetism} of
  {Bloch} {Electrons} {I}. {General} {Formula}},\ }\href
  {https://doi.org/10.7566/JPSJ.84.124708} {\bibfield  {journal} {\bibinfo
  {journal} {Journal of the Physical Society of Japan}\ }\textbf {\bibinfo
  {volume} {84}},\ \bibinfo {pages} {124708} (\bibinfo {year}
  {2015})}\BibitemShut {NoStop}%
\bibitem [{\citenamefont {Pi{\'e}chon}\ \emph {et~al.}(2016)\citenamefont
  {Pi{\'e}chon}, \citenamefont {Raoux}, \citenamefont {Fuchs},\ and\
  \citenamefont {Montambaux}}]{piechon_geometric_2016}%
  \BibitemOpen
  \bibfield  {author} {\bibinfo {author} {\bibfnamefont {F.}~\bibnamefont
  {Pi{\'e}chon}}, \bibinfo {author} {\bibfnamefont {A.}~\bibnamefont {Raoux}},
  \bibinfo {author} {\bibfnamefont {J.-N.}\ \bibnamefont {Fuchs}},\ and\
  \bibinfo {author} {\bibfnamefont {G.}~\bibnamefont {Montambaux}},\ }\bibfield
   {title} {\bibinfo {title} {Geometric orbital susceptibility: {Quantum}
  metric without {Berry} curvature},\ }\href
  {https://doi.org/10.1103/PhysRevB.94.134423} {\bibfield  {journal} {\bibinfo
  {journal} {Physical Review B}\ }\textbf {\bibinfo {volume} {94}},\ \bibinfo
  {pages} {134423} (\bibinfo {year} {2016})}\BibitemShut {NoStop}%
\bibitem [{\citenamefont {Rhim}\ \emph {et~al.}(2020)\citenamefont {Rhim},
  \citenamefont {Kim},\ and\ \citenamefont {Yang}}]{rhim_quantum_2020}%
  \BibitemOpen
  \bibfield  {author} {\bibinfo {author} {\bibfnamefont {J.-W.}\ \bibnamefont
  {Rhim}}, \bibinfo {author} {\bibfnamefont {K.}~\bibnamefont {Kim}},\ and\
  \bibinfo {author} {\bibfnamefont {B.-J.}\ \bibnamefont {Yang}},\ }\bibfield
  {title} {\bibinfo {title} {Quantum distance and anomalous {Landau} levels of
  flat bands},\ }\href {https://doi.org/10.1038/s41586-020-2540-1} {\bibfield
  {journal} {\bibinfo  {journal} {Nature}\ }\textbf {\bibinfo {volume} {584}},\
  \bibinfo {pages} {59} (\bibinfo {year} {2020})}\BibitemShut {NoStop}%
\bibitem [{\citenamefont {Freimuth}\ \emph {et~al.}(2017)\citenamefont
  {Freimuth}, \citenamefont {Bl\"ugel},\ and\ \citenamefont
  {Mokrousov}}]{PhysRevB.95.184428}%
  \BibitemOpen
  \bibfield  {author} {\bibinfo {author} {\bibfnamefont {F.}~\bibnamefont
  {Freimuth}}, \bibinfo {author} {\bibfnamefont {S.}~\bibnamefont {Bl\"ugel}},\
  and\ \bibinfo {author} {\bibfnamefont {Y.}~\bibnamefont {Mokrousov}},\
  }\bibfield  {title} {\bibinfo {title} {Geometrical contributions to the
  exchange constants: Free electrons with spin-orbit interaction},\ }\href
  {https://doi.org/10.1103/PhysRevB.95.184428} {\bibfield  {journal} {\bibinfo
  {journal} {Phys. Rev. B}\ }\textbf {\bibinfo {volume} {95}},\ \bibinfo
  {pages} {184428} (\bibinfo {year} {2017})}\BibitemShut {NoStop}%
\bibitem [{\citenamefont {Srivastava}\ and\ \citenamefont
  {Imamo\u{g}lu}(2015)}]{Srivastava:2015}%
  \BibitemOpen
  \bibfield  {author} {\bibinfo {author} {\bibfnamefont {A.}~\bibnamefont
  {Srivastava}}\ and\ \bibinfo {author} {\bibfnamefont {A.}~\bibnamefont
  {Imamo\u{g}lu}},\ }\bibfield  {title} {\bibinfo {title} {Signatures of
  {Bloch}-band geometry on excitons: {Nonhydrogenic} spectra in
  transition-metal dichalcogenides},\ }\href
  {http://link.aps.org/doi/10.1103/PhysRevLett.115.166802} {\bibfield
  {journal} {\bibinfo  {journal} {Phys. Rev. Lett.}\ }\textbf {\bibinfo
  {volume} {115}} (\bibinfo {year} {2015})}\BibitemShut {NoStop}%
\bibitem [{\citenamefont {Bleu}\ \emph {et~al.}(2018)\citenamefont {Bleu},
  \citenamefont {Solnyshkov},\ and\ \citenamefont {Malpuech}}]{Bleu2018}%
  \BibitemOpen
  \bibfield  {author} {\bibinfo {author} {\bibfnamefont {O.}~\bibnamefont
  {Bleu}}, \bibinfo {author} {\bibfnamefont {D.~D.}\ \bibnamefont
  {Solnyshkov}},\ and\ \bibinfo {author} {\bibfnamefont {G.}~\bibnamefont
  {Malpuech}},\ }\bibfield  {title} {\bibinfo {title} {{Measuring the Quantum
  Geometric Tensor in Two-Dimensional Photonic and Exciton-Polariton
  Systems}},\ }\href {https://doi.org/10.1103/PhysRevB.97.195422} {\bibfield
  {journal} {\bibinfo  {journal} {Phys. Rev. B}\ }\textbf {\bibinfo {volume}
  {97}},\ \bibinfo {pages} {195422} (\bibinfo {year} {2018})}\BibitemShut
  {NoStop}%
\bibitem [{\citenamefont {{Gianfrate}}\ \emph {et~al.}(2020)\citenamefont
  {{Gianfrate}}, \citenamefont {{Bleu}}, \citenamefont {{Dominici}},
  \citenamefont {{Ardizzone}}, \citenamefont {{De Giorgi}}, \citenamefont
  {{Ballarini}}, \citenamefont {{Lerario}}, \citenamefont {{West}},
  \citenamefont {{Pfeiffer}}, \citenamefont {{Solnyshkov}}, \citenamefont
  {{Sanvitto}},\ and\ \citenamefont {{Malpuech}}}]{Gianfrate:2019}%
  \BibitemOpen
  \bibfield  {author} {\bibinfo {author} {\bibfnamefont {A.}~\bibnamefont
  {{Gianfrate}}}, \bibinfo {author} {\bibfnamefont {O.}~\bibnamefont {{Bleu}}},
  \bibinfo {author} {\bibfnamefont {L.}~\bibnamefont {{Dominici}}}, \bibinfo
  {author} {\bibfnamefont {V.}~\bibnamefont {{Ardizzone}}}, \bibinfo {author}
  {\bibfnamefont {M.}~\bibnamefont {{De Giorgi}}}, \bibinfo {author}
  {\bibfnamefont {D.}~\bibnamefont {{Ballarini}}}, \bibinfo {author}
  {\bibfnamefont {G.}~\bibnamefont {{Lerario}}}, \bibinfo {author}
  {\bibfnamefont {K.}~\bibnamefont {{West}}}, \bibinfo {author} {\bibfnamefont
  {L.~N.}\ \bibnamefont {{Pfeiffer}}}, \bibinfo {author} {\bibfnamefont
  {D.~D.}\ \bibnamefont {{Solnyshkov}}}, \bibinfo {author} {\bibfnamefont
  {D.}~\bibnamefont {{Sanvitto}}},\ and\ \bibinfo {author} {\bibfnamefont
  {G.}~\bibnamefont {{Malpuech}}},\ }\bibfield  {title} {\bibinfo {title}
  {{Measurement of the Quantum Geometric Tensor and of the Anomalous Hall
  Drift}},\ }\href {https://doi.org/https://doi.org/10.1038/s41586-020-1989-2}
  {\bibfield  {journal} {\bibinfo  {journal} {Nature}\ }\textbf {\bibinfo
  {volume} {578}},\ \bibinfo {pages} {381} (\bibinfo {year}
  {2020})}\BibitemShut {NoStop}%
\bibitem [{\citenamefont {Tan}\ \emph {et~al.}(2019)\citenamefont {Tan},
  \citenamefont {Zhang}, \citenamefont {Yang}, \citenamefont {Chu},
  \citenamefont {Zhu}, \citenamefont {Li}, \citenamefont {Yang}, \citenamefont
  {Song}, \citenamefont {Han}, \citenamefont {Li}, \citenamefont {Dong},
  \citenamefont {Yu}, \citenamefont {Yan}, \citenamefont {Zhu},\ and\
  \citenamefont {Yu}}]{Tan2019ExpQubit}%
  \BibitemOpen
  \bibfield  {author} {\bibinfo {author} {\bibfnamefont {X.}~\bibnamefont
  {Tan}}, \bibinfo {author} {\bibfnamefont {D.-W.}\ \bibnamefont {Zhang}},
  \bibinfo {author} {\bibfnamefont {Z.}~\bibnamefont {Yang}}, \bibinfo {author}
  {\bibfnamefont {J.}~\bibnamefont {Chu}}, \bibinfo {author} {\bibfnamefont
  {Y.-Q.}\ \bibnamefont {Zhu}}, \bibinfo {author} {\bibfnamefont
  {D.}~\bibnamefont {Li}}, \bibinfo {author} {\bibfnamefont {X.}~\bibnamefont
  {Yang}}, \bibinfo {author} {\bibfnamefont {S.}~\bibnamefont {Song}}, \bibinfo
  {author} {\bibfnamefont {Z.}~\bibnamefont {Han}}, \bibinfo {author}
  {\bibfnamefont {Z.}~\bibnamefont {Li}}, \bibinfo {author} {\bibfnamefont
  {Y.}~\bibnamefont {Dong}}, \bibinfo {author} {\bibfnamefont {H.~F.}\
  \bibnamefont {Yu}}, \bibinfo {author} {\bibfnamefont {H.}~\bibnamefont
  {Yan}}, \bibinfo {author} {\bibfnamefont {S.-L.}\ \bibnamefont {Zhu}},\ and\
  \bibinfo {author} {\bibfnamefont {Y.}~\bibnamefont {Yu}},\ }\bibfield
  {title} {\bibinfo {title} {{Experimental Measurement of the Quantum Metric
  Tensor and Related Topological Phase Transition with a Superconducting
  Qubit}},\ }\href {https://doi.org/10.1103/physrevlett.122.210401} {\bibfield
  {journal} {\bibinfo  {journal} {Phys. Rev. Lett.}\ }\textbf {\bibinfo
  {volume} {122}},\ \bibinfo {pages} {210401} (\bibinfo {year}
  {2019})}\BibitemShut {NoStop}%
\bibitem [{\citenamefont {{Asteria}}\ \emph {et~al.}(2019)\citenamefont
  {{Asteria}}, \citenamefont {{Tran}}, \citenamefont {{Ozawa}}, \citenamefont
  {{Tarnowski}}, \citenamefont {{Rem}}, \citenamefont {{Fl{\"a}schner}},
  \citenamefont {{Sengstock}}, \citenamefont {{Goldman}},\ and\ \citenamefont
  {{Weitenberg}}}]{Weitenberg:2019}%
  \BibitemOpen
  \bibfield  {author} {\bibinfo {author} {\bibfnamefont {L.}~\bibnamefont
  {{Asteria}}}, \bibinfo {author} {\bibfnamefont {D.~T.}\ \bibnamefont
  {{Tran}}}, \bibinfo {author} {\bibfnamefont {T.}~\bibnamefont {{Ozawa}}},
  \bibinfo {author} {\bibfnamefont {M.}~\bibnamefont {{Tarnowski}}}, \bibinfo
  {author} {\bibfnamefont {B.~S.}\ \bibnamefont {{Rem}}}, \bibinfo {author}
  {\bibfnamefont {N.}~\bibnamefont {{Fl{\"a}schner}}}, \bibinfo {author}
  {\bibfnamefont {K.}~\bibnamefont {{Sengstock}}}, \bibinfo {author}
  {\bibfnamefont {N.}~\bibnamefont {{Goldman}}},\ and\ \bibinfo {author}
  {\bibfnamefont {C.}~\bibnamefont {{Weitenberg}}},\ }\bibfield  {title}
  {\bibinfo {title} {{Measuring Quantized Circular Dichroism in Ultracold
  Topological Matter}},\ }\href {https://doi.org/10.1038/s41567-019-0417-8}
  {\bibfield  {journal} {\bibinfo  {journal} {Nature Phys.}\ }\textbf {\bibinfo
  {volume} {15}},\ \bibinfo {pages} {449} (\bibinfo {year} {2019})}\BibitemShut
  {NoStop}%
\bibitem [{\citenamefont {Yu}\ \emph {et~al.}(2019)\citenamefont {Yu},
  \citenamefont {Yang}, \citenamefont {Gong}, \citenamefont {Cao},
  \citenamefont {Lu}, \citenamefont {Liu}, \citenamefont {Zhang}, \citenamefont
  {Plenio}, \citenamefont {Jelezko}, \citenamefont {Ozawa},\ and\ \citenamefont
  {et~al.}}]{Yu2020expdiamond}%
  \BibitemOpen
  \bibfield  {author} {\bibinfo {author} {\bibfnamefont {M.}~\bibnamefont
  {Yu}}, \bibinfo {author} {\bibfnamefont {P.}~\bibnamefont {Yang}}, \bibinfo
  {author} {\bibfnamefont {M.}~\bibnamefont {Gong}}, \bibinfo {author}
  {\bibfnamefont {Q.}~\bibnamefont {Cao}}, \bibinfo {author} {\bibfnamefont
  {Q.}~\bibnamefont {Lu}}, \bibinfo {author} {\bibfnamefont {H.}~\bibnamefont
  {Liu}}, \bibinfo {author} {\bibfnamefont {S.}~\bibnamefont {Zhang}}, \bibinfo
  {author} {\bibfnamefont {M.~B.}\ \bibnamefont {Plenio}}, \bibinfo {author}
  {\bibfnamefont {F.}~\bibnamefont {Jelezko}}, \bibinfo {author} {\bibfnamefont
  {T.}~\bibnamefont {Ozawa}},\ and\ \bibinfo {author} {\bibnamefont {et~al.}},\
  }\bibfield  {title} {\bibinfo {title} {{Experimental Measurement of the
  Quantum Geometric Tensor Using Coupled Qubits in Diamond}},\ }\href
  {https://doi.org/10.1093/nsr/nwz193} {\bibfield  {journal} {\bibinfo
  {journal} {Natl. Sci. Rev.}\ }\textbf {\bibinfo {volume} {7}},\ \bibinfo
  {pages} {254–260} (\bibinfo {year} {2019})}\BibitemShut {NoStop}%
\bibitem [{\citenamefont {Provost}\ and\ \citenamefont
  {Vallee}(1980)}]{Provost80}%
  \BibitemOpen
  \bibfield  {author} {\bibinfo {author} {\bibfnamefont {J.~P.}\ \bibnamefont
  {Provost}}\ and\ \bibinfo {author} {\bibfnamefont {G.}~\bibnamefont
  {Vallee}},\ }\bibfield  {title} {\bibinfo {title} {Riemannian structure on
  manifolds of quantum states},\ }\href
  {https://doi.org/https://doi.org/10.1007/BF02193559} {\bibfield  {journal}
  {\bibinfo  {journal} {Communications in Mathematical Physics}\ }\textbf
  {\bibinfo {volume} {76}},\ \bibinfo {pages} {289} (\bibinfo {year}
  {1980})}\BibitemShut {NoStop}%
\bibitem [{\citenamefont {Liang}\ \emph {et~al.}(2017)\citenamefont {Liang},
  \citenamefont {Vanhala}, \citenamefont {Peotta}, \citenamefont {Siro},
  \citenamefont {Harju},\ and\ \citenamefont {T\"orm\"a}}]{liang_band_2017}%
  \BibitemOpen
  \bibfield  {author} {\bibinfo {author} {\bibfnamefont {L.}~\bibnamefont
  {Liang}}, \bibinfo {author} {\bibfnamefont {T.~I.}\ \bibnamefont {Vanhala}},
  \bibinfo {author} {\bibfnamefont {S.}~\bibnamefont {Peotta}}, \bibinfo
  {author} {\bibfnamefont {T.}~\bibnamefont {Siro}}, \bibinfo {author}
  {\bibfnamefont {A.}~\bibnamefont {Harju}},\ and\ \bibinfo {author}
  {\bibfnamefont {P.}~\bibnamefont {T\"orm\"a}},\ }\bibfield  {title} {\bibinfo
  {title} {Band geometry, {Berry} curvature, and superfluid weight},\ }\href
  {https://doi.org/10.1103/PhysRevB.95.024515} {\bibfield  {journal} {\bibinfo
  {journal} {Physical Review B}\ }\textbf {\bibinfo {volume} {95}},\ \bibinfo
  {pages} {024515} (\bibinfo {year} {2017})}\BibitemShut {NoStop}%
\bibitem [{\citenamefont {Lopes~dos Santos}\ \emph {et~al.}(2007)\citenamefont
  {Lopes~dos Santos}, \citenamefont {Peres},\ and\ \citenamefont
  {Castro~Neto}}]{Neto07}%
  \BibitemOpen
  \bibfield  {author} {\bibinfo {author} {\bibfnamefont {J.~M.~B.}\
  \bibnamefont {Lopes~dos Santos}}, \bibinfo {author} {\bibfnamefont
  {N.~M.~R.}\ \bibnamefont {Peres}},\ and\ \bibinfo {author} {\bibfnamefont
  {A.~H.}\ \bibnamefont {Castro~Neto}},\ }\bibfield  {title} {\bibinfo {title}
  {Graphene bilayer with a twist: Electronic structure},\ }\href
  {https://doi.org/10.1103/PhysRevLett.99.256802} {\bibfield  {journal}
  {\bibinfo  {journal} {Phys. Rev. Lett.}\ }\textbf {\bibinfo {volume} {99}},\
  \bibinfo {pages} {256802} (\bibinfo {year} {2007})}\BibitemShut {NoStop}%
\bibitem [{\citenamefont {Su\'arez~Morell}\ \emph {et~al.}(2010)\citenamefont
  {Su\'arez~Morell}, \citenamefont {Correa}, \citenamefont {Vargas},
  \citenamefont {Pacheco},\ and\ \citenamefont {Barticevic}}]{Morell2010}%
  \BibitemOpen
  \bibfield  {author} {\bibinfo {author} {\bibfnamefont {E.}~\bibnamefont
  {Su\'arez~Morell}}, \bibinfo {author} {\bibfnamefont {J.~D.}\ \bibnamefont
  {Correa}}, \bibinfo {author} {\bibfnamefont {P.}~\bibnamefont {Vargas}},
  \bibinfo {author} {\bibfnamefont {M.}~\bibnamefont {Pacheco}},\ and\ \bibinfo
  {author} {\bibfnamefont {Z.}~\bibnamefont {Barticevic}},\ }\bibfield  {title}
  {\bibinfo {title} {Flat bands in slightly twisted bilayer graphene:
  Tight-binding calculations},\ }\href
  {https://doi.org/10.1103/PhysRevB.82.121407} {\bibfield  {journal} {\bibinfo
  {journal} {Phys. Rev. B}\ }\textbf {\bibinfo {volume} {82}},\ \bibinfo
  {pages} {121407} (\bibinfo {year} {2010})}\BibitemShut {NoStop}%
\bibitem [{\citenamefont {Bistritzer}\ and\ \citenamefont
  {MacDonald}(2011)}]{Bistritzer12233}%
  \BibitemOpen
  \bibfield  {author} {\bibinfo {author} {\bibfnamefont {R.}~\bibnamefont
  {Bistritzer}}\ and\ \bibinfo {author} {\bibfnamefont {A.~H.}\ \bibnamefont
  {MacDonald}},\ }\bibfield  {title} {\bibinfo {title} {Moir{\'e} bands in
  twisted double-layer graphene},\ }\href
  {https://doi.org/10.1073/pnas.1108174108} {\bibfield  {journal} {\bibinfo
  {journal} {Proceedings of the National Academy of Sciences}\ }\textbf
  {\bibinfo {volume} {108}},\ \bibinfo {pages} {12233} (\bibinfo {year}
  {2011})}\BibitemShut {NoStop}%
\bibitem [{\citenamefont {Li}\ \emph {et~al.}(2010)\citenamefont {Li},
  \citenamefont {Luican}, \citenamefont {Lopes~dos Santos}, \citenamefont
  {Castro~Neto}, \citenamefont {Reina}, \citenamefont {Kong},\ and\
  \citenamefont {Andrei}}]{Li2010}%
  \BibitemOpen
  \bibfield  {author} {\bibinfo {author} {\bibfnamefont {G.}~\bibnamefont
  {Li}}, \bibinfo {author} {\bibfnamefont {A.}~\bibnamefont {Luican}}, \bibinfo
  {author} {\bibfnamefont {J.~M.~B.}\ \bibnamefont {Lopes~dos Santos}},
  \bibinfo {author} {\bibfnamefont {A.~H.}\ \bibnamefont {Castro~Neto}},
  \bibinfo {author} {\bibfnamefont {A.}~\bibnamefont {Reina}}, \bibinfo
  {author} {\bibfnamefont {J.}~\bibnamefont {Kong}},\ and\ \bibinfo {author}
  {\bibfnamefont {E.~Y.}\ \bibnamefont {Andrei}},\ }\bibfield  {title}
  {\bibinfo {title} {Observation of {V}an {H}ove singularities in twisted
  graphene layers},\ }\href {https://doi.org/10.1038/nphys1463} {\bibfield
  {journal} {\bibinfo  {journal} {Nature Physics}\ }\textbf {\bibinfo {volume}
  {6}},\ \bibinfo {pages} {109} (\bibinfo {year} {2010})}\BibitemShut {NoStop}%
\bibitem [{\citenamefont {Cao}\ \emph {et~al.}(2018{\natexlab{a}})\citenamefont
  {Cao}, \citenamefont {Fatemi}, \citenamefont {Fang}, \citenamefont
  {Watanabe}, \citenamefont {Taniguchi}, \citenamefont {Kaxiras},\ and\
  \citenamefont {Jarillo-Herrero}}]{Cao2018a}%
  \BibitemOpen
  \bibfield  {author} {\bibinfo {author} {\bibfnamefont {Y.}~\bibnamefont
  {Cao}}, \bibinfo {author} {\bibfnamefont {V.}~\bibnamefont {Fatemi}},
  \bibinfo {author} {\bibfnamefont {S.}~\bibnamefont {Fang}}, \bibinfo {author}
  {\bibfnamefont {K.}~\bibnamefont {Watanabe}}, \bibinfo {author}
  {\bibfnamefont {T.}~\bibnamefont {Taniguchi}}, \bibinfo {author}
  {\bibfnamefont {E.}~\bibnamefont {Kaxiras}},\ and\ \bibinfo {author}
  {\bibfnamefont {P.}~\bibnamefont {Jarillo-Herrero}},\ }\bibfield  {title}
  {\bibinfo {title} {{Unconventional superconductivity in magic-angle graphene
  superlattices}},\ }\href {https://doi.org/10.1038/nature26160} {\bibfield
  {journal} {\bibinfo  {journal} {Nature}\ }\textbf {\bibinfo {volume} {556}},\
  \bibinfo {pages} {43} (\bibinfo {year} {2018}{\natexlab{a}})}\BibitemShut
  {NoStop}%
\bibitem [{\citenamefont {Cao}\ \emph {et~al.}(2018{\natexlab{b}})\citenamefont
  {Cao}, \citenamefont {Fatemi}, \citenamefont {Demir}, \citenamefont {Fang},
  \citenamefont {Tomarken}, \citenamefont {Luo}, \citenamefont
  {Sanchez-Yamagishi}, \citenamefont {Watanabe}, \citenamefont {Taniguchi},
  \citenamefont {Kaxiras}, \citenamefont {Ashoori},\ and\ \citenamefont
  {Jarillo-Herrero}}]{Cao2018}%
  \BibitemOpen
  \bibfield  {author} {\bibinfo {author} {\bibfnamefont {Y.}~\bibnamefont
  {Cao}}, \bibinfo {author} {\bibfnamefont {V.}~\bibnamefont {Fatemi}},
  \bibinfo {author} {\bibfnamefont {A.}~\bibnamefont {Demir}}, \bibinfo
  {author} {\bibfnamefont {S.}~\bibnamefont {Fang}}, \bibinfo {author}
  {\bibfnamefont {S.~L.}\ \bibnamefont {Tomarken}}, \bibinfo {author}
  {\bibfnamefont {J.~Y.}\ \bibnamefont {Luo}}, \bibinfo {author} {\bibfnamefont
  {J.~D.}\ \bibnamefont {Sanchez-Yamagishi}}, \bibinfo {author} {\bibfnamefont
  {K.}~\bibnamefont {Watanabe}}, \bibinfo {author} {\bibfnamefont
  {T.}~\bibnamefont {Taniguchi}}, \bibinfo {author} {\bibfnamefont
  {E.}~\bibnamefont {Kaxiras}}, \bibinfo {author} {\bibfnamefont {R.~C.}\
  \bibnamefont {Ashoori}},\ and\ \bibinfo {author} {\bibfnamefont
  {P.}~\bibnamefont {Jarillo-Herrero}},\ }\bibfield  {title} {\bibinfo {title}
  {{Correlated insulator behaviour at half-filling in magic-angle graphene
  superlattices}},\ }\href {https://doi.org/10.1038/nature26154} {\bibfield
  {journal} {\bibinfo  {journal} {Nature}\ }\textbf {\bibinfo {volume} {556}},\
  \bibinfo {pages} {80} (\bibinfo {year} {2018}{\natexlab{b}})}\BibitemShut
  {NoStop}%
\bibitem [{\citenamefont {Yankowitz}\ \emph {et~al.}(2019)\citenamefont
  {Yankowitz}, \citenamefont {Chen}, \citenamefont {Polshyn}, \citenamefont
  {Zhang}, \citenamefont {Watanabe}, \citenamefont {Taniguchi}, \citenamefont
  {Graf}, \citenamefont {Young},\ and\ \citenamefont {Dean}}]{Yankowitz2019}%
  \BibitemOpen
  \bibfield  {author} {\bibinfo {author} {\bibfnamefont {M.}~\bibnamefont
  {Yankowitz}}, \bibinfo {author} {\bibfnamefont {S.}~\bibnamefont {Chen}},
  \bibinfo {author} {\bibfnamefont {H.}~\bibnamefont {Polshyn}}, \bibinfo
  {author} {\bibfnamefont {Y.}~\bibnamefont {Zhang}}, \bibinfo {author}
  {\bibfnamefont {K.}~\bibnamefont {Watanabe}}, \bibinfo {author}
  {\bibfnamefont {T.}~\bibnamefont {Taniguchi}}, \bibinfo {author}
  {\bibfnamefont {D.}~\bibnamefont {Graf}}, \bibinfo {author} {\bibfnamefont
  {A.~F.}\ \bibnamefont {Young}},\ and\ \bibinfo {author} {\bibfnamefont
  {C.~R.}\ \bibnamefont {Dean}},\ }\bibfield  {title} {\bibinfo {title}
  {{Tuning superconductivity in twisted bilayer graphene}},\ }\href
  {https://doi.org/10.1126/science.aav1910} {\bibfield  {journal} {\bibinfo
  {journal} {Science}\ }\textbf {\bibinfo {volume} {363}},\ \bibinfo {pages}
  {1059} (\bibinfo {year} {2019})}\BibitemShut {NoStop}%
\bibitem [{\citenamefont {Kerelsky}\ \emph {et~al.}(2019)\citenamefont
  {Kerelsky}, \citenamefont {McGilly}, \citenamefont {Kennes}, \citenamefont
  {Xian}, \citenamefont {Yankowitz}, \citenamefont {Chen}, \citenamefont
  {Watanabe}, \citenamefont {Taniguchi}, \citenamefont {Hone}, \citenamefont
  {Dean}, \citenamefont {Rubio},\ and\ \citenamefont
  {Pasupathy}}]{Kerelsky2019}%
  \BibitemOpen
  \bibfield  {author} {\bibinfo {author} {\bibfnamefont {A.}~\bibnamefont
  {Kerelsky}}, \bibinfo {author} {\bibfnamefont {L.~J.}\ \bibnamefont
  {McGilly}}, \bibinfo {author} {\bibfnamefont {D.~M.}\ \bibnamefont {Kennes}},
  \bibinfo {author} {\bibfnamefont {L.}~\bibnamefont {Xian}}, \bibinfo {author}
  {\bibfnamefont {M.}~\bibnamefont {Yankowitz}}, \bibinfo {author}
  {\bibfnamefont {S.}~\bibnamefont {Chen}}, \bibinfo {author} {\bibfnamefont
  {K.}~\bibnamefont {Watanabe}}, \bibinfo {author} {\bibfnamefont
  {T.}~\bibnamefont {Taniguchi}}, \bibinfo {author} {\bibfnamefont
  {J.}~\bibnamefont {Hone}}, \bibinfo {author} {\bibfnamefont {C.}~\bibnamefont
  {Dean}}, \bibinfo {author} {\bibfnamefont {A.}~\bibnamefont {Rubio}},\ and\
  \bibinfo {author} {\bibfnamefont {A.~N.}\ \bibnamefont {Pasupathy}},\
  }\bibfield  {title} {\bibinfo {title} {Maximized electron interactions at the
  magic angle in twisted bilayer graphene},\ }\href
  {https://doi.org/10.1038/s41586-019-1431-9} {\bibfield  {journal} {\bibinfo
  {journal} {Nature}\ }\textbf {\bibinfo {volume} {572}},\ \bibinfo {pages}
  {95} (\bibinfo {year} {2019})}\BibitemShut {NoStop}%
\bibitem [{\citenamefont {Sharpe}\ \emph {et~al.}(2019)\citenamefont {Sharpe},
  \citenamefont {Fox}, \citenamefont {Barnard}, \citenamefont {Finney},
  \citenamefont {Watanabe}, \citenamefont {Taniguchi}, \citenamefont
  {Kastner},\ and\ \citenamefont {Goldhaber-Gordon}}]{Sharpe605}%
  \BibitemOpen
  \bibfield  {author} {\bibinfo {author} {\bibfnamefont {A.~L.}\ \bibnamefont
  {Sharpe}}, \bibinfo {author} {\bibfnamefont {E.~J.}\ \bibnamefont {Fox}},
  \bibinfo {author} {\bibfnamefont {A.~W.}\ \bibnamefont {Barnard}}, \bibinfo
  {author} {\bibfnamefont {J.}~\bibnamefont {Finney}}, \bibinfo {author}
  {\bibfnamefont {K.}~\bibnamefont {Watanabe}}, \bibinfo {author}
  {\bibfnamefont {T.}~\bibnamefont {Taniguchi}}, \bibinfo {author}
  {\bibfnamefont {M.~A.}\ \bibnamefont {Kastner}},\ and\ \bibinfo {author}
  {\bibfnamefont {D.}~\bibnamefont {Goldhaber-Gordon}},\ }\bibfield  {title}
  {\bibinfo {title} {Emergent ferromagnetism near three-quarters filling in
  twisted bilayer graphene},\ }\href {https://doi.org/10.1126/science.aaw3780}
  {\bibfield  {journal} {\bibinfo  {journal} {Science}\ }\textbf {\bibinfo
  {volume} {365}},\ \bibinfo {pages} {605} (\bibinfo {year}
  {2019})}\BibitemShut {NoStop}%
\bibitem [{\citenamefont {{Lu}}\ \emph {et~al.}(2019)\citenamefont {{Lu}},
  \citenamefont {{Stepanov}}, \citenamefont {{Yang}}, \citenamefont {{Xie}},
  \citenamefont {{Aamir}}, \citenamefont {{Das}}, \citenamefont {{Urgell}},
  \citenamefont {{Watanabe}}, \citenamefont {{Taniguchi}}, \citenamefont
  {{Zhang}}, \citenamefont {{Bachtold}}, \citenamefont {{MacDonald}},\ and\
  \citenamefont {{Efetov}}}]{lu2019superconductors}%
  \BibitemOpen
  \bibfield  {author} {\bibinfo {author} {\bibfnamefont {X.}~\bibnamefont
  {{Lu}}}, \bibinfo {author} {\bibfnamefont {P.}~\bibnamefont {{Stepanov}}},
  \bibinfo {author} {\bibfnamefont {W.}~\bibnamefont {{Yang}}}, \bibinfo
  {author} {\bibfnamefont {M.}~\bibnamefont {{Xie}}}, \bibinfo {author}
  {\bibfnamefont {M.~A.}\ \bibnamefont {{Aamir}}}, \bibinfo {author}
  {\bibfnamefont {I.}~\bibnamefont {{Das}}}, \bibinfo {author} {\bibfnamefont
  {C.}~\bibnamefont {{Urgell}}}, \bibinfo {author} {\bibfnamefont
  {K.}~\bibnamefont {{Watanabe}}}, \bibinfo {author} {\bibfnamefont
  {T.}~\bibnamefont {{Taniguchi}}}, \bibinfo {author} {\bibfnamefont
  {G.}~\bibnamefont {{Zhang}}}, \bibinfo {author} {\bibfnamefont
  {A.}~\bibnamefont {{Bachtold}}}, \bibinfo {author} {\bibfnamefont {A.~H.}\
  \bibnamefont {{MacDonald}}},\ and\ \bibinfo {author} {\bibfnamefont {D.~K.}\
  \bibnamefont {{Efetov}}},\ }\bibfield  {title} {\bibinfo {title}
  {{Superconductors , orbital magnets and correlated states in magic-angle
  bilayer graphene}},\ }\href {https://doi.org/10.1038/s41586-019-1695-0}
  {\bibfield  {journal} {\bibinfo  {journal} {Nature}\ }\textbf {\bibinfo
  {volume} {574}},\ \bibinfo {pages} {20} (\bibinfo {year} {2019})}\BibitemShut
  {NoStop}%
\bibitem [{\citenamefont {Serlin}\ \emph {et~al.}(2020)\citenamefont {Serlin},
  \citenamefont {Tschirhart}, \citenamefont {Polshyn}, \citenamefont {Zhang},
  \citenamefont {Zhu}, \citenamefont {Watanabe}, \citenamefont {Taniguchi},
  \citenamefont {Balents},\ and\ \citenamefont {Young}}]{Serlin2019}%
  \BibitemOpen
  \bibfield  {author} {\bibinfo {author} {\bibfnamefont {M.}~\bibnamefont
  {Serlin}}, \bibinfo {author} {\bibfnamefont {C.~L.}\ \bibnamefont
  {Tschirhart}}, \bibinfo {author} {\bibfnamefont {H.}~\bibnamefont {Polshyn}},
  \bibinfo {author} {\bibfnamefont {Y.}~\bibnamefont {Zhang}}, \bibinfo
  {author} {\bibfnamefont {J.}~\bibnamefont {Zhu}}, \bibinfo {author}
  {\bibfnamefont {K.}~\bibnamefont {Watanabe}}, \bibinfo {author}
  {\bibfnamefont {T.}~\bibnamefont {Taniguchi}}, \bibinfo {author}
  {\bibfnamefont {L.}~\bibnamefont {Balents}},\ and\ \bibinfo {author}
  {\bibfnamefont {A.~F.}\ \bibnamefont {Young}},\ }\bibfield  {title} {\bibinfo
  {title} {Intrinsic quantized anomalous {H}all effect in a moir{\'e}
  heterostructure},\ }\href {https://doi.org/10.1126/science.aay5533}
  {\bibfield  {journal} {\bibinfo  {journal} {Science}\ }\textbf {\bibinfo
  {volume} {367}},\ \bibinfo {pages} {900} (\bibinfo {year}
  {2020})}\BibitemShut {NoStop}%
\bibitem [{\citenamefont {MacDonald}(2019)}]{MacDonald2019}%
  \BibitemOpen
  \bibfield  {author} {\bibinfo {author} {\bibfnamefont {A.~H.}\ \bibnamefont
  {MacDonald}},\ }\bibfield  {title} {\bibinfo {title} {Bilayer graphene’s
  wicked, twisted road},\ }\bibfield  {journal} {\bibinfo  {journal} {Physics}\
  }\href {https://doi.org/10.1103/Physics.12.12} {10.1103/Physics.12.12}
  (\bibinfo {year} {2019})\BibitemShut {NoStop}%
\bibitem [{\citenamefont {Andrei}\ and\ \citenamefont
  {MacDonald}(2020)}]{Andrei2020}%
  \BibitemOpen
  \bibfield  {author} {\bibinfo {author} {\bibfnamefont {E.~Y.}\ \bibnamefont
  {Andrei}}\ and\ \bibinfo {author} {\bibfnamefont {A.~H.}\ \bibnamefont
  {MacDonald}},\ }\bibfield  {title} {\bibinfo {title} {Graphene bilayers with
  a twist},\ }\href {https://doi.org/10.1038/s41563-020-00840-0} {\bibfield
  {journal} {\bibinfo  {journal} {Nature Materials}\ }\textbf {\bibinfo
  {volume} {19}},\ \bibinfo {pages} {1265} (\bibinfo {year}
  {2020})}\BibitemShut {NoStop}%
\bibitem [{\citenamefont {Balents}\ \emph {et~al.}(2020)\citenamefont
  {Balents}, \citenamefont {Dean}, \citenamefont {Efetov},\ and\ \citenamefont
  {Young}}]{Balents2020}%
  \BibitemOpen
  \bibfield  {author} {\bibinfo {author} {\bibfnamefont {L.}~\bibnamefont
  {Balents}}, \bibinfo {author} {\bibfnamefont {C.~R.}\ \bibnamefont {Dean}},
  \bibinfo {author} {\bibfnamefont {D.~K.}\ \bibnamefont {Efetov}},\ and\
  \bibinfo {author} {\bibfnamefont {A.~F.}\ \bibnamefont {Young}},\ }\bibfield
  {title} {\bibinfo {title} {Superconductivity and strong correlations in
  moir{\'e} flat bands},\ }\href {https://doi.org/10.1038/s41567-020-0906-9}
  {\bibfield  {journal} {\bibinfo  {journal} {Nature Physics}\ }\textbf
  {\bibinfo {volume} {16}},\ \bibinfo {pages} {725} (\bibinfo {year}
  {2020})}\BibitemShut {NoStop}%
\bibitem [{\citenamefont {Kennes}\ \emph {et~al.}(2021)\citenamefont {Kennes},
  \citenamefont {Claassen}, \citenamefont {Xian}, \citenamefont {Georges},
  \citenamefont {Millis}, \citenamefont {Hone}, \citenamefont {Dean},
  \citenamefont {Basov}, \citenamefont {Pasupathy},\ and\ \citenamefont
  {Rubio}}]{Kennes2021}%
  \BibitemOpen
  \bibfield  {author} {\bibinfo {author} {\bibfnamefont {D.~M.}\ \bibnamefont
  {Kennes}}, \bibinfo {author} {\bibfnamefont {M.}~\bibnamefont {Claassen}},
  \bibinfo {author} {\bibfnamefont {L.}~\bibnamefont {Xian}}, \bibinfo {author}
  {\bibfnamefont {A.}~\bibnamefont {Georges}}, \bibinfo {author} {\bibfnamefont
  {A.~J.}\ \bibnamefont {Millis}}, \bibinfo {author} {\bibfnamefont
  {J.}~\bibnamefont {Hone}}, \bibinfo {author} {\bibfnamefont {C.~R.}\
  \bibnamefont {Dean}}, \bibinfo {author} {\bibfnamefont {D.~N.}\ \bibnamefont
  {Basov}}, \bibinfo {author} {\bibfnamefont {A.~N.}\ \bibnamefont
  {Pasupathy}},\ and\ \bibinfo {author} {\bibfnamefont {A.}~\bibnamefont
  {Rubio}},\ }\bibfield  {title} {\bibinfo {title} {Moir{\'e} heterostructures
  as a condensed-matter quantum simulator},\ }\href
  {https://doi.org/10.1038/s41567-020-01154-3} {\bibfield  {journal} {\bibinfo
  {journal} {Nature Physics}\ }\textbf {\bibinfo {volume} {17}},\ \bibinfo
  {pages} {155} (\bibinfo {year} {2021})}\BibitemShut {NoStop}%
\bibitem [{\citenamefont {Wu}\ \emph {et~al.}(2018{\natexlab{a}})\citenamefont
  {Wu}, \citenamefont {MacDonald},\ and\ \citenamefont
  {Martin}}]{PhysRevLett.121.257001}%
  \BibitemOpen
  \bibfield  {author} {\bibinfo {author} {\bibfnamefont {F.}~\bibnamefont
  {Wu}}, \bibinfo {author} {\bibfnamefont {A.~H.}\ \bibnamefont {MacDonald}},\
  and\ \bibinfo {author} {\bibfnamefont {I.}~\bibnamefont {Martin}},\
  }\bibfield  {title} {\bibinfo {title} {Theory of phonon-mediated
  superconductivity in twisted bilayer graphene},\ }\href
  {https://doi.org/10.1103/PhysRevLett.121.257001} {\bibfield  {journal}
  {\bibinfo  {journal} {Phys. Rev. Lett.}\ }\textbf {\bibinfo {volume} {121}},\
  \bibinfo {pages} {257001} (\bibinfo {year} {2018}{\natexlab{a}})}\BibitemShut
  {NoStop}%
\bibitem [{\citenamefont {Choi}\ and\ \citenamefont
  {Choi}(2018)}]{PhysRevB.98.241412}%
  \BibitemOpen
  \bibfield  {author} {\bibinfo {author} {\bibfnamefont {Y.~W.}\ \bibnamefont
  {Choi}}\ and\ \bibinfo {author} {\bibfnamefont {H.~J.}\ \bibnamefont
  {Choi}},\ }\bibfield  {title} {\bibinfo {title} {Strong electron-phonon
  coupling, electron-hole asymmetry, and nonadiabaticity in magic-angle twisted
  bilayer graphene},\ }\href {https://doi.org/10.1103/PhysRevB.98.241412}
  {\bibfield  {journal} {\bibinfo  {journal} {Phys. Rev. B}\ }\textbf {\bibinfo
  {volume} {98}},\ \bibinfo {pages} {241412} (\bibinfo {year}
  {2018})}\BibitemShut {NoStop}%
\bibitem [{\citenamefont {Peltonen}\ \emph {et~al.}(2018)\citenamefont
  {Peltonen}, \citenamefont {Ojaj\"arvi},\ and\ \citenamefont
  {Heikkil\"a}}]{PhysRevB.98.220504}%
  \BibitemOpen
  \bibfield  {author} {\bibinfo {author} {\bibfnamefont {T.~J.}\ \bibnamefont
  {Peltonen}}, \bibinfo {author} {\bibfnamefont {R.}~\bibnamefont
  {Ojaj\"arvi}},\ and\ \bibinfo {author} {\bibfnamefont {T.~T.}\ \bibnamefont
  {Heikkil\"a}},\ }\bibfield  {title} {\bibinfo {title} {Mean-field theory for
  superconductivity in twisted bilayer graphene},\ }\href
  {https://doi.org/10.1103/PhysRevB.98.220504} {\bibfield  {journal} {\bibinfo
  {journal} {Phys. Rev. B}\ }\textbf {\bibinfo {volume} {98}},\ \bibinfo
  {pages} {220504} (\bibinfo {year} {2018})}\BibitemShut {NoStop}%
\bibitem [{\citenamefont {Angeli}\ \emph {et~al.}(2019)\citenamefont {Angeli},
  \citenamefont {Tosatti},\ and\ \citenamefont {Fabrizio}}]{PhysRevX.9.041010}%
  \BibitemOpen
  \bibfield  {author} {\bibinfo {author} {\bibfnamefont {M.}~\bibnamefont
  {Angeli}}, \bibinfo {author} {\bibfnamefont {E.}~\bibnamefont {Tosatti}},\
  and\ \bibinfo {author} {\bibfnamefont {M.}~\bibnamefont {Fabrizio}},\
  }\bibfield  {title} {\bibinfo {title} {Valley {J}ahn-{T}eller effect in
  twisted bilayer graphene},\ }\href
  {https://doi.org/10.1103/PhysRevX.9.041010} {\bibfield  {journal} {\bibinfo
  {journal} {Phys. Rev. X}\ }\textbf {\bibinfo {volume} {9}},\ \bibinfo {pages}
  {041010} (\bibinfo {year} {2019})}\BibitemShut {NoStop}%
\bibitem [{\citenamefont {Lian}\ \emph {et~al.}(2019)\citenamefont {Lian},
  \citenamefont {Wang},\ and\ \citenamefont
  {Bernevig}}]{PhysRevLett.122.257002}%
  \BibitemOpen
  \bibfield  {author} {\bibinfo {author} {\bibfnamefont {B.}~\bibnamefont
  {Lian}}, \bibinfo {author} {\bibfnamefont {Z.}~\bibnamefont {Wang}},\ and\
  \bibinfo {author} {\bibfnamefont {B.~A.}\ \bibnamefont {Bernevig}},\
  }\bibfield  {title} {\bibinfo {title} {Twisted bilayer graphene: A
  phonon-driven superconductor},\ }\href
  {https://doi.org/10.1103/PhysRevLett.122.257002} {\bibfield  {journal}
  {\bibinfo  {journal} {Phys. Rev. Lett.}\ }\textbf {\bibinfo {volume} {122}},\
  \bibinfo {pages} {257002} (\bibinfo {year} {2019})}\BibitemShut {NoStop}%
\bibitem [{\citenamefont {Samajdar}\ and\ \citenamefont
  {Scheurer}(2020)}]{PhysRevB.102.064501}%
  \BibitemOpen
  \bibfield  {author} {\bibinfo {author} {\bibfnamefont {R.}~\bibnamefont
  {Samajdar}}\ and\ \bibinfo {author} {\bibfnamefont {M.~S.}\ \bibnamefont
  {Scheurer}},\ }\bibfield  {title} {\bibinfo {title} {Microscopic pairing
  mechanism, order parameter, and disorder sensitivity in moir\'e
  superlattices: Applications to twisted double-bilayer graphene},\ }\href
  {https://doi.org/10.1103/PhysRevB.102.064501} {\bibfield  {journal} {\bibinfo
   {journal} {Phys. Rev. B}\ }\textbf {\bibinfo {volume} {102}},\ \bibinfo
  {pages} {064501} (\bibinfo {year} {2020})}\BibitemShut {NoStop}%
\bibitem [{\citenamefont {Venderbos}\ and\ \citenamefont
  {Fernandes}(2018)}]{PhysRevB.98.245103}%
  \BibitemOpen
  \bibfield  {author} {\bibinfo {author} {\bibfnamefont {J.~W.~F.}\
  \bibnamefont {Venderbos}}\ and\ \bibinfo {author} {\bibfnamefont {R.~M.}\
  \bibnamefont {Fernandes}},\ }\bibfield  {title} {\bibinfo {title}
  {Correlations and electronic order in a two-orbital honeycomb lattice model
  for twisted bilayer graphene},\ }\href
  {https://doi.org/10.1103/PhysRevB.98.245103} {\bibfield  {journal} {\bibinfo
  {journal} {Phys. Rev. B}\ }\textbf {\bibinfo {volume} {98}},\ \bibinfo
  {pages} {245103} (\bibinfo {year} {2018})}\BibitemShut {NoStop}%
\bibitem [{\citenamefont {Isobe}\ \emph {et~al.}(2018)\citenamefont {Isobe},
  \citenamefont {Yuan},\ and\ \citenamefont {Fu}}]{PhysRevX.8.041041}%
  \BibitemOpen
  \bibfield  {author} {\bibinfo {author} {\bibfnamefont {H.}~\bibnamefont
  {Isobe}}, \bibinfo {author} {\bibfnamefont {N.~F.~Q.}\ \bibnamefont {Yuan}},\
  and\ \bibinfo {author} {\bibfnamefont {L.}~\bibnamefont {Fu}},\ }\bibfield
  {title} {\bibinfo {title} {Unconventional superconductivity and density waves
  in twisted bilayer graphene},\ }\href
  {https://doi.org/10.1103/PhysRevX.8.041041} {\bibfield  {journal} {\bibinfo
  {journal} {Phys. Rev. X}\ }\textbf {\bibinfo {volume} {8}},\ \bibinfo {pages}
  {041041} (\bibinfo {year} {2018})}\BibitemShut {NoStop}%
\bibitem [{\citenamefont {Sherkunov}\ and\ \citenamefont
  {Betouras}(2018)}]{PhysRevB.98.205151}%
  \BibitemOpen
  \bibfield  {author} {\bibinfo {author} {\bibfnamefont {Y.}~\bibnamefont
  {Sherkunov}}\ and\ \bibinfo {author} {\bibfnamefont {J.~J.}\ \bibnamefont
  {Betouras}},\ }\bibfield  {title} {\bibinfo {title} {Electronic phases in
  twisted bilayer graphene at magic angles as a result of van hove
  singularities and interactions},\ }\href
  {https://doi.org/10.1103/PhysRevB.98.205151} {\bibfield  {journal} {\bibinfo
  {journal} {Phys. Rev. B}\ }\textbf {\bibinfo {volume} {98}},\ \bibinfo
  {pages} {205151} (\bibinfo {year} {2018})}\BibitemShut {NoStop}%
\bibitem [{\citenamefont {Lin}\ and\ \citenamefont
  {Nandkishore}(2018)}]{PhysRevB.98.214521}%
  \BibitemOpen
  \bibfield  {author} {\bibinfo {author} {\bibfnamefont {Y.-P.}\ \bibnamefont
  {Lin}}\ and\ \bibinfo {author} {\bibfnamefont {R.~M.}\ \bibnamefont
  {Nandkishore}},\ }\bibfield  {title} {\bibinfo {title} {Kohn-{L}uttinger
  superconductivity on two orbital honeycomb lattice},\ }\href
  {https://doi.org/10.1103/PhysRevB.98.214521} {\bibfield  {journal} {\bibinfo
  {journal} {Phys. Rev. B}\ }\textbf {\bibinfo {volume} {98}},\ \bibinfo
  {pages} {214521} (\bibinfo {year} {2018})}\BibitemShut {NoStop}%
\bibitem [{\citenamefont {Dodaro}\ \emph {et~al.}(2018)\citenamefont {Dodaro},
  \citenamefont {Kivelson}, \citenamefont {Schattner}, \citenamefont {Sun},\
  and\ \citenamefont {Wang}}]{PhysRevB.98.075154}%
  \BibitemOpen
  \bibfield  {author} {\bibinfo {author} {\bibfnamefont {J.~F.}\ \bibnamefont
  {Dodaro}}, \bibinfo {author} {\bibfnamefont {S.~A.}\ \bibnamefont
  {Kivelson}}, \bibinfo {author} {\bibfnamefont {Y.}~\bibnamefont {Schattner}},
  \bibinfo {author} {\bibfnamefont {X.~Q.}\ \bibnamefont {Sun}},\ and\ \bibinfo
  {author} {\bibfnamefont {C.}~\bibnamefont {Wang}},\ }\bibfield  {title}
  {\bibinfo {title} {Phases of a phenomenological model of twisted bilayer
  graphene},\ }\href {https://doi.org/10.1103/PhysRevB.98.075154} {\bibfield
  {journal} {\bibinfo  {journal} {Phys. Rev. B}\ }\textbf {\bibinfo {volume}
  {98}},\ \bibinfo {pages} {075154} (\bibinfo {year} {2018})}\BibitemShut
  {NoStop}%
\bibitem [{\citenamefont {Xu}\ and\ \citenamefont
  {Balents}(2018)}]{PhysRevLett.121.087001}%
  \BibitemOpen
  \bibfield  {author} {\bibinfo {author} {\bibfnamefont {C.}~\bibnamefont
  {Xu}}\ and\ \bibinfo {author} {\bibfnamefont {L.}~\bibnamefont {Balents}},\
  }\bibfield  {title} {\bibinfo {title} {Topological superconductivity in
  twisted multilayer graphene},\ }\href
  {https://doi.org/10.1103/PhysRevLett.121.087001} {\bibfield  {journal}
  {\bibinfo  {journal} {Phys. Rev. Lett.}\ }\textbf {\bibinfo {volume} {121}},\
  \bibinfo {pages} {087001} (\bibinfo {year} {2018})}\BibitemShut {NoStop}%
\bibitem [{\citenamefont {Fidrysiak}\ \emph {et~al.}(2018)\citenamefont
  {Fidrysiak}, \citenamefont {Zegrodnik},\ and\ \citenamefont
  {Spa\l{}ek}}]{PhysRevB.98.085436}%
  \BibitemOpen
  \bibfield  {author} {\bibinfo {author} {\bibfnamefont {M.}~\bibnamefont
  {Fidrysiak}}, \bibinfo {author} {\bibfnamefont {M.}~\bibnamefont
  {Zegrodnik}},\ and\ \bibinfo {author} {\bibfnamefont {J.}~\bibnamefont
  {Spa\l{}ek}},\ }\bibfield  {title} {\bibinfo {title} {Unconventional
  topological superconductivity and phase diagram for an effective two-orbital
  model as applied to twisted bilayer graphene},\ }\href
  {https://doi.org/10.1103/PhysRevB.98.085436} {\bibfield  {journal} {\bibinfo
  {journal} {Phys. Rev. B}\ }\textbf {\bibinfo {volume} {98}},\ \bibinfo
  {pages} {085436} (\bibinfo {year} {2018})}\BibitemShut {NoStop}%
\bibitem [{\citenamefont {Liu}\ \emph {et~al.}(2018)\citenamefont {Liu},
  \citenamefont {Zhang}, \citenamefont {Chen},\ and\ \citenamefont
  {Yang}}]{PhysRevLett.121.217001}%
  \BibitemOpen
  \bibfield  {author} {\bibinfo {author} {\bibfnamefont {C.-C.}\ \bibnamefont
  {Liu}}, \bibinfo {author} {\bibfnamefont {L.-D.}\ \bibnamefont {Zhang}},
  \bibinfo {author} {\bibfnamefont {W.-Q.}\ \bibnamefont {Chen}},\ and\
  \bibinfo {author} {\bibfnamefont {F.}~\bibnamefont {Yang}},\ }\bibfield
  {title} {\bibinfo {title} {Chiral spin density wave and $d+id$
  superconductivity in the magic-angle-twisted bilayer graphene},\ }\href
  {https://doi.org/10.1103/PhysRevLett.121.217001} {\bibfield  {journal}
  {\bibinfo  {journal} {Phys. Rev. Lett.}\ }\textbf {\bibinfo {volume} {121}},\
  \bibinfo {pages} {217001} (\bibinfo {year} {2018})}\BibitemShut {NoStop}%
\bibitem [{\citenamefont {Su}\ and\ \citenamefont
  {Lin}(2018{\natexlab{a}})}]{PhysRevB.98.195101}%
  \BibitemOpen
  \bibfield  {author} {\bibinfo {author} {\bibfnamefont {Y.}~\bibnamefont
  {Su}}\ and\ \bibinfo {author} {\bibfnamefont {S.-Z.}\ \bibnamefont {Lin}},\
  }\bibfield  {title} {\bibinfo {title} {Pairing symmetry and spontaneous
  vortex-antivortex lattice in superconducting twisted-bilayer graphene:
  {B}ogoliubov-de {G}ennes approach},\ }\href
  {https://doi.org/10.1103/PhysRevB.98.195101} {\bibfield  {journal} {\bibinfo
  {journal} {Phys. Rev. B}\ }\textbf {\bibinfo {volume} {98}},\ \bibinfo
  {pages} {195101} (\bibinfo {year} {2018}{\natexlab{a}})}\BibitemShut
  {NoStop}%
\bibitem [{\citenamefont {Kennes}\ \emph {et~al.}(2018)\citenamefont {Kennes},
  \citenamefont {Lischner},\ and\ \citenamefont
  {Karrasch}}]{PhysRevB.98.241407}%
  \BibitemOpen
  \bibfield  {author} {\bibinfo {author} {\bibfnamefont {D.~M.}\ \bibnamefont
  {Kennes}}, \bibinfo {author} {\bibfnamefont {J.}~\bibnamefont {Lischner}},\
  and\ \bibinfo {author} {\bibfnamefont {C.}~\bibnamefont {Karrasch}},\
  }\bibfield  {title} {\bibinfo {title} {Strong correlations and
  $d+\mathit{id}$ superconductivity in twisted bilayer graphene},\ }\href
  {https://doi.org/10.1103/PhysRevB.98.241407} {\bibfield  {journal} {\bibinfo
  {journal} {Phys. Rev. B}\ }\textbf {\bibinfo {volume} {98}},\ \bibinfo
  {pages} {241407} (\bibinfo {year} {2018})}\BibitemShut {NoStop}%
\bibitem [{\citenamefont {Po}\ \emph {et~al.}(2018{\natexlab{a}})\citenamefont
  {Po}, \citenamefont {Zou}, \citenamefont {Vishwanath},\ and\ \citenamefont
  {Senthil}}]{PhysRevX.8.031089}%
  \BibitemOpen
  \bibfield  {author} {\bibinfo {author} {\bibfnamefont {H.~C.}\ \bibnamefont
  {Po}}, \bibinfo {author} {\bibfnamefont {L.}~\bibnamefont {Zou}}, \bibinfo
  {author} {\bibfnamefont {A.}~\bibnamefont {Vishwanath}},\ and\ \bibinfo
  {author} {\bibfnamefont {T.}~\bibnamefont {Senthil}},\ }\bibfield  {title}
  {\bibinfo {title} {Origin of {M}ott insulating behavior and superconductivity
  in twisted bilayer graphene},\ }\href
  {https://doi.org/10.1103/PhysRevX.8.031089} {\bibfield  {journal} {\bibinfo
  {journal} {Phys. Rev. X}\ }\textbf {\bibinfo {volume} {8}},\ \bibinfo {pages}
  {031089} (\bibinfo {year} {2018}{\natexlab{a}})}\BibitemShut {NoStop}%
\bibitem [{\citenamefont {You}\ and\ \citenamefont
  {Vishwanath}(2019)}]{You2019}%
  \BibitemOpen
  \bibfield  {author} {\bibinfo {author} {\bibfnamefont {Y.-Z.}\ \bibnamefont
  {You}}\ and\ \bibinfo {author} {\bibfnamefont {A.}~\bibnamefont
  {Vishwanath}},\ }\bibfield  {title} {\bibinfo {title} {Superconductivity from
  valley fluctuations and approximate {SO}(4) symmetry in a weak coupling
  theory of twisted bilayer graphene},\ }\href
  {https://doi.org/10.1038/s41535-019-0153-4} {\bibfield  {journal} {\bibinfo
  {journal} {npj Quantum Materials}\ }\textbf {\bibinfo {volume} {4}},\
  \bibinfo {pages} {16} (\bibinfo {year} {2019})}\BibitemShut {NoStop}%
\bibitem [{\citenamefont {Classen}\ \emph {et~al.}(2019)\citenamefont
  {Classen}, \citenamefont {Honerkamp},\ and\ \citenamefont
  {Scherer}}]{PhysRevB.99.195120}%
  \BibitemOpen
  \bibfield  {author} {\bibinfo {author} {\bibfnamefont {L.}~\bibnamefont
  {Classen}}, \bibinfo {author} {\bibfnamefont {C.}~\bibnamefont {Honerkamp}},\
  and\ \bibinfo {author} {\bibfnamefont {M.~M.}\ \bibnamefont {Scherer}},\
  }\bibfield  {title} {\bibinfo {title} {Competing phases of interacting
  electrons on triangular lattices in moir\'e heterostructures},\ }\href
  {https://doi.org/10.1103/PhysRevB.99.195120} {\bibfield  {journal} {\bibinfo
  {journal} {Phys. Rev. B}\ }\textbf {\bibinfo {volume} {99}},\ \bibinfo
  {pages} {195120} (\bibinfo {year} {2019})}\BibitemShut {NoStop}%
\bibitem [{\citenamefont {Ray}\ \emph {et~al.}(2019)\citenamefont {Ray},
  \citenamefont {Jung},\ and\ \citenamefont {Das}}]{PhysRevB.99.134515}%
  \BibitemOpen
  \bibfield  {author} {\bibinfo {author} {\bibfnamefont {S.}~\bibnamefont
  {Ray}}, \bibinfo {author} {\bibfnamefont {J.}~\bibnamefont {Jung}},\ and\
  \bibinfo {author} {\bibfnamefont {T.}~\bibnamefont {Das}},\ }\bibfield
  {title} {\bibinfo {title} {Wannier pairs in superconducting twisted bilayer
  graphene and related systems},\ }\href
  {https://doi.org/10.1103/PhysRevB.99.134515} {\bibfield  {journal} {\bibinfo
  {journal} {Phys. Rev. B}\ }\textbf {\bibinfo {volume} {99}},\ \bibinfo
  {pages} {134515} (\bibinfo {year} {2019})}\BibitemShut {NoStop}%
\bibitem [{\citenamefont {Gonz\'alez}\ and\ \citenamefont
  {Stauber}(2019)}]{PhysRevLett.122.026801}%
  \BibitemOpen
  \bibfield  {author} {\bibinfo {author} {\bibfnamefont {J.}~\bibnamefont
  {Gonz\'alez}}\ and\ \bibinfo {author} {\bibfnamefont {T.}~\bibnamefont
  {Stauber}},\ }\bibfield  {title} {\bibinfo {title} {Kohn-{L}uttinger
  superconductivity in twisted bilayer graphene},\ }\href
  {https://doi.org/10.1103/PhysRevLett.122.026801} {\bibfield  {journal}
  {\bibinfo  {journal} {Phys. Rev. Lett.}\ }\textbf {\bibinfo {volume} {122}},\
  \bibinfo {pages} {026801} (\bibinfo {year} {2019})}\BibitemShut {NoStop}%
\bibitem [{\citenamefont {Lin}\ and\ \citenamefont
  {Nandkishore}(2019)}]{PhysRevB.100.085136}%
  \BibitemOpen
  \bibfield  {author} {\bibinfo {author} {\bibfnamefont {Y.-P.}\ \bibnamefont
  {Lin}}\ and\ \bibinfo {author} {\bibfnamefont {R.~M.}\ \bibnamefont
  {Nandkishore}},\ }\bibfield  {title} {\bibinfo {title} {Chiral twist on the
  high-${T}_{c}$ phase diagram in moir\'e heterostructures},\ }\href
  {https://doi.org/10.1103/PhysRevB.100.085136} {\bibfield  {journal} {\bibinfo
   {journal} {Phys. Rev. B}\ }\textbf {\bibinfo {volume} {100}},\ \bibinfo
  {pages} {085136} (\bibinfo {year} {2019})}\BibitemShut {NoStop}%
\bibitem [{\citenamefont {Claassen}\ \emph {et~al.}(2019)\citenamefont
  {Claassen}, \citenamefont {Kennes}, \citenamefont {Zingl}, \citenamefont
  {Sentef},\ and\ \citenamefont {Rubio}}]{Claassen2019}%
  \BibitemOpen
  \bibfield  {author} {\bibinfo {author} {\bibfnamefont {M.}~\bibnamefont
  {Claassen}}, \bibinfo {author} {\bibfnamefont {D.~M.}\ \bibnamefont
  {Kennes}}, \bibinfo {author} {\bibfnamefont {M.}~\bibnamefont {Zingl}},
  \bibinfo {author} {\bibfnamefont {M.~A.}\ \bibnamefont {Sentef}},\ and\
  \bibinfo {author} {\bibfnamefont {A.}~\bibnamefont {Rubio}},\ }\bibfield
  {title} {\bibinfo {title} {Universal optical control of chiral
  superconductors and {M}ajorana modes},\ }\href
  {https://doi.org/10.1038/s41567-019-0532-6} {\bibfield  {journal} {\bibinfo
  {journal} {Nature Physics}\ }\textbf {\bibinfo {volume} {15}},\ \bibinfo
  {pages} {766} (\bibinfo {year} {2019})}\BibitemShut {NoStop}%
\bibitem [{\citenamefont {Chen}\ \emph
  {et~al.}(2020{\natexlab{a}})\citenamefont {Chen}, \citenamefont {Chu},
  \citenamefont {Huang},\ and\ \citenamefont {Ma}}]{PhysRevB.101.155413}%
  \BibitemOpen
  \bibfield  {author} {\bibinfo {author} {\bibfnamefont {W.}~\bibnamefont
  {Chen}}, \bibinfo {author} {\bibfnamefont {Y.}~\bibnamefont {Chu}}, \bibinfo
  {author} {\bibfnamefont {T.}~\bibnamefont {Huang}},\ and\ \bibinfo {author}
  {\bibfnamefont {T.}~\bibnamefont {Ma}},\ }\bibfield  {title} {\bibinfo
  {title} {Metal-insulator transition and dominant $d+id$ pairing symmetry in
  twisted bilayer graphene},\ }\href
  {https://doi.org/10.1103/PhysRevB.101.155413} {\bibfield  {journal} {\bibinfo
   {journal} {Phys. Rev. B}\ }\textbf {\bibinfo {volume} {101}},\ \bibinfo
  {pages} {155413} (\bibinfo {year} {2020}{\natexlab{a}})}\BibitemShut
  {NoStop}%
\bibitem [{\citenamefont {Fischer}\ \emph {et~al.}(2021)\citenamefont
  {Fischer}, \citenamefont {Klebl}, \citenamefont {Honerkamp},\ and\
  \citenamefont {Kennes}}]{PhysRevB.103.L041103}%
  \BibitemOpen
  \bibfield  {author} {\bibinfo {author} {\bibfnamefont {A.}~\bibnamefont
  {Fischer}}, \bibinfo {author} {\bibfnamefont {L.}~\bibnamefont {Klebl}},
  \bibinfo {author} {\bibfnamefont {C.}~\bibnamefont {Honerkamp}},\ and\
  \bibinfo {author} {\bibfnamefont {D.~M.}\ \bibnamefont {Kennes}},\ }\bibfield
   {title} {\bibinfo {title} {Spin-fluctuation-induced pairing in twisted
  bilayer graphene},\ }\href {https://doi.org/10.1103/PhysRevB.103.L041103}
  {\bibfield  {journal} {\bibinfo  {journal} {Phys. Rev. B}\ }\textbf {\bibinfo
  {volume} {103}},\ \bibinfo {pages} {L041103} (\bibinfo {year}
  {2021})}\BibitemShut {NoStop}%
\bibitem [{\citenamefont {Wang}\ \emph {et~al.}(2021)\citenamefont {Wang},
  \citenamefont {Kang},\ and\ \citenamefont {Fernandes}}]{PhysRevB.103.024506}%
  \BibitemOpen
  \bibfield  {author} {\bibinfo {author} {\bibfnamefont {Y.}~\bibnamefont
  {Wang}}, \bibinfo {author} {\bibfnamefont {J.}~\bibnamefont {Kang}},\ and\
  \bibinfo {author} {\bibfnamefont {R.~M.}\ \bibnamefont {Fernandes}},\
  }\bibfield  {title} {\bibinfo {title} {Topological and nematic
  superconductivity mediated by ferro-{SU}(4) fluctuations in twisted bilayer
  graphene},\ }\href {https://doi.org/10.1103/PhysRevB.103.024506} {\bibfield
  {journal} {\bibinfo  {journal} {Phys. Rev. B}\ }\textbf {\bibinfo {volume}
  {103}},\ \bibinfo {pages} {024506} (\bibinfo {year} {2021})}\BibitemShut
  {NoStop}%
\bibitem [{\citenamefont {Cao}\ \emph {et~al.}(2020)\citenamefont {Cao},
  \citenamefont {Rodan-Legrain}, \citenamefont {Park}, \citenamefont {Yuan},
  \citenamefont {Watanabe}, \citenamefont {Taniguchi}, \citenamefont
  {Fernandes}, \citenamefont {Fu},\ and\ \citenamefont
  {Jarillo-Herrero}}]{NematicSCExp}%
  \BibitemOpen
  \bibfield  {author} {\bibinfo {author} {\bibfnamefont {Y.}~\bibnamefont
  {Cao}}, \bibinfo {author} {\bibfnamefont {D.}~\bibnamefont {Rodan-Legrain}},
  \bibinfo {author} {\bibfnamefont {J.~M.}\ \bibnamefont {Park}}, \bibinfo
  {author} {\bibfnamefont {F.~N.}\ \bibnamefont {Yuan}}, \bibinfo {author}
  {\bibfnamefont {K.}~\bibnamefont {Watanabe}}, \bibinfo {author}
  {\bibfnamefont {T.}~\bibnamefont {Taniguchi}}, \bibinfo {author}
  {\bibfnamefont {R.~M.}\ \bibnamefont {Fernandes}}, \bibinfo {author}
  {\bibfnamefont {L.}~\bibnamefont {Fu}},\ and\ \bibinfo {author}
  {\bibfnamefont {P.}~\bibnamefont {Jarillo-Herrero}},\ }\bibfield  {title}
  {\bibinfo {title} {Nematicity and competing orders in superconducting
  magic-angle graphene},\ }\href {https://arxiv.org/abs/2004.04148} {\bibfield
  {journal} {\bibinfo  {journal} {arXiv:2004.04148}\ } (\bibinfo {year}
  {2020})}\BibitemShut {NoStop}%
\bibitem [{\citenamefont {Fernandes}\ and\ \citenamefont
  {Fu}(2021)}]{NematicSCRaf}%
  \BibitemOpen
  \bibfield  {author} {\bibinfo {author} {\bibfnamefont {R.~M.}\ \bibnamefont
  {Fernandes}}\ and\ \bibinfo {author} {\bibfnamefont {L.}~\bibnamefont {Fu}},\
  }\bibfield  {title} {\bibinfo {title} {Charge-$4e$ superconductivity from
  multi-component nematic pairing: Application to twisted bilayer graphene},\
  }\href {https://arxiv.org/abs/2101.07943} {\bibfield  {journal} {\bibinfo
  {journal} {arXiv:2101.07943}\ } (\bibinfo {year} {2021})}\BibitemShut
  {NoStop}%
\bibitem [{\citenamefont {Yu}\ \emph {et~al.}(2021)\citenamefont {Yu},
  \citenamefont {Kennes}, \citenamefont {Rubio},\ and\ \citenamefont
  {Sentef}}]{TaoNemSC}%
  \BibitemOpen
  \bibfield  {author} {\bibinfo {author} {\bibfnamefont {T.}~\bibnamefont
  {Yu}}, \bibinfo {author} {\bibfnamefont {D.~M.}\ \bibnamefont {Kennes}},
  \bibinfo {author} {\bibfnamefont {A.}~\bibnamefont {Rubio}},\ and\ \bibinfo
  {author} {\bibfnamefont {M.~A.}\ \bibnamefont {Sentef}},\ }\bibfield  {title}
  {\bibinfo {title} {Nematicity arising from a chiral superconducting ground
  state in magic-angle twisted bilayer graphene under in-plane magnetic
  fields},\ }\href {https://arxiv.org/abs/2101.01426} {\bibfield  {journal}
  {\bibinfo  {journal} {arXiv:2101.01426}\ } (\bibinfo {year}
  {2021})}\BibitemShut {NoStop}%
\bibitem [{\citenamefont {Qin}\ \emph {et~al.}(2021)\citenamefont {Qin},
  \citenamefont {Zou},\ and\ \citenamefont {MacDonald}}]{ScMacDoEL}%
  \BibitemOpen
  \bibfield  {author} {\bibinfo {author} {\bibfnamefont {W.}~\bibnamefont
  {Qin}}, \bibinfo {author} {\bibfnamefont {B.}~\bibnamefont {Zou}},\ and\
  \bibinfo {author} {\bibfnamefont {A.~H.}\ \bibnamefont {MacDonald}},\
  }\bibfield  {title} {\bibinfo {title} {Critical magnetic fields and
  electron-pairing in magic-angle twisted bilayer graphene},\ }\href
  {https://arxiv.org/abs/2102.10504} {\bibfield  {journal} {\bibinfo  {journal}
  {arXiv:2102.10504}\ } (\bibinfo {year} {2021})}\BibitemShut {NoStop}%
\bibitem [{\citenamefont {Löthman}\ \emph {et~al.}(2021)\citenamefont
  {Löthman}, \citenamefont {Schmidt}, \citenamefont {Parhizgar},\ and\
  \citenamefont {Black-Schaffer}}]{BlackSchafferSC}%
  \BibitemOpen
  \bibfield  {author} {\bibinfo {author} {\bibfnamefont {T.}~\bibnamefont
  {Löthman}}, \bibinfo {author} {\bibfnamefont {J.}~\bibnamefont {Schmidt}},
  \bibinfo {author} {\bibfnamefont {F.}~\bibnamefont {Parhizgar}},\ and\
  \bibinfo {author} {\bibfnamefont {A.~M.}\ \bibnamefont {Black-Schaffer}},\
  }\bibfield  {title} {\bibinfo {title} {Nematic superconductivity in
  magic-angle twisted bilayer graphene from atomistic modeling},\ }\href
  {https://arxiv.org/abs/2101.11555} {\bibfield  {journal} {\bibinfo  {journal}
  {arXiv:2101.11555}\ } (\bibinfo {year} {2021})}\BibitemShut {NoStop}%
\bibitem [{\citenamefont {Stepanov}\ \emph {et~al.}(2020)\citenamefont
  {Stepanov}, \citenamefont {Das}, \citenamefont {Lu}, \citenamefont
  {Fahimniya}, \citenamefont {Watanabe}, \citenamefont {Taniguchi},
  \citenamefont {Koppens}, \citenamefont {Lischner}, \citenamefont {Levitov},\
  and\ \citenamefont {Efetov}}]{Stepanov2020unt}%
  \BibitemOpen
  \bibfield  {author} {\bibinfo {author} {\bibfnamefont {P.}~\bibnamefont
  {Stepanov}}, \bibinfo {author} {\bibfnamefont {I.}~\bibnamefont {Das}},
  \bibinfo {author} {\bibfnamefont {X.}~\bibnamefont {Lu}}, \bibinfo {author}
  {\bibfnamefont {A.}~\bibnamefont {Fahimniya}}, \bibinfo {author}
  {\bibfnamefont {K.}~\bibnamefont {Watanabe}}, \bibinfo {author}
  {\bibfnamefont {T.}~\bibnamefont {Taniguchi}}, \bibinfo {author}
  {\bibfnamefont {F.~H.~L.}\ \bibnamefont {Koppens}}, \bibinfo {author}
  {\bibfnamefont {J.}~\bibnamefont {Lischner}}, \bibinfo {author}
  {\bibfnamefont {L.}~\bibnamefont {Levitov}},\ and\ \bibinfo {author}
  {\bibfnamefont {D.~K.}\ \bibnamefont {Efetov}},\ }\bibfield  {title}
  {\bibinfo {title} {Untying the insulating and superconducting orders in
  magic-angle graphene},\ }\href {https://doi.org/10.1038/s41586-020-2459-6}
  {\bibfield  {journal} {\bibinfo  {journal} {Nature}\ }\textbf {\bibinfo
  {volume} {583}},\ \bibinfo {pages} {375} (\bibinfo {year}
  {2020})}\BibitemShut {NoStop}%
\bibitem [{\citenamefont {Julku}\ \emph {et~al.}(2020)\citenamefont {Julku},
  \citenamefont {Peltonen}, \citenamefont {Liang}, \citenamefont {Heikkilä},\
  and\ \citenamefont {T\"orm\"a}}]{julku_superfluid_2020}%
  \BibitemOpen
  \bibfield  {author} {\bibinfo {author} {\bibfnamefont {A.}~\bibnamefont
  {Julku}}, \bibinfo {author} {\bibfnamefont {T.~J.}\ \bibnamefont {Peltonen}},
  \bibinfo {author} {\bibfnamefont {L.}~\bibnamefont {Liang}}, \bibinfo
  {author} {\bibfnamefont {T.~T.}\ \bibnamefont {Heikkilä}},\ and\ \bibinfo
  {author} {\bibfnamefont {P.}~\bibnamefont {T\"orm\"a}},\ }\bibfield  {title}
  {\bibinfo {title} {Superfluid weight and
  {Berezinskii}-{Kosterlitz}-{Thouless} transition temperature of twisted
  bilayer graphene},\ }\href {https://doi.org/10.1103/PhysRevB.101.060505}
  {\bibfield  {journal} {\bibinfo  {journal} {Physical Review B}\ }\textbf
  {\bibinfo {volume} {101}},\ \bibinfo {pages} {060505} (\bibinfo {year}
  {2020})},\ \bibinfo {note} {publisher: American Physical Society}\BibitemShut
  {NoStop}%
\bibitem [{\citenamefont {{Hu}}\ \emph {et~al.}(2019)\citenamefont {{Hu}},
  \citenamefont {{Hyart}}, \citenamefont {{Pikulin}},\ and\ \citenamefont
  {{Rossi}}}]{hu:2019}%
  \BibitemOpen
  \bibfield  {author} {\bibinfo {author} {\bibfnamefont {X.}~\bibnamefont
  {{Hu}}}, \bibinfo {author} {\bibfnamefont {T.}~\bibnamefont {{Hyart}}},
  \bibinfo {author} {\bibfnamefont {D.}~\bibnamefont {{Pikulin}}},\ and\
  \bibinfo {author} {\bibfnamefont {E.}~\bibnamefont {{Rossi}}},\ }\bibfield
  {title} {\bibinfo {title} {{Geometric and Conventional Contribution to
  Superfluid Weight in Twisted Bilayer Graphene}},\ }\href
  {https://doi.org/https://doi.org/10.1103/PhysRevLett.123.237002} {\bibfield
  {journal} {\bibinfo  {journal} {Phys. Rev. Lett.}\ }\textbf {\bibinfo
  {volume} {123}},\ \bibinfo {pages} {237002} (\bibinfo {year}
  {2019})}\BibitemShut {NoStop}%
\bibitem [{\citenamefont {Xie}\ \emph {et~al.}(2020)\citenamefont {Xie},
  \citenamefont {Song}, \citenamefont {Lian},\ and\ \citenamefont
  {Bernevig}}]{xie_topology-bounded_2020}%
  \BibitemOpen
  \bibfield  {author} {\bibinfo {author} {\bibfnamefont {F.}~\bibnamefont
  {Xie}}, \bibinfo {author} {\bibfnamefont {Z.}~\bibnamefont {Song}}, \bibinfo
  {author} {\bibfnamefont {B.}~\bibnamefont {Lian}},\ and\ \bibinfo {author}
  {\bibfnamefont {B.~A.}\ \bibnamefont {Bernevig}},\ }\bibfield  {title}
  {\bibinfo {title} {Topology-{Bounded} {Superfluid} {Weight} in {Twisted}
  {Bilayer} {Graphene}},\ }\href
  {https://doi.org/10.1103/PhysRevLett.124.167002} {\bibfield  {journal}
  {\bibinfo  {journal} {Physical Review Letters}\ }\textbf {\bibinfo {volume}
  {124}},\ \bibinfo {pages} {167002} (\bibinfo {year} {2020})},\ \bibinfo
  {note} {publisher: American Physical Society}\BibitemShut {NoStop}%
\bibitem [{\citenamefont {Liu}\ \emph {et~al.}(2019)\citenamefont {Liu},
  \citenamefont {Hao}, \citenamefont {Khalaf}, \citenamefont {Lee},
  \citenamefont {Watanabe}, \citenamefont {Taniguchi}, \citenamefont
  {Vishwanath},\ and\ \citenamefont {Kim}}]{Liu2019}%
  \BibitemOpen
  \bibfield  {author} {\bibinfo {author} {\bibfnamefont {X.}~\bibnamefont
  {Liu}}, \bibinfo {author} {\bibfnamefont {Z.}~\bibnamefont {Hao}}, \bibinfo
  {author} {\bibfnamefont {E.}~\bibnamefont {Khalaf}}, \bibinfo {author}
  {\bibfnamefont {J.~Y.}\ \bibnamefont {Lee}}, \bibinfo {author} {\bibfnamefont
  {K.}~\bibnamefont {Watanabe}}, \bibinfo {author} {\bibfnamefont
  {T.}~\bibnamefont {Taniguchi}}, \bibinfo {author} {\bibfnamefont
  {A.}~\bibnamefont {Vishwanath}},\ and\ \bibinfo {author} {\bibfnamefont
  {P.}~\bibnamefont {Kim}},\ }\bibfield  {title} {\bibinfo {title}
  {{Spin-polarized Correlated Insulator and Superconductor in Twisted Double
  Bilayer Graphene}},\ }\href {http://arxiv.org/abs/1903.08130} {\  (\bibinfo
  {year} {2019})},\ \Eprint {https://arxiv.org/abs/1903.08130}
  {arXiv:1903.08130} \BibitemShut {NoStop}%
\bibitem [{\citenamefont {Cao}\ \emph {et~al.}(2019)\citenamefont {Cao},
  \citenamefont {Rodan-Legrain}, \citenamefont {Rubies-Bigorda}, \citenamefont
  {Park}, \citenamefont {Watanabe}, \citenamefont {Taniguchi},\ and\
  \citenamefont {Jarillo-Herrero}}]{Cao2019}%
  \BibitemOpen
  \bibfield  {author} {\bibinfo {author} {\bibfnamefont {Y.}~\bibnamefont
  {Cao}}, \bibinfo {author} {\bibfnamefont {D.}~\bibnamefont {Rodan-Legrain}},
  \bibinfo {author} {\bibfnamefont {O.}~\bibnamefont {Rubies-Bigorda}},
  \bibinfo {author} {\bibfnamefont {J.~M.}\ \bibnamefont {Park}}, \bibinfo
  {author} {\bibfnamefont {K.}~\bibnamefont {Watanabe}}, \bibinfo {author}
  {\bibfnamefont {T.}~\bibnamefont {Taniguchi}},\ and\ \bibinfo {author}
  {\bibfnamefont {P.}~\bibnamefont {Jarillo-Herrero}},\ }\bibfield  {title}
  {\bibinfo {title} {{Electric Field Tunable Correlated States and Magnetic
  Phase Transitions in Twisted Bilayer-Bilayer Graphene}},\ }\href
  {http://arxiv.org/abs/1903.08596} {\  (\bibinfo {year} {2019})},\ \Eprint
  {https://arxiv.org/abs/1903.08596} {arXiv:1903.08596} \BibitemShut {NoStop}%
\bibitem [{\citenamefont {Shen}\ \emph {et~al.}()\citenamefont {Shen},
  \citenamefont {Chu}, \citenamefont {Wu}, \citenamefont {Li}, \citenamefont
  {Wang}, \citenamefont {Zhao}, \citenamefont {Tang}, \citenamefont {Liu},
  \citenamefont {Tian}, \citenamefont {Watanabe}, \citenamefont {Taniguchi},
  \citenamefont {Yang}, \citenamefont {Meng}, \citenamefont {Shi},
  \citenamefont {Yazyev},\ and\ \citenamefont {Zhang}}]{Shen2019}%
  \BibitemOpen
  \bibfield  {author} {\bibinfo {author} {\bibfnamefont {C.}~\bibnamefont
  {Shen}}, \bibinfo {author} {\bibfnamefont {Y.}~\bibnamefont {Chu}}, \bibinfo
  {author} {\bibfnamefont {Q.}~\bibnamefont {Wu}}, \bibinfo {author}
  {\bibfnamefont {N.}~\bibnamefont {Li}}, \bibinfo {author} {\bibfnamefont
  {S.}~\bibnamefont {Wang}}, \bibinfo {author} {\bibfnamefont {Y.}~\bibnamefont
  {Zhao}}, \bibinfo {author} {\bibfnamefont {J.}~\bibnamefont {Tang}}, \bibinfo
  {author} {\bibfnamefont {J.}~\bibnamefont {Liu}}, \bibinfo {author}
  {\bibfnamefont {J.}~\bibnamefont {Tian}}, \bibinfo {author} {\bibfnamefont
  {K.}~\bibnamefont {Watanabe}}, \bibinfo {author} {\bibfnamefont
  {T.}~\bibnamefont {Taniguchi}}, \bibinfo {author} {\bibfnamefont
  {R.}~\bibnamefont {Yang}}, \bibinfo {author} {\bibfnamefont {Z.~Y.}\
  \bibnamefont {Meng}}, \bibinfo {author} {\bibfnamefont {D.}~\bibnamefont
  {Shi}}, \bibinfo {author} {\bibfnamefont {O.~V.}\ \bibnamefont {Yazyev}},\
  and\ \bibinfo {author} {\bibfnamefont {G.}~\bibnamefont {Zhang}},\ }\bibfield
   {title} {\bibinfo {title} {Correlated states in twisted double bilayer
  graphene},\ }\href {https://doi.org/10.1038/s41567-020-0825-9} {\bibfield
  {journal} {\bibinfo  {journal} {Nature Physics}\ }\textbf {\bibinfo {volume}
  {16}},\ \bibinfo {pages} {520}}\BibitemShut {NoStop}%
\bibitem [{\citenamefont {Chen}\ \emph
  {et~al.}(2019{\natexlab{a}})\citenamefont {Chen}, \citenamefont {Jiang},
  \citenamefont {Wu}, \citenamefont {Lyu}, \citenamefont {Li}, \citenamefont
  {Chittari}, \citenamefont {Watanabe}, \citenamefont {Taniguchi},
  \citenamefont {Shi}, \citenamefont {Jung}, \citenamefont {Zhang},\ and\
  \citenamefont {Wang}}]{Chen2019a}%
  \BibitemOpen
  \bibfield  {author} {\bibinfo {author} {\bibfnamefont {G.}~\bibnamefont
  {Chen}}, \bibinfo {author} {\bibfnamefont {L.}~\bibnamefont {Jiang}},
  \bibinfo {author} {\bibfnamefont {S.}~\bibnamefont {Wu}}, \bibinfo {author}
  {\bibfnamefont {B.}~\bibnamefont {Lyu}}, \bibinfo {author} {\bibfnamefont
  {H.}~\bibnamefont {Li}}, \bibinfo {author} {\bibfnamefont {B.~L.}\
  \bibnamefont {Chittari}}, \bibinfo {author} {\bibfnamefont {K.}~\bibnamefont
  {Watanabe}}, \bibinfo {author} {\bibfnamefont {T.}~\bibnamefont {Taniguchi}},
  \bibinfo {author} {\bibfnamefont {Z.}~\bibnamefont {Shi}}, \bibinfo {author}
  {\bibfnamefont {J.}~\bibnamefont {Jung}}, \bibinfo {author} {\bibfnamefont
  {Y.}~\bibnamefont {Zhang}},\ and\ \bibinfo {author} {\bibfnamefont
  {F.}~\bibnamefont {Wang}},\ }\bibfield  {title} {\bibinfo {title} {{Evidence
  of a gate-tunable {M}ott insulator in a trilayer graphene moir{\'{e}}
  superlattice}},\ }\href {https://doi.org/10.1038/s41567-018-0387-2}
  {\bibfield  {journal} {\bibinfo  {journal} {Nature Physics}\ }\textbf
  {\bibinfo {volume} {15}},\ \bibinfo {pages} {237} (\bibinfo {year}
  {2019}{\natexlab{a}})}\BibitemShut {NoStop}%
\bibitem [{\citenamefont {Chen}\ \emph
  {et~al.}(2019{\natexlab{b}})\citenamefont {Chen}, \citenamefont {Sharpe},
  \citenamefont {Gallagher}, \citenamefont {Rosen}, \citenamefont {Fox},
  \citenamefont {Jiang}, \citenamefont {Lyu}, \citenamefont {Li}, \citenamefont
  {Watanabe}, \citenamefont {Taniguchi}, \citenamefont {Jung}, \citenamefont
  {Shi}, \citenamefont {Goldhaber-Gordon}, \citenamefont {Zhang},\ and\
  \citenamefont {Wang}}]{Chen2019}%
  \BibitemOpen
  \bibfield  {author} {\bibinfo {author} {\bibfnamefont {G.}~\bibnamefont
  {Chen}}, \bibinfo {author} {\bibfnamefont {A.~L.}\ \bibnamefont {Sharpe}},
  \bibinfo {author} {\bibfnamefont {P.}~\bibnamefont {Gallagher}}, \bibinfo
  {author} {\bibfnamefont {I.~T.}\ \bibnamefont {Rosen}}, \bibinfo {author}
  {\bibfnamefont {E.~J.}\ \bibnamefont {Fox}}, \bibinfo {author} {\bibfnamefont
  {L.}~\bibnamefont {Jiang}}, \bibinfo {author} {\bibfnamefont
  {B.}~\bibnamefont {Lyu}}, \bibinfo {author} {\bibfnamefont {H.}~\bibnamefont
  {Li}}, \bibinfo {author} {\bibfnamefont {K.}~\bibnamefont {Watanabe}},
  \bibinfo {author} {\bibfnamefont {T.}~\bibnamefont {Taniguchi}}, \bibinfo
  {author} {\bibfnamefont {J.}~\bibnamefont {Jung}}, \bibinfo {author}
  {\bibfnamefont {Z.}~\bibnamefont {Shi}}, \bibinfo {author} {\bibfnamefont
  {D.}~\bibnamefont {Goldhaber-Gordon}}, \bibinfo {author} {\bibfnamefont
  {Y.}~\bibnamefont {Zhang}},\ and\ \bibinfo {author} {\bibfnamefont
  {F.}~\bibnamefont {Wang}},\ }\bibfield  {title} {\bibinfo {title} {Signatures
  of tunable superconductivity in a trilayer graphene moir{\'e} superlattice},\
  }\href {https://doi.org/10.1038/s41586-019-1393-y} {\bibfield  {journal}
  {\bibinfo  {journal} {Nature}\ }\textbf {\bibinfo {volume} {572}},\ \bibinfo
  {pages} {215} (\bibinfo {year} {2019}{\natexlab{b}})}\BibitemShut {NoStop}%
\bibitem [{\citenamefont {Chen}\ \emph
  {et~al.}(2020{\natexlab{b}})\citenamefont {Chen}, \citenamefont {Sharpe},
  \citenamefont {Fox}, \citenamefont {Zhang}, \citenamefont {Wang},
  \citenamefont {Jiang}, \citenamefont {Lyu}, \citenamefont {Li}, \citenamefont
  {Watanabe}, \citenamefont {Taniguchi}, \citenamefont {Shi}, \citenamefont
  {Senthil}, \citenamefont {Goldhaber-Gordon}, \citenamefont {Zhang},\ and\
  \citenamefont {Wang}}]{chen2019tunable}%
  \BibitemOpen
  \bibfield  {author} {\bibinfo {author} {\bibfnamefont {G.}~\bibnamefont
  {Chen}}, \bibinfo {author} {\bibfnamefont {A.~L.}\ \bibnamefont {Sharpe}},
  \bibinfo {author} {\bibfnamefont {E.~J.}\ \bibnamefont {Fox}}, \bibinfo
  {author} {\bibfnamefont {Y.-H.}\ \bibnamefont {Zhang}}, \bibinfo {author}
  {\bibfnamefont {S.}~\bibnamefont {Wang}}, \bibinfo {author} {\bibfnamefont
  {L.}~\bibnamefont {Jiang}}, \bibinfo {author} {\bibfnamefont
  {B.}~\bibnamefont {Lyu}}, \bibinfo {author} {\bibfnamefont {H.}~\bibnamefont
  {Li}}, \bibinfo {author} {\bibfnamefont {K.}~\bibnamefont {Watanabe}},
  \bibinfo {author} {\bibfnamefont {T.}~\bibnamefont {Taniguchi}}, \bibinfo
  {author} {\bibfnamefont {Z.}~\bibnamefont {Shi}}, \bibinfo {author}
  {\bibfnamefont {T.}~\bibnamefont {Senthil}}, \bibinfo {author} {\bibfnamefont
  {D.}~\bibnamefont {Goldhaber-Gordon}}, \bibinfo {author} {\bibfnamefont
  {Y.}~\bibnamefont {Zhang}},\ and\ \bibinfo {author} {\bibfnamefont
  {F.}~\bibnamefont {Wang}},\ }\bibfield  {title} {\bibinfo {title} {{Tunable
  correlated Chern insulator and ferromagnetism in a moir{\'{e}}
  superlattice}},\ }\href@noop {} {\bibfield  {journal} {\bibinfo  {journal}
  {Nature}\ }\textbf {\bibinfo {volume} {579}} (\bibinfo {year}
  {2020}{\natexlab{b}})}\BibitemShut {NoStop}%
\bibitem [{\citenamefont {Burg}\ \emph {et~al.}(2019)\citenamefont {Burg},
  \citenamefont {Zhu}, \citenamefont {Taniguchi}, \citenamefont {Watanabe},
  \citenamefont {MacDonald},\ and\ \citenamefont {Tutuc}}]{TutucBi}%
  \BibitemOpen
  \bibfield  {author} {\bibinfo {author} {\bibfnamefont {G.~W.}\ \bibnamefont
  {Burg}}, \bibinfo {author} {\bibfnamefont {J.}~\bibnamefont {Zhu}}, \bibinfo
  {author} {\bibfnamefont {T.}~\bibnamefont {Taniguchi}}, \bibinfo {author}
  {\bibfnamefont {K.}~\bibnamefont {Watanabe}}, \bibinfo {author}
  {\bibfnamefont {A.~H.}\ \bibnamefont {MacDonald}},\ and\ \bibinfo {author}
  {\bibfnamefont {E.}~\bibnamefont {Tutuc}},\ }\bibfield  {title} {\bibinfo
  {title} {Correlated insulating states in twisted double bilayer graphene},\
  }\href {https://doi.org/10.1103/PhysRevLett.123.197702} {\bibfield  {journal}
  {\bibinfo  {journal} {Phys. Rev. Lett.}\ }\textbf {\bibinfo {volume} {123}},\
  \bibinfo {pages} {197702} (\bibinfo {year} {2019})}\BibitemShut {NoStop}%
\bibitem [{\citenamefont {Rubio-Verdú}\ \emph {et~al.}(2020)\citenamefont
  {Rubio-Verdú}, \citenamefont {Turkel}, \citenamefont {Song}, \citenamefont
  {Klebl}, \citenamefont {Samajdar}, \citenamefont {Scheurer}, \citenamefont
  {Venderbos}, \citenamefont {Watanabe}, \citenamefont {Taniguchi},
  \citenamefont {Ochoa}, \citenamefont {Xian}, \citenamefont {Kennes},
  \citenamefont {Fernandes}, \citenamefont {Ángel Rubio},\ and\ \citenamefont
  {Pasupathy}}]{rubioverdu2020universal}%
  \BibitemOpen
  \bibfield  {author} {\bibinfo {author} {\bibfnamefont {C.}~\bibnamefont
  {Rubio-Verdú}}, \bibinfo {author} {\bibfnamefont {S.}~\bibnamefont
  {Turkel}}, \bibinfo {author} {\bibfnamefont {L.}~\bibnamefont {Song}},
  \bibinfo {author} {\bibfnamefont {L.}~\bibnamefont {Klebl}}, \bibinfo
  {author} {\bibfnamefont {R.}~\bibnamefont {Samajdar}}, \bibinfo {author}
  {\bibfnamefont {M.~S.}\ \bibnamefont {Scheurer}}, \bibinfo {author}
  {\bibfnamefont {J.~W.~F.}\ \bibnamefont {Venderbos}}, \bibinfo {author}
  {\bibfnamefont {K.}~\bibnamefont {Watanabe}}, \bibinfo {author}
  {\bibfnamefont {T.}~\bibnamefont {Taniguchi}}, \bibinfo {author}
  {\bibfnamefont {H.}~\bibnamefont {Ochoa}}, \bibinfo {author} {\bibfnamefont
  {L.}~\bibnamefont {Xian}}, \bibinfo {author} {\bibfnamefont {D.}~\bibnamefont
  {Kennes}}, \bibinfo {author} {\bibfnamefont {R.~M.}\ \bibnamefont
  {Fernandes}}, \bibinfo {author} {\bibnamefont {Ángel Rubio}},\ and\ \bibinfo
  {author} {\bibfnamefont {A.~N.}\ \bibnamefont {Pasupathy}},\ }\bibfield
  {title} {\bibinfo {title} {Universal moir\'e nematic phase in twisted
  graphitic systems},\ }\href {https://arxiv.org/abs/2009.11645} {\bibfield
  {journal} {\bibinfo  {journal} {arXiv:2009.11645}\ } (\bibinfo {year}
  {2020})}\BibitemShut {NoStop}%
\bibitem [{\citenamefont {Tritsaris}\ \emph {et~al.}(2020)\citenamefont
  {Tritsaris}, \citenamefont {Carr}, \citenamefont {Zhu}, \citenamefont {Xie},
  \citenamefont {Torrisi}, \citenamefont {Tang}, \citenamefont {Mattheakis},
  \citenamefont {Larson},\ and\ \citenamefont {Kaxiras}}]{Tritsaris_2020}%
  \BibitemOpen
  \bibfield  {author} {\bibinfo {author} {\bibfnamefont {G.~A.}\ \bibnamefont
  {Tritsaris}}, \bibinfo {author} {\bibfnamefont {S.}~\bibnamefont {Carr}},
  \bibinfo {author} {\bibfnamefont {Z.}~\bibnamefont {Zhu}}, \bibinfo {author}
  {\bibfnamefont {Y.}~\bibnamefont {Xie}}, \bibinfo {author} {\bibfnamefont
  {S.~B.}\ \bibnamefont {Torrisi}}, \bibinfo {author} {\bibfnamefont
  {J.}~\bibnamefont {Tang}}, \bibinfo {author} {\bibfnamefont {M.}~\bibnamefont
  {Mattheakis}}, \bibinfo {author} {\bibfnamefont {D.~T.}\ \bibnamefont
  {Larson}},\ and\ \bibinfo {author} {\bibfnamefont {E.}~\bibnamefont
  {Kaxiras}},\ }\bibfield  {title} {\bibinfo {title} {Electronic structure
  calculations of twisted multi-layer graphene superlattices},\ }\href
  {https://doi.org/10.1088/2053-1583/ab8f62} {\bibfield  {journal} {\bibinfo
  {journal} {2D Materials}\ }\textbf {\bibinfo {volume} {7}},\ \bibinfo {pages}
  {035028} (\bibinfo {year} {2020})}\BibitemShut {NoStop}%
\bibitem [{\citenamefont {Xian}\ \emph {et~al.}(2019)\citenamefont {Xian},
  \citenamefont {Kennes}, \citenamefont {Tancogne-Dejean}, \citenamefont
  {Altarelli},\ and\ \citenamefont {Rubio}}]{Xian2019BN}%
  \BibitemOpen
  \bibfield  {author} {\bibinfo {author} {\bibfnamefont {L.}~\bibnamefont
  {Xian}}, \bibinfo {author} {\bibfnamefont {D.~M.}\ \bibnamefont {Kennes}},
  \bibinfo {author} {\bibfnamefont {N.}~\bibnamefont {Tancogne-Dejean}},
  \bibinfo {author} {\bibfnamefont {M.}~\bibnamefont {Altarelli}},\ and\
  \bibinfo {author} {\bibfnamefont {A.}~\bibnamefont {Rubio}},\ }\bibfield
  {title} {\bibinfo {title} {Multiflat bands and strong correlations in twisted
  bilayer boron nitride: Doping-induced correlated insulator and
  superconductor},\ }\href {https://doi.org/10.1021/acs.nanolett.9b00986}
  {\bibfield  {journal} {\bibinfo  {journal} {Nano Letters}\ }\textbf {\bibinfo
  {volume} {19}},\ \bibinfo {pages} {4934} (\bibinfo {year}
  {2019})}\BibitemShut {NoStop}%
\bibitem [{\citenamefont {Wu}\ \emph {et~al.}(2018{\natexlab{b}})\citenamefont
  {Wu}, \citenamefont {Lovorn}, \citenamefont {Tutuc},\ and\ \citenamefont
  {Macdonald}}]{Wu2018}%
  \BibitemOpen
  \bibfield  {author} {\bibinfo {author} {\bibfnamefont {F.}~\bibnamefont
  {Wu}}, \bibinfo {author} {\bibfnamefont {T.}~\bibnamefont {Lovorn}}, \bibinfo
  {author} {\bibfnamefont {E.}~\bibnamefont {Tutuc}},\ and\ \bibinfo {author}
  {\bibfnamefont {A.~H.}\ \bibnamefont {Macdonald}},\ }\bibfield  {title}
  {\bibinfo {title} {{Hubbard Model Physics in Transition Metal Dichalcogenide
  Moir{\'{e}} Bands}},\ }\href {https://doi.org/10.1103/PhysRevLett.121.026402}
  {\bibfield  {journal} {\bibinfo  {journal} {Physical Review Letters}\
  }\textbf {\bibinfo {volume} {121}},\ \bibinfo {pages} {26402} (\bibinfo
  {year} {2018}{\natexlab{b}})}\BibitemShut {NoStop}%
\bibitem [{\citenamefont {Wu}\ \emph {et~al.}(2019)\citenamefont {Wu},
  \citenamefont {Lovorn}, \citenamefont {Tutuc}, \citenamefont {Martin},\ and\
  \citenamefont {Macdonald}}]{Wu2019}%
  \BibitemOpen
  \bibfield  {author} {\bibinfo {author} {\bibfnamefont {F.}~\bibnamefont
  {Wu}}, \bibinfo {author} {\bibfnamefont {T.}~\bibnamefont {Lovorn}}, \bibinfo
  {author} {\bibfnamefont {E.}~\bibnamefont {Tutuc}}, \bibinfo {author}
  {\bibfnamefont {I.}~\bibnamefont {Martin}},\ and\ \bibinfo {author}
  {\bibfnamefont {A.~H.}\ \bibnamefont {Macdonald}},\ }\bibfield  {title}
  {\bibinfo {title} {{Topological Insulators in Twisted Transition Metal
  Dichalcogenide Homobilayers}},\ }\href
  {https://doi.org/10.1103/PhysRevLett.122.086402} {\bibfield  {journal}
  {\bibinfo  {journal} {Physical Review Letters}\ }\textbf {\bibinfo {volume}
  {122}},\ \bibinfo {pages} {86402} (\bibinfo {year} {2019})}\BibitemShut
  {NoStop}%
\bibitem [{\citenamefont {Naik}\ and\ \citenamefont {Jain}(2018)}]{Naik2018}%
  \BibitemOpen
  \bibfield  {author} {\bibinfo {author} {\bibfnamefont {M.~H.}\ \bibnamefont
  {Naik}}\ and\ \bibinfo {author} {\bibfnamefont {M.}~\bibnamefont {Jain}},\
  }\bibfield  {title} {\bibinfo {title} {{Ultraflatbands and Shear Solitons in
  Moir{\'{e}} Patterns of Twisted Bilayer Transition Metal Dichalcogenides}},\
  }\href {https://doi.org/10.1103/PhysRevLett.121.266401} {\bibfield  {journal}
  {\bibinfo  {journal} {Physical Review Letters}\ }\textbf {\bibinfo {volume}
  {121}},\ \bibinfo {pages} {266401} (\bibinfo {year} {2018})}\BibitemShut
  {NoStop}%
\bibitem [{\citenamefont {Ruiz-Tijerina}\ and\ \citenamefont
  {Fal'Ko}(2019)}]{Ruiz-Tijerina2019}%
  \BibitemOpen
  \bibfield  {author} {\bibinfo {author} {\bibfnamefont {D.~A.}\ \bibnamefont
  {Ruiz-Tijerina}}\ and\ \bibinfo {author} {\bibfnamefont {V.~I.}\ \bibnamefont
  {Fal'Ko}},\ }\bibfield  {title} {\bibinfo {title} {{Interlayer hybridization
  and moir{\'{e}} superlattice minibands for electrons and excitons in
  heterobilayers of transition-metal dichalcogenides}},\ }\href
  {https://doi.org/10.1103/PhysRevB.99.125424} {\bibfield  {journal} {\bibinfo
  {journal} {Physical Review B}\ }\textbf {\bibinfo {volume} {99}},\ \bibinfo
  {pages} {30} (\bibinfo {year} {2019})}\BibitemShut {NoStop}%
\bibitem [{\citenamefont {Schrade}\ and\ \citenamefont
  {Fu}(2019)}]{Schrade2019}%
  \BibitemOpen
  \bibfield  {author} {\bibinfo {author} {\bibfnamefont {C.}~\bibnamefont
  {Schrade}}\ and\ \bibinfo {author} {\bibfnamefont {L.}~\bibnamefont {Fu}},\
  }\bibfield  {title} {\bibinfo {title} {{Spin-valley density wave in
  moir{\'{e}} materials}},\ }\href
  {https://doi.org/10.1103/PhysRevB.100.035413} {\bibfield  {journal} {\bibinfo
   {journal} {Physical Review B}\ }\textbf {\bibinfo {volume} {100}},\ \bibinfo
  {pages} {035413} (\bibinfo {year} {2019})}\BibitemShut {NoStop}%
\bibitem [{\citenamefont {Wang}\ \emph {et~al.}(2020)\citenamefont {Wang},
  \citenamefont {Shih}, \citenamefont {Ghiotto}, \citenamefont {Xian},
  \citenamefont {Rhodes}, \citenamefont {Tan}, \citenamefont {Claassen},
  \citenamefont {Kennes}, \citenamefont {Bai}, \citenamefont {Kim},
  \citenamefont {Watanabe}, \citenamefont {Taniguchi}, \citenamefont {Zhu},
  \citenamefont {Hone}, \citenamefont {Rubio}, \citenamefont {Pasupathy},\ and\
  \citenamefont {Dean}}]{Wang2020WSe2}%
  \BibitemOpen
  \bibfield  {author} {\bibinfo {author} {\bibfnamefont {L.}~\bibnamefont
  {Wang}}, \bibinfo {author} {\bibfnamefont {E.-M.}\ \bibnamefont {Shih}},
  \bibinfo {author} {\bibfnamefont {A.}~\bibnamefont {Ghiotto}}, \bibinfo
  {author} {\bibfnamefont {L.}~\bibnamefont {Xian}}, \bibinfo {author}
  {\bibfnamefont {D.~A.}\ \bibnamefont {Rhodes}}, \bibinfo {author}
  {\bibfnamefont {C.}~\bibnamefont {Tan}}, \bibinfo {author} {\bibfnamefont
  {M.}~\bibnamefont {Claassen}}, \bibinfo {author} {\bibfnamefont {D.~M.}\
  \bibnamefont {Kennes}}, \bibinfo {author} {\bibfnamefont {Y.}~\bibnamefont
  {Bai}}, \bibinfo {author} {\bibfnamefont {B.}~\bibnamefont {Kim}}, \bibinfo
  {author} {\bibfnamefont {K.}~\bibnamefont {Watanabe}}, \bibinfo {author}
  {\bibfnamefont {T.}~\bibnamefont {Taniguchi}}, \bibinfo {author}
  {\bibfnamefont {X.}~\bibnamefont {Zhu}}, \bibinfo {author} {\bibfnamefont
  {J.}~\bibnamefont {Hone}}, \bibinfo {author} {\bibfnamefont {A.}~\bibnamefont
  {Rubio}}, \bibinfo {author} {\bibfnamefont {A.~N.}\ \bibnamefont
  {Pasupathy}},\ and\ \bibinfo {author} {\bibfnamefont {C.~R.}\ \bibnamefont
  {Dean}},\ }\bibfield  {title} {\bibinfo {title} {Correlated electronic phases
  in twisted bilayer transition metal dichalcogenides},\ }\href
  {https://doi.org/10.1038/s41563-020-0708-6} {\bibfield  {journal} {\bibinfo
  {journal} {Nature Materials}\ }\textbf {\bibinfo {volume} {19}},\ \bibinfo
  {pages} {861} (\bibinfo {year} {2020})}\BibitemShut {NoStop}%
\bibitem [{\citenamefont {Xian}\ \emph {et~al.}(2020)\citenamefont {Xian},
  \citenamefont {Claassen}, \citenamefont {Kiese}, \citenamefont {Scherer},
  \citenamefont {Trebst}, \citenamefont {Kennes},\ and\ \citenamefont
  {Rubio}}]{XianMoS2}%
  \BibitemOpen
  \bibfield  {author} {\bibinfo {author} {\bibfnamefont {L.}~\bibnamefont
  {Xian}}, \bibinfo {author} {\bibfnamefont {M.}~\bibnamefont {Claassen}},
  \bibinfo {author} {\bibfnamefont {D.}~\bibnamefont {Kiese}}, \bibinfo
  {author} {\bibfnamefont {M.~M.}\ \bibnamefont {Scherer}}, \bibinfo {author}
  {\bibfnamefont {S.}~\bibnamefont {Trebst}}, \bibinfo {author} {\bibfnamefont
  {D.~M.}\ \bibnamefont {Kennes}},\ and\ \bibinfo {author} {\bibfnamefont
  {A.}~\bibnamefont {Rubio}},\ }\bibfield  {title} {\bibinfo {title}
  {Realization of nearly dispersionless bands with strong orbital anisotropy
  from destructive interference in twisted bilayer {MoS}$_2$},\ }\href
  {https://arxiv.org/abs/2004.02964} {\bibfield  {journal} {\bibinfo  {journal}
  {arXiv:2004.02964}\ } (\bibinfo {year} {2020})}\BibitemShut {NoStop}%
\bibitem [{\citenamefont {Vitale}\ \emph {et~al.}(2021)\citenamefont {Vitale},
  \citenamefont {Atalar}, \citenamefont {Mostofi},\ and\ \citenamefont
  {Lischner}}]{LischnerTMDs}%
  \BibitemOpen
  \bibfield  {author} {\bibinfo {author} {\bibfnamefont {V.}~\bibnamefont
  {Vitale}}, \bibinfo {author} {\bibfnamefont {K.}~\bibnamefont {Atalar}},
  \bibinfo {author} {\bibfnamefont {A.~A.}\ \bibnamefont {Mostofi}},\ and\
  \bibinfo {author} {\bibfnamefont {J.}~\bibnamefont {Lischner}},\ }\bibfield
  {title} {\bibinfo {title} {Flat band properties of twisted transition metal
  dichalcogenide homo- and heterobilayers of {MoS}$_2$, {MoSe}$_2$, {WS}$_2$
  and {WSe}$_2$},\ }\href {https://arxiv.org/abs/2102.03259} {\bibfield
  {journal} {\bibinfo  {journal} {arXiv:2102.03259}\ } (\bibinfo {year}
  {2021})}\BibitemShut {NoStop}%
\bibitem [{\citenamefont {Kennes}\ \emph {et~al.}(2020)\citenamefont {Kennes},
  \citenamefont {Xian}, \citenamefont {Claassen},\ and\ \citenamefont
  {Rubio}}]{KennesGeSe}%
  \BibitemOpen
  \bibfield  {author} {\bibinfo {author} {\bibfnamefont {D.~M.}\ \bibnamefont
  {Kennes}}, \bibinfo {author} {\bibfnamefont {L.}~\bibnamefont {Xian}},
  \bibinfo {author} {\bibfnamefont {M.}~\bibnamefont {Claassen}},\ and\
  \bibinfo {author} {\bibfnamefont {A.}~\bibnamefont {Rubio}},\ }\bibfield
  {title} {\bibinfo {title} {One-dimensional flat bands in twisted bilayer
  germanium selenide},\ }\href {https://doi.org/10.1038/s41467-020-14947-0}
  {\bibfield  {journal} {\bibinfo  {journal} {Nature Communications}\ }\textbf
  {\bibinfo {volume} {11}},\ \bibinfo {pages} {1124} (\bibinfo {year}
  {2020})}\BibitemShut {NoStop}%
\bibitem [{\citenamefont {Lian}\ \emph {et~al.}(2020)\citenamefont {Lian},
  \citenamefont {Liu}, \citenamefont {Zhang},\ and\ \citenamefont
  {Wang}}]{Lian20}%
  \BibitemOpen
  \bibfield  {author} {\bibinfo {author} {\bibfnamefont {B.}~\bibnamefont
  {Lian}}, \bibinfo {author} {\bibfnamefont {Z.}~\bibnamefont {Liu}}, \bibinfo
  {author} {\bibfnamefont {Y.}~\bibnamefont {Zhang}},\ and\ \bibinfo {author}
  {\bibfnamefont {J.}~\bibnamefont {Wang}},\ }\bibfield  {title} {\bibinfo
  {title} {Flat {C}hern band from twisted bilayer {MnBi$_{2}$Te$_{4}$}},\
  }\href {https://doi.org/10.1103/PhysRevLett.124.126402} {\bibfield  {journal}
  {\bibinfo  {journal} {Phys. Rev. Lett.}\ }\textbf {\bibinfo {volume} {124}},\
  \bibinfo {pages} {126402} (\bibinfo {year} {2020})}\BibitemShut {NoStop}%
\bibitem [{\citenamefont {Wang}\ \emph {et~al.}(2012)\citenamefont {Wang},
  \citenamefont {Kalantar-Zadeh}, \citenamefont {Kis}, \citenamefont
  {Coleman},\ and\ \citenamefont {Strano}}]{wang_electronics_2012}%
  \BibitemOpen
  \bibfield  {author} {\bibinfo {author} {\bibfnamefont {Q.~H.}\ \bibnamefont
  {Wang}}, \bibinfo {author} {\bibfnamefont {K.}~\bibnamefont
  {Kalantar-Zadeh}}, \bibinfo {author} {\bibfnamefont {A.}~\bibnamefont {Kis}},
  \bibinfo {author} {\bibfnamefont {J.~N.}\ \bibnamefont {Coleman}},\ and\
  \bibinfo {author} {\bibfnamefont {M.~S.}\ \bibnamefont {Strano}},\ }\bibfield
   {title} {\bibinfo {title} {Electronics and optoelectronics of
  two-dimensional transition metal dichalcogenides},\ }\href
  {https://doi.org/10.1038/nnano.2012.193} {\bibfield  {journal} {\bibinfo
  {journal} {Nature Nanotechnology}\ }\textbf {\bibinfo {volume} {7}},\
  \bibinfo {pages} {699} (\bibinfo {year} {2012})}\BibitemShut {NoStop}%
\bibitem [{\citenamefont {Liu}\ \emph {et~al.}(2015)\citenamefont {Liu},
  \citenamefont {Galfsky}, \citenamefont {Sun}, \citenamefont {Xia},
  \citenamefont {Lin}, \citenamefont {Lee}, \citenamefont {K{\'e}na-Cohen},\
  and\ \citenamefont {Menon}}]{liu_strong_2015}%
  \BibitemOpen
  \bibfield  {author} {\bibinfo {author} {\bibfnamefont {X.}~\bibnamefont
  {Liu}}, \bibinfo {author} {\bibfnamefont {T.}~\bibnamefont {Galfsky}},
  \bibinfo {author} {\bibfnamefont {Z.}~\bibnamefont {Sun}}, \bibinfo {author}
  {\bibfnamefont {F.}~\bibnamefont {Xia}}, \bibinfo {author} {\bibfnamefont
  {E.-c.}\ \bibnamefont {Lin}}, \bibinfo {author} {\bibfnamefont {Y.-H.}\
  \bibnamefont {Lee}}, \bibinfo {author} {\bibfnamefont {S.}~\bibnamefont
  {K{\'e}na-Cohen}},\ and\ \bibinfo {author} {\bibfnamefont {V.~M.}\
  \bibnamefont {Menon}},\ }\bibfield  {title} {\bibinfo {title} {Strong
  light{\textendash}matter coupling in two-dimensional atomic crystals},\
  }\href {https://doi.org/10.1038/nphoton.2014.304} {\bibfield  {journal}
  {\bibinfo  {journal} {Nature Photonics}\ }\textbf {\bibinfo {volume} {9}},\
  \bibinfo {pages} {30} (\bibinfo {year} {2015})}\BibitemShut {NoStop}%
\bibitem [{\citenamefont {Carusotto}\ and\ \citenamefont
  {Ciuti}(2013)}]{carusotto_quantum_2013}%
  \BibitemOpen
  \bibfield  {author} {\bibinfo {author} {\bibfnamefont {I.}~\bibnamefont
  {Carusotto}}\ and\ \bibinfo {author} {\bibfnamefont {C.}~\bibnamefont
  {Ciuti}},\ }\bibfield  {title} {\bibinfo {title} {{Quantum Fluids of
  Light}},\ }\href {https://doi.org/10.1103/RevModPhys.85.299} {\bibfield
  {journal} {\bibinfo  {journal} {Rev. Mod. Phys.}\ }\textbf {\bibinfo {volume}
  {85}},\ \bibinfo {pages} {299} (\bibinfo {year} {2013})}\BibitemShut
  {NoStop}%
\bibitem [{\citenamefont {T\"orm\"a}\ and\ \citenamefont
  {Barnes}(2015)}]{torma_strong_2015}%
  \BibitemOpen
  \bibfield  {author} {\bibinfo {author} {\bibfnamefont {P.}~\bibnamefont
  {T\"orm\"a}}\ and\ \bibinfo {author} {\bibfnamefont {W.~L.}\ \bibnamefont
  {Barnes}},\ }\bibfield  {title} {\bibinfo {title} {{S}trong {C}oupling
  between {S}urface {P}lasmon {P}olaritons and {E}mitters: {A} {R}eview},\
  }\href {https://doi.org/10.1088/0034-4885/78/1/013901} {\bibfield  {journal}
  {\bibinfo  {journal} {Reports on Progress in Physics}\ }\textbf {\bibinfo
  {volume} {78}},\ \bibinfo {pages} {013901} (\bibinfo {year}
  {2015})}\BibitemShut {NoStop}%
\bibitem [{\citenamefont {Flick}\ \emph {et~al.}(2017)\citenamefont {Flick},
  \citenamefont {Ruggenthaler}, \citenamefont {Appel},\ and\ \citenamefont
  {Rubio}}]{flick_atoms_2017}%
  \BibitemOpen
  \bibfield  {author} {\bibinfo {author} {\bibfnamefont {J.}~\bibnamefont
  {Flick}}, \bibinfo {author} {\bibfnamefont {M.}~\bibnamefont {Ruggenthaler}},
  \bibinfo {author} {\bibfnamefont {H.}~\bibnamefont {Appel}},\ and\ \bibinfo
  {author} {\bibfnamefont {A.}~\bibnamefont {Rubio}},\ }\bibfield  {title}
  {\bibinfo {title} {Atoms and molecules in cavities, from weak to strong
  coupling in quantum-electrodynamics ({QED}) chemistry},\ }\href
  {https://doi.org/10.1073/pnas.1615509114} {\bibfield  {journal} {\bibinfo
  {journal} {Proceedings of the National Academy of Sciences}\ }\textbf
  {\bibinfo {volume} {114}},\ \bibinfo {pages} {3026} (\bibinfo {year}
  {2017})}\BibitemShut {NoStop}%
\bibitem [{\citenamefont {Flick}\ \emph {et~al.}(2018)\citenamefont {Flick},
  \citenamefont {Rivera},\ and\ \citenamefont {Narang}}]{flick_strong_2018}%
  \BibitemOpen
  \bibfield  {author} {\bibinfo {author} {\bibfnamefont {J.}~\bibnamefont
  {Flick}}, \bibinfo {author} {\bibfnamefont {N.}~\bibnamefont {Rivera}},\ and\
  \bibinfo {author} {\bibfnamefont {P.}~\bibnamefont {Narang}},\ }\bibfield
  {title} {\bibinfo {title} {Strong light-matter coupling in quantum chemistry
  and quantum photonics},\ }\href {https://doi.org/10.1515/nanoph-2018-0067}
  {\bibfield  {journal} {\bibinfo  {journal} {Nanophotonics}\ }\textbf
  {\bibinfo {volume} {7}},\ \bibinfo {pages} {1479} (\bibinfo {year}
  {2018})}\BibitemShut {NoStop}%
\bibitem [{\citenamefont {Frisk~Kockum}\ \emph {et~al.}(2019)\citenamefont
  {Frisk~Kockum}, \citenamefont {Miranowicz}, \citenamefont {De~Liberato},
  \citenamefont {Savasta},\ and\ \citenamefont
  {Nori}}]{frisk_kockum_ultrastrong_2019}%
  \BibitemOpen
  \bibfield  {author} {\bibinfo {author} {\bibfnamefont {A.}~\bibnamefont
  {Frisk~Kockum}}, \bibinfo {author} {\bibfnamefont {A.}~\bibnamefont
  {Miranowicz}}, \bibinfo {author} {\bibfnamefont {S.}~\bibnamefont
  {De~Liberato}}, \bibinfo {author} {\bibfnamefont {S.}~\bibnamefont
  {Savasta}},\ and\ \bibinfo {author} {\bibfnamefont {F.}~\bibnamefont
  {Nori}},\ }\bibfield  {title} {\bibinfo {title} {Ultrastrong coupling between
  light and matter},\ }\href {https://doi.org/10.1038/s42254-018-0006-2}
  {\bibfield  {journal} {\bibinfo  {journal} {Nature Reviews Physics}\ }\textbf
  {\bibinfo {volume} {1}},\ \bibinfo {pages} {19} (\bibinfo {year}
  {2019})}\BibitemShut {NoStop}%
\bibitem [{\citenamefont {Forn-D{\'i}az}\ \emph {et~al.}(2019)\citenamefont
  {Forn-D{\'i}az}, \citenamefont {Lamata}, \citenamefont {Rico}, \citenamefont
  {Kono},\ and\ \citenamefont {Solano}}]{forn-diaz_ultrastrong_2019}%
  \BibitemOpen
  \bibfield  {author} {\bibinfo {author} {\bibfnamefont {P.}~\bibnamefont
  {Forn-D{\'i}az}}, \bibinfo {author} {\bibfnamefont {L.}~\bibnamefont
  {Lamata}}, \bibinfo {author} {\bibfnamefont {E.}~\bibnamefont {Rico}},
  \bibinfo {author} {\bibfnamefont {J.}~\bibnamefont {Kono}},\ and\ \bibinfo
  {author} {\bibfnamefont {E.}~\bibnamefont {Solano}},\ }\bibfield  {title}
  {\bibinfo {title} {Ultrastrong coupling regimes of light-matter
  interaction},\ }\bibfield  {journal} {\bibinfo  {journal} {Reviews of Modern
  Physics}\ }\textbf {\bibinfo {volume} {91}},\ \href
  {https://doi.org/10.1103/RevModPhys.91.025005} {10.1103/RevModPhys.91.025005}
  (\bibinfo {year} {2019})\BibitemShut {NoStop}%
\bibitem [{\citenamefont {Wang}\ \emph {et~al.}(2013)\citenamefont {Wang},
  \citenamefont {Steinberg}, \citenamefont {Jarillo-Herrero},\ and\
  \citenamefont {Gedik}}]{wang_observation_2013}%
  \BibitemOpen
  \bibfield  {author} {\bibinfo {author} {\bibfnamefont {Y.~H.}\ \bibnamefont
  {Wang}}, \bibinfo {author} {\bibfnamefont {H.}~\bibnamefont {Steinberg}},
  \bibinfo {author} {\bibfnamefont {P.}~\bibnamefont {Jarillo-Herrero}},\ and\
  \bibinfo {author} {\bibfnamefont {N.}~\bibnamefont {Gedik}},\ }\bibfield
  {title} {\bibinfo {title} {Observation of {F}loquet-{B}loch states on the
  surface of a {T}opological {I}nsulator},\ }\href
  {https://doi.org/10.1126/science.1239834} {\bibfield  {journal} {\bibinfo
  {journal} {Science}\ }\textbf {\bibinfo {volume} {342}},\ \bibinfo {pages}
  {453} (\bibinfo {year} {2013})}\BibitemShut {NoStop}%
\bibitem [{\citenamefont {Goldman}\ and\ \citenamefont
  {Dalibard}(2014)}]{PhysRevX.4.031027}%
  \BibitemOpen
  \bibfield  {author} {\bibinfo {author} {\bibfnamefont {N.}~\bibnamefont
  {Goldman}}\ and\ \bibinfo {author} {\bibfnamefont {J.}~\bibnamefont
  {Dalibard}},\ }\bibfield  {title} {\bibinfo {title} {Periodically driven
  quantum systems: Effective {H}amiltonians and engineered gauge fields},\
  }\href {https://doi.org/10.1103/PhysRevX.4.031027} {\bibfield  {journal}
  {\bibinfo  {journal} {Phys. Rev. X}\ }\textbf {\bibinfo {volume} {4}},\
  \bibinfo {pages} {031027} (\bibinfo {year} {2014})}\BibitemShut {NoStop}%
\bibitem [{\citenamefont {Mahmood}\ \emph {et~al.}(2016)\citenamefont
  {Mahmood}, \citenamefont {Chan}, \citenamefont {Alpichshev}, \citenamefont
  {Gardner}, \citenamefont {Lee}, \citenamefont {Lee},\ and\ \citenamefont
  {Gedik}}]{mahmood_selective_2016}%
  \BibitemOpen
  \bibfield  {author} {\bibinfo {author} {\bibfnamefont {F.}~\bibnamefont
  {Mahmood}}, \bibinfo {author} {\bibfnamefont {C.-K.}\ \bibnamefont {Chan}},
  \bibinfo {author} {\bibfnamefont {Z.}~\bibnamefont {Alpichshev}}, \bibinfo
  {author} {\bibfnamefont {D.}~\bibnamefont {Gardner}}, \bibinfo {author}
  {\bibfnamefont {Y.}~\bibnamefont {Lee}}, \bibinfo {author} {\bibfnamefont
  {P.~A.}\ \bibnamefont {Lee}},\ and\ \bibinfo {author} {\bibfnamefont
  {N.}~\bibnamefont {Gedik}},\ }\bibfield  {title} {\bibinfo {title} {Selective
  scattering between {F}loquet–{B}loch and {V}olkov states in a {T}opological
  {I}nsulator},\ }\href {https://doi.org/10.1038/nphys3609} {\bibfield
  {journal} {\bibinfo  {journal} {Nature Physics}\ }\textbf {\bibinfo {volume}
  {12}},\ \bibinfo {pages} {306} (\bibinfo {year} {2016})}\BibitemShut
  {NoStop}%
\bibitem [{\citenamefont {Salerno}\ \emph {et~al.}(2019)\citenamefont
  {Salerno}, \citenamefont {Price}, \citenamefont {Lebrat}, \citenamefont
  {H\"ausler}, \citenamefont {Esslinger}, \citenamefont {Corman}, \citenamefont
  {Brantut},\ and\ \citenamefont {Goldman}}]{PhysRevX.9.041001}%
  \BibitemOpen
  \bibfield  {author} {\bibinfo {author} {\bibfnamefont {G.}~\bibnamefont
  {Salerno}}, \bibinfo {author} {\bibfnamefont {H.~M.}\ \bibnamefont {Price}},
  \bibinfo {author} {\bibfnamefont {M.}~\bibnamefont {Lebrat}}, \bibinfo
  {author} {\bibfnamefont {S.}~\bibnamefont {H\"ausler}}, \bibinfo {author}
  {\bibfnamefont {T.}~\bibnamefont {Esslinger}}, \bibinfo {author}
  {\bibfnamefont {L.}~\bibnamefont {Corman}}, \bibinfo {author} {\bibfnamefont
  {J.-P.}\ \bibnamefont {Brantut}},\ and\ \bibinfo {author} {\bibfnamefont
  {N.}~\bibnamefont {Goldman}},\ }\bibfield  {title} {\bibinfo {title}
  {Quantized {H}all conductance of a single atomic wire: A proposal based on
  synthetic dimensions},\ }\href {https://doi.org/10.1103/PhysRevX.9.041001}
  {\bibfield  {journal} {\bibinfo  {journal} {Phys. Rev. X}\ }\textbf {\bibinfo
  {volume} {9}},\ \bibinfo {pages} {041001} (\bibinfo {year}
  {2019})}\BibitemShut {NoStop}%
\bibitem [{\citenamefont {Rudner}\ and\ \citenamefont
  {Lindner}(2020)}]{rudner_band_2020}%
  \BibitemOpen
  \bibfield  {author} {\bibinfo {author} {\bibfnamefont {M.~S.}\ \bibnamefont
  {Rudner}}\ and\ \bibinfo {author} {\bibfnamefont {N.~H.}\ \bibnamefont
  {Lindner}},\ }\bibfield  {title} {\bibinfo {title} {Band structure
  engineering and non-equilibrium dynamics in {Floquet} topological
  insulators},\ }\href {https://doi.org/10.1038/s42254-020-0170-z} {\bibfield
  {journal} {\bibinfo  {journal} {Nature Reviews Physics}\ }\textbf {\bibinfo
  {volume} {2}},\ \bibinfo {pages} {229} (\bibinfo {year} {2020})}\BibitemShut
  {NoStop}%
\bibitem [{\citenamefont {Oka}\ and\ \citenamefont
  {Kitamura}(2019)}]{oka_floquet_2019}%
  \BibitemOpen
  \bibfield  {author} {\bibinfo {author} {\bibfnamefont {T.}~\bibnamefont
  {Oka}}\ and\ \bibinfo {author} {\bibfnamefont {S.}~\bibnamefont {Kitamura}},\
  }\bibfield  {title} {\bibinfo {title} {Floquet {Engineering} of {Quantum}
  {Materials}},\ }\bibfield  {journal} {\bibinfo  {journal} {Annual Review of
  Condensed Matter Physics}\ }\textbf {\bibinfo {volume} {10}},\ \href
  {https://doi.org/10.1146/annurev-conmatphys-031218-013423}
  {10.1146/annurev-conmatphys-031218-013423} (\bibinfo {year}
  {2019})\BibitemShut {NoStop}%
\bibitem [{\citenamefont {Laussy}\ \emph {et~al.}(2010)\citenamefont {Laussy},
  \citenamefont {Kavokin},\ and\ \citenamefont
  {Shelykh}}]{laussy_exciton-polariton_2010}%
  \BibitemOpen
  \bibfield  {author} {\bibinfo {author} {\bibfnamefont {F.~P.}\ \bibnamefont
  {Laussy}}, \bibinfo {author} {\bibfnamefont {A.~V.}\ \bibnamefont
  {Kavokin}},\ and\ \bibinfo {author} {\bibfnamefont {I.~A.}\ \bibnamefont
  {Shelykh}},\ }\bibfield  {title} {\bibinfo {title} {Exciton-{Polariton}
  {Mediated} {Superconductivity}},\ }\href
  {https://doi.org/10.1103/PhysRevLett.104.106402} {\bibfield  {journal}
  {\bibinfo  {journal} {Physical Review Letters}\ }\textbf {\bibinfo {volume}
  {104}},\ \bibinfo {pages} {106402} (\bibinfo {year} {2010})}\BibitemShut
  {NoStop}%
\bibitem [{\citenamefont {Cotle{\c t}}\ \emph {et~al.}(2016)\citenamefont
  {Cotle{\c t}}, \citenamefont {Zeytino{\v g}lu}, \citenamefont {Sigrist},
  \citenamefont {Demler},\ and\ \citenamefont {Imamo{\v
  g}lu}}]{cotlet_superconductivity_2016}%
  \BibitemOpen
  \bibfield  {author} {\bibinfo {author} {\bibfnamefont {O.}~\bibnamefont
  {Cotle{\c t}}}, \bibinfo {author} {\bibfnamefont {S.}~\bibnamefont
  {Zeytino{\v g}lu}}, \bibinfo {author} {\bibfnamefont {M.}~\bibnamefont
  {Sigrist}}, \bibinfo {author} {\bibfnamefont {E.}~\bibnamefont {Demler}},\
  and\ \bibinfo {author} {\bibfnamefont {A.}~\bibnamefont {Imamo{\v g}lu}},\
  }\bibfield  {title} {\bibinfo {title} {Superconductivity and other collective
  phenomena in a hybrid {Bose}-{Fermi} mixture formed by a polariton condensate
  and an electron system in two dimensions},\ }\href
  {https://doi.org/10.1103/PhysRevB.93.054510} {\bibfield  {journal} {\bibinfo
  {journal} {Physical Review B}\ }\textbf {\bibinfo {volume} {93}},\ \bibinfo
  {pages} {054510} (\bibinfo {year} {2016})}\BibitemShut {NoStop}%
\bibitem [{\citenamefont {Kavokin}\ and\ \citenamefont
  {Lagoudakis}(2016)}]{kavokin_exciton-polariton_2016}%
  \BibitemOpen
  \bibfield  {author} {\bibinfo {author} {\bibfnamefont {A.}~\bibnamefont
  {Kavokin}}\ and\ \bibinfo {author} {\bibfnamefont {P.}~\bibnamefont
  {Lagoudakis}},\ }\bibfield  {title} {\bibinfo {title} {Exciton-polariton
  condensates: {Exciton}-mediated superconductivity},\ }\href
  {https://doi.org/10.1038/nmat4646} {\bibfield  {journal} {\bibinfo  {journal}
  {Nature Materials}\ }\textbf {\bibinfo {volume} {15}},\ \bibinfo {pages}
  {599} (\bibinfo {year} {2016})}\BibitemShut {NoStop}%
\bibitem [{\citenamefont {Sentef}\ \emph {et~al.}(2018)\citenamefont {Sentef},
  \citenamefont {Ruggenthaler},\ and\ \citenamefont
  {Rubio}}]{sentef_cavity_2018}%
  \BibitemOpen
  \bibfield  {author} {\bibinfo {author} {\bibfnamefont {M.~A.}\ \bibnamefont
  {Sentef}}, \bibinfo {author} {\bibfnamefont {M.}~\bibnamefont
  {Ruggenthaler}},\ and\ \bibinfo {author} {\bibfnamefont {A.}~\bibnamefont
  {Rubio}},\ }\bibfield  {title} {\bibinfo {title} {Cavity
  quantum-electrodynamical polaritonically enhanced electron-phonon coupling
  and its influence on superconductivity},\ }\href
  {https://doi.org/10.1126/sciadv.aau6969} {\bibfield  {journal} {\bibinfo
  {journal} {Science Advances}\ }\textbf {\bibinfo {volume} {4}},\ \bibinfo
  {pages} {eaau6969} (\bibinfo {year} {2018})}\BibitemShut {NoStop}%
\bibitem [{\citenamefont {Schlawin}\ \emph {et~al.}(2019)\citenamefont
  {Schlawin}, \citenamefont {Cavalleri},\ and\ \citenamefont
  {Jaksch}}]{schlawin_cavity-mediated_2019}%
  \BibitemOpen
  \bibfield  {author} {\bibinfo {author} {\bibfnamefont {F.}~\bibnamefont
  {Schlawin}}, \bibinfo {author} {\bibfnamefont {A.}~\bibnamefont
  {Cavalleri}},\ and\ \bibinfo {author} {\bibfnamefont {D.}~\bibnamefont
  {Jaksch}},\ }\bibfield  {title} {\bibinfo {title} {Cavity-{Mediated}
  {Electron}-{Photon} {Superconductivity}},\ }\href
  {https://doi.org/10.1103/PhysRevLett.122.133602} {\bibfield  {journal}
  {\bibinfo  {journal} {Physical Review Letters}\ }\textbf {\bibinfo {volume}
  {122}},\ \bibinfo {pages} {133602} (\bibinfo {year} {2019})}\BibitemShut
  {NoStop}%
\bibitem [{\citenamefont {Hagenm{\"u}ller}\ \emph {et~al.}(2019)\citenamefont
  {Hagenm{\"u}ller}, \citenamefont {Schachenmayer}, \citenamefont {Genet},
  \citenamefont {Ebbesen},\ and\ \citenamefont
  {Pupillo}}]{hagenmuller_enhancement_2019}%
  \BibitemOpen
  \bibfield  {author} {\bibinfo {author} {\bibfnamefont {D.}~\bibnamefont
  {Hagenm{\"u}ller}}, \bibinfo {author} {\bibfnamefont {J.}~\bibnamefont
  {Schachenmayer}}, \bibinfo {author} {\bibfnamefont {C.}~\bibnamefont
  {Genet}}, \bibinfo {author} {\bibfnamefont {T.~W.}\ \bibnamefont {Ebbesen}},\
  and\ \bibinfo {author} {\bibfnamefont {G.}~\bibnamefont {Pupillo}},\
  }\bibfield  {title} {\bibinfo {title} {Enhancement of the electron-phonon
  scattering induced by intrinsic surface plasmon-phonon polaritons},\
  }\bibfield  {journal} {\bibinfo  {journal} {ACS Photonics}\ }\textbf
  {\bibinfo {volume} {6}},\ \href
  {https://doi.org/10.1021/acsphotonics.9b00268} {10.1021/acsphotonics.9b00268}
  (\bibinfo {year} {2019})\BibitemShut {NoStop}%
\bibitem [{\citenamefont {Curtis}\ \emph {et~al.}(2019)\citenamefont {Curtis},
  \citenamefont {Raines}, \citenamefont {Allocca}, \citenamefont {Hafezi},\
  and\ \citenamefont {Galitski}}]{curtis_cavity_2019}%
  \BibitemOpen
  \bibfield  {author} {\bibinfo {author} {\bibfnamefont {J.~B.}\ \bibnamefont
  {Curtis}}, \bibinfo {author} {\bibfnamefont {Z.~M.}\ \bibnamefont {Raines}},
  \bibinfo {author} {\bibfnamefont {A.~A.}\ \bibnamefont {Allocca}}, \bibinfo
  {author} {\bibfnamefont {M.}~\bibnamefont {Hafezi}},\ and\ \bibinfo {author}
  {\bibfnamefont {V.~M.}\ \bibnamefont {Galitski}},\ }\bibfield  {title}
  {\bibinfo {title} {Cavity {Quantum} {Eliashberg} {Enhancement} of
  {Superconductivity}},\ }\href
  {https://doi.org/10.1103/PhysRevLett.122.167002} {\bibfield  {journal}
  {\bibinfo  {journal} {Physical Review Letters}\ }\textbf {\bibinfo {volume}
  {122}},\ \bibinfo {pages} {167002} (\bibinfo {year} {2019})}\BibitemShut
  {NoStop}%
\bibitem [{\citenamefont {Wang}\ \emph {et~al.}(2019)\citenamefont {Wang},
  \citenamefont {Ronca},\ and\ \citenamefont {Sentef}}]{wang_cavity_2019}%
  \BibitemOpen
  \bibfield  {author} {\bibinfo {author} {\bibfnamefont {X.}~\bibnamefont
  {Wang}}, \bibinfo {author} {\bibfnamefont {E.}~\bibnamefont {Ronca}},\ and\
  \bibinfo {author} {\bibfnamefont {M.~A.}\ \bibnamefont {Sentef}},\ }\bibfield
   {title} {\bibinfo {title} {Cavity quantum electrodynamical {Chern}
  insulator: {Towards} light-induced quantized anomalous {Hall} effect in
  graphene},\ }\href {https://doi.org/10.1103/PhysRevB.99.235156} {\bibfield
  {journal} {\bibinfo  {journal} {Physical Review B}\ }\textbf {\bibinfo
  {volume} {99}},\ \bibinfo {pages} {235156} (\bibinfo {year}
  {2019})}\BibitemShut {NoStop}%
\bibitem [{\citenamefont {Sentef}\ \emph {et~al.}(2020)\citenamefont {Sentef},
  \citenamefont {Li}, \citenamefont {K{\"u}nzel},\ and\ \citenamefont
  {Eckstein}}]{sentef_quantum_2020}%
  \BibitemOpen
  \bibfield  {author} {\bibinfo {author} {\bibfnamefont {M.~A.}\ \bibnamefont
  {Sentef}}, \bibinfo {author} {\bibfnamefont {J.}~\bibnamefont {Li}}, \bibinfo
  {author} {\bibfnamefont {F.}~\bibnamefont {K{\"u}nzel}},\ and\ \bibinfo
  {author} {\bibfnamefont {M.}~\bibnamefont {Eckstein}},\ }\bibfield  {title}
  {\bibinfo {title} {Quantum to classical crossover of {Floquet} engineering in
  correlated quantum systems},\ }\href
  {https://doi.org/10.1103/PhysRevResearch.2.033033} {\bibfield  {journal}
  {\bibinfo  {journal} {Physical Review Research}\ }\textbf {\bibinfo {volume}
  {2}},\ \bibinfo {pages} {033033} (\bibinfo {year} {2020})}\BibitemShut
  {NoStop}%
\bibitem [{\citenamefont {Thomas}\ \emph {et~al.}(2019)\citenamefont {Thomas},
  \citenamefont {Devaux}, \citenamefont {Nagarajan}, \citenamefont {Chervy},
  \citenamefont {Seidel}, \citenamefont {Hagenm{\"u}ller}, \citenamefont
  {Sch{\"u}tz}, \citenamefont {Schachenmayer}, \citenamefont {Genet},
  \citenamefont {Pupillo},\ and\ \citenamefont
  {Ebbesen}}]{thomas_exploring_2019}%
  \BibitemOpen
  \bibfield  {author} {\bibinfo {author} {\bibfnamefont {A.}~\bibnamefont
  {Thomas}}, \bibinfo {author} {\bibfnamefont {E.}~\bibnamefont {Devaux}},
  \bibinfo {author} {\bibfnamefont {K.}~\bibnamefont {Nagarajan}}, \bibinfo
  {author} {\bibfnamefont {T.}~\bibnamefont {Chervy}}, \bibinfo {author}
  {\bibfnamefont {M.}~\bibnamefont {Seidel}}, \bibinfo {author} {\bibfnamefont
  {D.}~\bibnamefont {Hagenm{\"u}ller}}, \bibinfo {author} {\bibfnamefont
  {S.}~\bibnamefont {Sch{\"u}tz}}, \bibinfo {author} {\bibfnamefont
  {J.}~\bibnamefont {Schachenmayer}}, \bibinfo {author} {\bibfnamefont
  {C.}~\bibnamefont {Genet}}, \bibinfo {author} {\bibfnamefont
  {G.}~\bibnamefont {Pupillo}},\ and\ \bibinfo {author} {\bibfnamefont {T.~W.}\
  \bibnamefont {Ebbesen}},\ }\bibfield  {title} {\bibinfo {title} {Exploring
  {Superconductivity} under {Strong} {Coupling} with the {Vacuum}
  {Electromagnetic} {Field}},\ }\href {http://arxiv.org/abs/1911.01459}
  {\bibfield  {journal} {\bibinfo  {journal} {arXiv:1911.01459 [cond-mat,
  physics:quant-ph]}\ } (\bibinfo {year} {2019})}\BibitemShut {NoStop}%
\bibitem [{\citenamefont {Kiffner}\ \emph {et~al.}(2019)\citenamefont
  {Kiffner}, \citenamefont {Coulthard}, \citenamefont {Schlawin}, \citenamefont
  {Ardavan},\ and\ \citenamefont {Jaksch}}]{kiffner_manipulating_2019}%
  \BibitemOpen
  \bibfield  {author} {\bibinfo {author} {\bibfnamefont {M.}~\bibnamefont
  {Kiffner}}, \bibinfo {author} {\bibfnamefont {J.~R.}\ \bibnamefont
  {Coulthard}}, \bibinfo {author} {\bibfnamefont {F.}~\bibnamefont {Schlawin}},
  \bibinfo {author} {\bibfnamefont {A.}~\bibnamefont {Ardavan}},\ and\ \bibinfo
  {author} {\bibfnamefont {D.}~\bibnamefont {Jaksch}},\ }\bibfield  {title}
  {\bibinfo {title} {Manipulating quantum materials with quantum light},\
  }\href {https://doi.org/10.1103/PhysRevB.99.085116} {\bibfield  {journal}
  {\bibinfo  {journal} {Physical Review B}\ }\textbf {\bibinfo {volume} {99}},\
  \bibinfo {pages} {085116} (\bibinfo {year} {2019})}\BibitemShut {NoStop}%
\bibitem [{\citenamefont {Mazza}\ and\ \citenamefont
  {Georges}(2019)}]{mazza_superradiant_2019}%
  \BibitemOpen
  \bibfield  {author} {\bibinfo {author} {\bibfnamefont {G.}~\bibnamefont
  {Mazza}}\ and\ \bibinfo {author} {\bibfnamefont {A.}~\bibnamefont
  {Georges}},\ }\bibfield  {title} {\bibinfo {title} {Superradiant {Quantum}
  {Materials}},\ }\href {https://doi.org/10.1103/PhysRevLett.122.017401}
  {\bibfield  {journal} {\bibinfo  {journal} {Physical Review Letters}\
  }\textbf {\bibinfo {volume} {122}},\ \bibinfo {pages} {017401} (\bibinfo
  {year} {2019})}\BibitemShut {NoStop}%
\bibitem [{\citenamefont {Andolina}\ \emph {et~al.}(2019)\citenamefont
  {Andolina}, \citenamefont {Pellegrino}, \citenamefont {Giovannetti},
  \citenamefont {MacDonald},\ and\ \citenamefont
  {Polini}}]{andolina_cavity_2019}%
  \BibitemOpen
  \bibfield  {author} {\bibinfo {author} {\bibfnamefont {G.~M.}\ \bibnamefont
  {Andolina}}, \bibinfo {author} {\bibfnamefont {F.~M.~D.}\ \bibnamefont
  {Pellegrino}}, \bibinfo {author} {\bibfnamefont {V.}~\bibnamefont
  {Giovannetti}}, \bibinfo {author} {\bibfnamefont {A.~H.}\ \bibnamefont
  {MacDonald}},\ and\ \bibinfo {author} {\bibfnamefont {M.}~\bibnamefont
  {Polini}},\ }\bibfield  {title} {\bibinfo {title} {Cavity quantum
  electrodynamics of strongly correlated electron systems: {A} no-go theorem
  for photon condensation},\ }\href
  {https://doi.org/10.1103/PhysRevB.100.121109} {\bibfield  {journal} {\bibinfo
   {journal} {Physical Review B}\ }\textbf {\bibinfo {volume} {100}},\ \bibinfo
  {pages} {121109} (\bibinfo {year} {2019})}\BibitemShut {NoStop}%
\bibitem [{\citenamefont {Gao}\ \emph {et~al.}(2020)\citenamefont {Gao},
  \citenamefont {Schlawin}, \citenamefont {Buzzi}, \citenamefont {Cavalleri},\
  and\ \citenamefont {Jaksch}}]{gao_photoinduced_2020}%
  \BibitemOpen
  \bibfield  {author} {\bibinfo {author} {\bibfnamefont {H.}~\bibnamefont
  {Gao}}, \bibinfo {author} {\bibfnamefont {F.}~\bibnamefont {Schlawin}},
  \bibinfo {author} {\bibfnamefont {M.}~\bibnamefont {Buzzi}}, \bibinfo
  {author} {\bibfnamefont {A.}~\bibnamefont {Cavalleri}},\ and\ \bibinfo
  {author} {\bibfnamefont {D.}~\bibnamefont {Jaksch}},\ }\bibfield  {title}
  {\bibinfo {title} {Photoinduced {Electron} {Pairing} in a {Driven}
  {Cavity}},\ }\href {https://doi.org/10.1103/PhysRevLett.125.053602}
  {\bibfield  {journal} {\bibinfo  {journal} {Physical Review Letters}\
  }\textbf {\bibinfo {volume} {125}},\ \bibinfo {pages} {053602} (\bibinfo
  {year} {2020})}\BibitemShut {NoStop}%
\bibitem [{\citenamefont {Chakraborty}\ and\ \citenamefont
  {Piazza}(2020)}]{chakraborty_non-bcs-type_2020}%
  \BibitemOpen
  \bibfield  {author} {\bibinfo {author} {\bibfnamefont {A.}~\bibnamefont
  {Chakraborty}}\ and\ \bibinfo {author} {\bibfnamefont {F.}~\bibnamefont
  {Piazza}},\ }\bibfield  {title} {\bibinfo {title} {Non-{BCS}-{Type}
  {Enhancement} of {Superconductivity} from {Long}-{Range} {Photon}
  {Fluctuations}},\ }\href {http://arxiv.org/abs/2008.06513} {\bibfield
  {journal} {\bibinfo  {journal} {arXiv:2008.06513 [cond-mat]}\ } (\bibinfo
  {year} {2020})}\BibitemShut {NoStop}%
\bibitem [{\citenamefont {Li}\ and\ \citenamefont
  {Eckstein}(2020)}]{li_manipulating_2020}%
  \BibitemOpen
  \bibfield  {author} {\bibinfo {author} {\bibfnamefont {J.}~\bibnamefont
  {Li}}\ and\ \bibinfo {author} {\bibfnamefont {M.}~\bibnamefont {Eckstein}},\
  }\bibfield  {title} {\bibinfo {title} {Manipulating {Intertwined} {Orders} in
  {Solids} with {Quantum} {Light}},\ }\href
  {https://doi.org/10.1103/PhysRevLett.125.217402} {\bibfield  {journal}
  {\bibinfo  {journal} {Physical Review Letters}\ }\textbf {\bibinfo {volume}
  {125}},\ \bibinfo {pages} {217402} (\bibinfo {year} {2020})}\BibitemShut
  {NoStop}%
\bibitem [{\citenamefont {H{\"u}bener}\ \emph {et~al.}(2020)\citenamefont
  {H{\"u}bener}, \citenamefont {De~Giovannini}, \citenamefont {Sch{\"a}fer},
  \citenamefont {Andberger}, \citenamefont {Ruggenthaler}, \citenamefont
  {Faist},\ and\ \citenamefont {Rubio}}]{hubener_engineering_2020}%
  \BibitemOpen
  \bibfield  {author} {\bibinfo {author} {\bibfnamefont {H.}~\bibnamefont
  {H{\"u}bener}}, \bibinfo {author} {\bibfnamefont {U.}~\bibnamefont
  {De~Giovannini}}, \bibinfo {author} {\bibfnamefont {C.}~\bibnamefont
  {Sch{\"a}fer}}, \bibinfo {author} {\bibfnamefont {J.}~\bibnamefont
  {Andberger}}, \bibinfo {author} {\bibfnamefont {M.}~\bibnamefont
  {Ruggenthaler}}, \bibinfo {author} {\bibfnamefont {J.}~\bibnamefont
  {Faist}},\ and\ \bibinfo {author} {\bibfnamefont {A.}~\bibnamefont {Rubio}},\
  }\bibfield  {title} {\bibinfo {title} {Engineering quantum materials with
  chiral optical cavities},\ }\href
  {https://doi.org/10.1038/s41563-020-00801-7} {\bibfield  {journal} {\bibinfo
  {journal} {Nature Materials}\ ,\ \bibinfo {pages} {1}} (\bibinfo {year}
  {2020})}\BibitemShut {NoStop}%
\bibitem [{\citenamefont {Ashida}\ \emph {et~al.}(2020)\citenamefont {Ashida},
  \citenamefont {{\.I}mamo{\u g}lu}, \citenamefont {Faist}, \citenamefont
  {Jaksch}, \citenamefont {Cavalleri},\ and\ \citenamefont
  {Demler}}]{ashida_quantum_2020}%
  \BibitemOpen
  \bibfield  {author} {\bibinfo {author} {\bibfnamefont {Y.}~\bibnamefont
  {Ashida}}, \bibinfo {author} {\bibfnamefont {A.}~\bibnamefont {{\.I}mamo{\u
  g}lu}}, \bibinfo {author} {\bibfnamefont {J.}~\bibnamefont {Faist}}, \bibinfo
  {author} {\bibfnamefont {D.}~\bibnamefont {Jaksch}}, \bibinfo {author}
  {\bibfnamefont {A.}~\bibnamefont {Cavalleri}},\ and\ \bibinfo {author}
  {\bibfnamefont {E.}~\bibnamefont {Demler}},\ }\bibfield  {title} {\bibinfo
  {title} {Quantum {Electrodynamic} {Control} of {Matter}: {Cavity}-{Enhanced}
  {Ferroelectric} {Phase} {Transition}},\ }\href
  {https://doi.org/10.1103/PhysRevX.10.041027} {\bibfield  {journal} {\bibinfo
  {journal} {Physical Review X}\ }\textbf {\bibinfo {volume} {10}},\ \bibinfo
  {pages} {041027} (\bibinfo {year} {2020})}\BibitemShut {NoStop}%
\bibitem [{\citenamefont {Latini}\ \emph {et~al.}(2021)\citenamefont {Latini},
  \citenamefont {Shin}, \citenamefont {Sato}, \citenamefont {Sch{\"a}fer},
  \citenamefont {De~Giovannini}, \citenamefont {H{\"u}bener},\ and\
  \citenamefont {Rubio}}]{latini_ferroelectric_2021}%
  \BibitemOpen
  \bibfield  {author} {\bibinfo {author} {\bibfnamefont {S.}~\bibnamefont
  {Latini}}, \bibinfo {author} {\bibfnamefont {D.}~\bibnamefont {Shin}},
  \bibinfo {author} {\bibfnamefont {S.~A.}\ \bibnamefont {Sato}}, \bibinfo
  {author} {\bibfnamefont {C.}~\bibnamefont {Sch{\"a}fer}}, \bibinfo {author}
  {\bibfnamefont {U.}~\bibnamefont {De~Giovannini}}, \bibinfo {author}
  {\bibfnamefont {H.}~\bibnamefont {H{\"u}bener}},\ and\ \bibinfo {author}
  {\bibfnamefont {A.}~\bibnamefont {Rubio}},\ }\bibfield  {title} {\bibinfo
  {title} {The {Ferroelectric} {Photo}-{Groundstate} of {SrTiO}\$\_3\$:
  {Cavity} {Materials} {Engineering}},\ }\href
  {http://arxiv.org/abs/2101.11313} {\bibfield  {journal} {\bibinfo  {journal}
  {arXiv:2101.11313 [cond-mat, physics:physics]}\ } (\bibinfo {year}
  {2021})}\BibitemShut {NoStop}%
\bibitem [{\citenamefont {Topp}\ \emph {et~al.}(2019)\citenamefont {Topp},
  \citenamefont {Jotzu}, \citenamefont {McIver}, \citenamefont {Xian},
  \citenamefont {Rubio},\ and\ \citenamefont {Sentef}}]{topp_topological_2019}%
  \BibitemOpen
  \bibfield  {author} {\bibinfo {author} {\bibfnamefont {G.~E.}\ \bibnamefont
  {Topp}}, \bibinfo {author} {\bibfnamefont {G.}~\bibnamefont {Jotzu}},
  \bibinfo {author} {\bibfnamefont {J.~W.}\ \bibnamefont {McIver}}, \bibinfo
  {author} {\bibfnamefont {L.}~\bibnamefont {Xian}}, \bibinfo {author}
  {\bibfnamefont {A.}~\bibnamefont {Rubio}},\ and\ \bibinfo {author}
  {\bibfnamefont {M.~A.}\ \bibnamefont {Sentef}},\ }\bibfield  {title}
  {\bibinfo {title} {Topological {Floquet} engineering of twisted bilayer
  graphene},\ }\href {https://doi.org/10.1103/PhysRevResearch.1.023031}
  {\bibfield  {journal} {\bibinfo  {journal} {Physical Review Research}\
  }\textbf {\bibinfo {volume} {1}},\ \bibinfo {pages} {023031} (\bibinfo {year}
  {2019})}\BibitemShut {NoStop}%
\bibitem [{\citenamefont {Vogl}\ \emph
  {et~al.}(2020{\natexlab{a}})\citenamefont {Vogl}, \citenamefont
  {Rodriguez-Vega},\ and\ \citenamefont {Fiete}}]{vogl_effective_2020}%
  \BibitemOpen
  \bibfield  {author} {\bibinfo {author} {\bibfnamefont {M.}~\bibnamefont
  {Vogl}}, \bibinfo {author} {\bibfnamefont {M.}~\bibnamefont
  {Rodriguez-Vega}},\ and\ \bibinfo {author} {\bibfnamefont {G.~A.}\
  \bibnamefont {Fiete}},\ }\bibfield  {title} {\bibinfo {title} {Effective
  {Floquet} {Hamiltonians} for periodically driven twisted bilayer graphene},\
  }\href {https://doi.org/10.1103/PhysRevB.101.235411} {\bibfield  {journal}
  {\bibinfo  {journal} {Physical Review B}\ }\textbf {\bibinfo {volume}
  {101}},\ \bibinfo {pages} {235411} (\bibinfo {year}
  {2020}{\natexlab{a}})}\BibitemShut {NoStop}%
\bibitem [{\citenamefont {Vogl}\ \emph
  {et~al.}(2020{\natexlab{b}})\citenamefont {Vogl}, \citenamefont
  {Rodriguez-Vega},\ and\ \citenamefont {Fiete}}]{vogl_floquet_2020}%
  \BibitemOpen
  \bibfield  {author} {\bibinfo {author} {\bibfnamefont {M.}~\bibnamefont
  {Vogl}}, \bibinfo {author} {\bibfnamefont {M.}~\bibnamefont
  {Rodriguez-Vega}},\ and\ \bibinfo {author} {\bibfnamefont {G.~A.}\
  \bibnamefont {Fiete}},\ }\bibfield  {title} {\bibinfo {title} {Floquet
  engineering of interlayer couplings: {Tuning} the magic angle of twisted
  bilayer graphene at the exit of a waveguide},\ }\bibfield  {journal}
  {\bibinfo  {journal} {Physical Review B}\ }\textbf {\bibinfo {volume}
  {101}},\ \href {https://doi.org/10.1103/PhysRevB.101.241408}
  {10.1103/PhysRevB.101.241408} (\bibinfo {year}
  {2020}{\natexlab{b}})\BibitemShut {NoStop}%
\bibitem [{\citenamefont {Katz}\ \emph {et~al.}(2020)\citenamefont {Katz},
  \citenamefont {Refael},\ and\ \citenamefont {Lindner}}]{katz_optically_2020}%
  \BibitemOpen
  \bibfield  {author} {\bibinfo {author} {\bibfnamefont {O.}~\bibnamefont
  {Katz}}, \bibinfo {author} {\bibfnamefont {G.}~\bibnamefont {Refael}},\ and\
  \bibinfo {author} {\bibfnamefont {N.~H.}\ \bibnamefont {Lindner}},\
  }\bibfield  {title} {\bibinfo {title} {Optically induced flat bands in
  twisted bilayer graphene},\ }\href
  {https://doi.org/10.1103/PhysRevB.102.155123} {\bibfield  {journal} {\bibinfo
   {journal} {Physical Review B}\ }\textbf {\bibinfo {volume} {102}},\ \bibinfo
  {pages} {155123} (\bibinfo {year} {2020})}\BibitemShut {NoStop}%
\bibitem [{\citenamefont {Rodriguez-Vega}\ \emph {et~al.}(2020)\citenamefont
  {Rodriguez-Vega}, \citenamefont {Vogl},\ and\ \citenamefont
  {Fiete}}]{rodriguez-vega_floquet_2020}%
  \BibitemOpen
  \bibfield  {author} {\bibinfo {author} {\bibfnamefont {M.}~\bibnamefont
  {Rodriguez-Vega}}, \bibinfo {author} {\bibfnamefont {M.}~\bibnamefont
  {Vogl}},\ and\ \bibinfo {author} {\bibfnamefont {G.~A.}\ \bibnamefont
  {Fiete}},\ }\bibfield  {title} {\bibinfo {title} {Floquet engineering of
  twisted double bilayer graphene},\ }\href
  {https://doi.org/10.1103/PhysRevResearch.2.033494} {\bibfield  {journal}
  {\bibinfo  {journal} {Physical Review Research}\ }\textbf {\bibinfo {volume}
  {2}},\ \bibinfo {pages} {033494} (\bibinfo {year} {2020})}\BibitemShut
  {NoStop}%
\bibitem [{\citenamefont {Li}\ \emph {et~al.}(2020{\natexlab{a}})\citenamefont
  {Li}, \citenamefont {Fertig},\ and\ \citenamefont
  {Seradjeh}}]{li_floquet-engineered_2020}%
  \BibitemOpen
  \bibfield  {author} {\bibinfo {author} {\bibfnamefont {Y.}~\bibnamefont
  {Li}}, \bibinfo {author} {\bibfnamefont {H.~A.}\ \bibnamefont {Fertig}},\
  and\ \bibinfo {author} {\bibfnamefont {B.}~\bibnamefont {Seradjeh}},\
  }\bibfield  {title} {\bibinfo {title} {Floquet-engineered topological flat
  bands in irradiated twisted bilayer graphene},\ }\href
  {https://doi.org/10.1103/PhysRevResearch.2.043275} {\bibfield  {journal}
  {\bibinfo  {journal} {Physical Review Research}\ }\textbf {\bibinfo {volume}
  {2}},\ \bibinfo {pages} {043275} (\bibinfo {year}
  {2020}{\natexlab{a}})}\BibitemShut {NoStop}%
\bibitem [{\citenamefont {Lu}\ \emph {et~al.}(2020)\citenamefont {Lu},
  \citenamefont {Zeng}, \citenamefont {Liu}, \citenamefont {Gao},\ and\
  \citenamefont {Xie}}]{lu_valley-selective_2020}%
  \BibitemOpen
  \bibfield  {author} {\bibinfo {author} {\bibfnamefont {M.}~\bibnamefont
  {Lu}}, \bibinfo {author} {\bibfnamefont {J.}~\bibnamefont {Zeng}}, \bibinfo
  {author} {\bibfnamefont {H.}~\bibnamefont {Liu}}, \bibinfo {author}
  {\bibfnamefont {J.-H.}\ \bibnamefont {Gao}},\ and\ \bibinfo {author}
  {\bibfnamefont {X.~C.}\ \bibnamefont {Xie}},\ }\bibfield  {title} {\bibinfo
  {title} {Valley-selective {Floquet} {Chern} {Flat} {Bands} in {Twisted}
  {Multilayer} {Graphene}},\ }\href {http://arxiv.org/abs/2007.15489}
  {\bibfield  {journal} {\bibinfo  {journal} {arXiv:2007.15489 [cond-mat]}\ }
  (\bibinfo {year} {2020})}\BibitemShut {NoStop}%
\bibitem [{\citenamefont {Vogl}\ \emph {et~al.}(2021)\citenamefont {Vogl},
  \citenamefont {Rodriguez-Vega}, \citenamefont {Flebus}, \citenamefont
  {MacDonald},\ and\ \citenamefont {Fiete}}]{vogl_floquet_2021}%
  \BibitemOpen
  \bibfield  {author} {\bibinfo {author} {\bibfnamefont {M.}~\bibnamefont
  {Vogl}}, \bibinfo {author} {\bibfnamefont {M.}~\bibnamefont
  {Rodriguez-Vega}}, \bibinfo {author} {\bibfnamefont {B.}~\bibnamefont
  {Flebus}}, \bibinfo {author} {\bibfnamefont {A.~H.}\ \bibnamefont
  {MacDonald}},\ and\ \bibinfo {author} {\bibfnamefont {G.~A.}\ \bibnamefont
  {Fiete}},\ }\bibfield  {title} {\bibinfo {title} {Floquet engineering of
  topological transitions in a twisted transition metal dichalcogenide
  homobilayer},\ }\href {https://doi.org/10.1103/PhysRevB.103.014310}
  {\bibfield  {journal} {\bibinfo  {journal} {Physical Review B}\ }\textbf
  {\bibinfo {volume} {103}},\ \bibinfo {pages} {014310} (\bibinfo {year}
  {2021})}\BibitemShut {NoStop}%
\bibitem [{\citenamefont {Rodriguez-Vega}\ \emph {et~al.}(2021)\citenamefont
  {Rodriguez-Vega}, \citenamefont {Vogl},\ and\ \citenamefont
  {Fiete}}]{rodriguez-vega_low-frequency_2021}%
  \BibitemOpen
  \bibfield  {author} {\bibinfo {author} {\bibfnamefont {M.}~\bibnamefont
  {Rodriguez-Vega}}, \bibinfo {author} {\bibfnamefont {M.}~\bibnamefont
  {Vogl}},\ and\ \bibinfo {author} {\bibfnamefont {G.~A.}\ \bibnamefont
  {Fiete}},\ }\bibfield  {title} {\bibinfo {title} {Low-frequency and moir{\'e}
  {Floquet} engineering: a review},\ }\href {http://arxiv.org/abs/2011.11079}
  {\bibfield  {journal} {\bibinfo  {journal} {arXiv:2011.11079 [cond-mat]}\ }
  (\bibinfo {year} {2021})}\BibitemShut {NoStop}%
\bibitem [{\citenamefont {Ikeda}(2020)}]{ikeda_high-order_2020}%
  \BibitemOpen
  \bibfield  {author} {\bibinfo {author} {\bibfnamefont {T.~N.}\ \bibnamefont
  {Ikeda}},\ }\bibfield  {title} {\bibinfo {title} {High-order nonlinear
  optical response of a twisted bilayer graphene},\ }\href
  {https://doi.org/10.1103/PhysRevResearch.2.032015} {\bibfield  {journal}
  {\bibinfo  {journal} {Physical Review Research}\ }\textbf {\bibinfo {volume}
  {2}},\ \bibinfo {pages} {032015} (\bibinfo {year} {2020})}\BibitemShut
  {NoStop}%
\bibitem [{\citenamefont {Deng}\ \emph {et~al.}(2020)\citenamefont {Deng},
  \citenamefont {Ma}, \citenamefont {Wang}, \citenamefont {Yuan}, \citenamefont
  {Watanabe}, \citenamefont {Taniguchi}, \citenamefont {Zhang},\ and\
  \citenamefont {Xia}}]{deng_strong_2020}%
  \BibitemOpen
  \bibfield  {author} {\bibinfo {author} {\bibfnamefont {B.}~\bibnamefont
  {Deng}}, \bibinfo {author} {\bibfnamefont {C.}~\bibnamefont {Ma}}, \bibinfo
  {author} {\bibfnamefont {Q.}~\bibnamefont {Wang}}, \bibinfo {author}
  {\bibfnamefont {S.}~\bibnamefont {Yuan}}, \bibinfo {author} {\bibfnamefont
  {K.}~\bibnamefont {Watanabe}}, \bibinfo {author} {\bibfnamefont
  {T.}~\bibnamefont {Taniguchi}}, \bibinfo {author} {\bibfnamefont
  {F.}~\bibnamefont {Zhang}},\ and\ \bibinfo {author} {\bibfnamefont
  {F.}~\bibnamefont {Xia}},\ }\bibfield  {title} {\bibinfo {title} {Strong
  mid-infrared photoresponse in small-twist-angle bilayer graphene},\ }\href
  {https://doi.org/10.1038/s41566-020-0644-7} {\bibfield  {journal} {\bibinfo
  {journal} {Nature Photonics}\ }\textbf {\bibinfo {volume} {14}},\ \bibinfo
  {pages} {549} (\bibinfo {year} {2020})}\BibitemShut {NoStop}%
\bibitem [{\citenamefont {Mauri}\ and\ \citenamefont
  {Louie}(1996)}]{mauri_magnetic_1996}%
  \BibitemOpen
  \bibfield  {author} {\bibinfo {author} {\bibfnamefont {F.}~\bibnamefont
  {Mauri}}\ and\ \bibinfo {author} {\bibfnamefont {S.~G.}\ \bibnamefont
  {Louie}},\ }\bibfield  {title} {\bibinfo {title} {Magnetic {Susceptibility}
  of {Insulators} from {First} {Principles}},\ }\href
  {https://doi.org/10.1103/PhysRevLett.76.4246} {\bibfield  {journal} {\bibinfo
   {journal} {Physical Review Letters}\ }\textbf {\bibinfo {volume} {76}},\
  \bibinfo {pages} {4246} (\bibinfo {year} {1996})}\BibitemShut {NoStop}%
\bibitem [{\citenamefont {Pickard}\ and\ \citenamefont
  {Mauri}(2002)}]{pickard_first-principles_2002}%
  \BibitemOpen
  \bibfield  {author} {\bibinfo {author} {\bibfnamefont {C.~J.}\ \bibnamefont
  {Pickard}}\ and\ \bibinfo {author} {\bibfnamefont {F.}~\bibnamefont
  {Mauri}},\ }\bibfield  {title} {\bibinfo {title} {First-{Principles} {Theory}
  of the {EPR} {\textit g} {Tensor} in {Solids}: {Defects} in {Quartz}},\
  }\href {https://doi.org/10.1103/PhysRevLett.88.086403} {\bibfield  {journal}
  {\bibinfo  {journal} {Physical Review Letters}\ }\textbf {\bibinfo {volume}
  {88}},\ \bibinfo {pages} {086403} (\bibinfo {year} {2002})}\BibitemShut
  {NoStop}%
\bibitem [{\citenamefont {Strubbe}\ \emph {et~al.}(2012)\citenamefont
  {Strubbe}, \citenamefont {Lehtovaara}, \citenamefont {Rubio}, \citenamefont
  {Marques},\ and\ \citenamefont {Louie}}]{marques_response_2012}%
  \BibitemOpen
  \bibfield  {author} {\bibinfo {author} {\bibfnamefont {D.~A.}\ \bibnamefont
  {Strubbe}}, \bibinfo {author} {\bibfnamefont {L.}~\bibnamefont {Lehtovaara}},
  \bibinfo {author} {\bibfnamefont {A.}~\bibnamefont {Rubio}}, \bibinfo
  {author} {\bibfnamefont {M.~A.~L.}\ \bibnamefont {Marques}},\ and\ \bibinfo
  {author} {\bibfnamefont {S.~G.}\ \bibnamefont {Louie}},\ }\bibfield  {title}
  {\bibinfo {title} {Response {Functions} in {TDDFT}: {Concepts} and
  {Implementation}},\ }in\ \href {https://doi.org/10.1007/978-3-642-23518-4_7}
  {\emph {\bibinfo {booktitle} {Fundamentals of {Time}-{Dependent} {Density}
  {Functional} {Theory}}}},\ Vol.\ \bibinfo {volume} {837},\ \bibinfo {editor}
  {edited by\ \bibinfo {editor} {\bibfnamefont {M.~A.}\ \bibnamefont
  {Marques}}, \bibinfo {editor} {\bibfnamefont {N.~T.}\ \bibnamefont {Maitra}},
  \bibinfo {editor} {\bibfnamefont {F.~M.}\ \bibnamefont {Nogueira}}, \bibinfo
  {editor} {\bibfnamefont {E.}~\bibnamefont {Gross}},\ and\ \bibinfo {editor}
  {\bibfnamefont {A.}~\bibnamefont {Rubio}}}\ (\bibinfo  {publisher} {Springer
  Berlin Heidelberg},\ \bibinfo {address} {Berlin, Heidelberg},\ \bibinfo
  {year} {2012})\ pp.\ \bibinfo {pages} {139--166},\ \bibinfo {note} {series
  Title: Lecture Notes in Physics}\BibitemShut {NoStop}%
\bibitem [{\citenamefont {Chen}\ and\ \citenamefont
  {Huang}(2021)}]{chen_probing_2021}%
  \BibitemOpen
  \bibfield  {author} {\bibinfo {author} {\bibfnamefont {W.}~\bibnamefont
  {Chen}}\ and\ \bibinfo {author} {\bibfnamefont {W.}~\bibnamefont {Huang}},\
  }\bibfield  {title} {\bibinfo {title} {Probing the quantum geometry of paired
  electrons in multiorbital superconductors through intrinsic optical
  conductivity},\ }\href {http://arxiv.org/abs/2106.00037} {\bibfield
  {journal} {\bibinfo  {journal} {{arXiv}:2106.00037 [cond-mat]}\ } (\bibinfo
  {year} {2021})}\BibitemShut {NoStop}%
\bibitem [{\citenamefont {Sipe}\ and\ \citenamefont
  {Shkrebtii}()}]{sipe_second-order_2000}%
  \BibitemOpen
  \bibfield  {author} {\bibinfo {author} {\bibfnamefont {J.~E.}\ \bibnamefont
  {Sipe}}\ and\ \bibinfo {author} {\bibfnamefont {A.~I.}\ \bibnamefont
  {Shkrebtii}},\ }\bibfield  {title} {\bibinfo {title} {Second-order optical
  response in semiconductors},\ }\href
  {https://doi.org/10.1103/PhysRevB.61.5337} {\ \textbf {\bibinfo {volume}
  {61}},\ \bibinfo {pages} {5337}},\ \bibinfo {note} {publisher: American
  Physical Society}\BibitemShut {NoStop}%
\bibitem [{\citenamefont {Li}\ \emph {et~al.}(2020{\natexlab{b}})\citenamefont
  {Li}, \citenamefont {Tohyama}, \citenamefont {Iitaka}, \citenamefont {Su},\
  and\ \citenamefont {Zeng}}]{li_detection_2020}%
  \BibitemOpen
  \bibfield  {author} {\bibinfo {author} {\bibfnamefont {Z.}~\bibnamefont
  {Li}}, \bibinfo {author} {\bibfnamefont {T.}~\bibnamefont {Tohyama}},
  \bibinfo {author} {\bibfnamefont {T.}~\bibnamefont {Iitaka}}, \bibinfo
  {author} {\bibfnamefont {H.}~\bibnamefont {Su}},\ and\ \bibinfo {author}
  {\bibfnamefont {H.}~\bibnamefont {Zeng}},\ }\bibfield  {title} {\bibinfo
  {title} {Detection of quantum geometric tensor by nonlinear optical
  response},\ }\href {http://arxiv.org/abs/2007.02481} {\bibfield  {journal}
  {\bibinfo  {journal} {arXiv:2007.02481 [cond-mat, physics:physics]}\ }
  (\bibinfo {year} {2020}{\natexlab{b}})}\BibitemShut {NoStop}%
\bibitem [{\citenamefont {Ahn}\ \emph {et~al.}(2020)\citenamefont {Ahn},
  \citenamefont {Guo},\ and\ \citenamefont {Nagaosa}}]{ahn_low-frequency_2020}%
  \BibitemOpen
  \bibfield  {author} {\bibinfo {author} {\bibfnamefont {J.}~\bibnamefont
  {Ahn}}, \bibinfo {author} {\bibfnamefont {G.-Y.}\ \bibnamefont {Guo}},\ and\
  \bibinfo {author} {\bibfnamefont {N.}~\bibnamefont {Nagaosa}},\ }\bibfield
  {title} {\bibinfo {title} {Low-{Frequency} {Divergence} and {Quantum}
  {Geometry} of the {Bulk} {Photovoltaic} {Effect} in {Topological}
  {Semimetals}},\ }\href {https://doi.org/10.1103/PhysRevX.10.041041}
  {\bibfield  {journal} {\bibinfo  {journal} {Physical Review X}\ }\textbf
  {\bibinfo {volume} {10}},\ \bibinfo {pages} {041041} (\bibinfo {year}
  {2020})}\BibitemShut {NoStop}%
\bibitem [{\citenamefont {Ahn}\ \emph {et~al.}(2021)\citenamefont {Ahn},
  \citenamefont {Guo}, \citenamefont {Nagaosa},\ and\ \citenamefont
  {Vishwanath}}]{ahn_riemannian_2021}%
  \BibitemOpen
  \bibfield  {author} {\bibinfo {author} {\bibfnamefont {J.}~\bibnamefont
  {Ahn}}, \bibinfo {author} {\bibfnamefont {G.-Y.}\ \bibnamefont {Guo}},
  \bibinfo {author} {\bibfnamefont {N.}~\bibnamefont {Nagaosa}},\ and\ \bibinfo
  {author} {\bibfnamefont {A.}~\bibnamefont {Vishwanath}},\ }\bibfield  {title}
  {\bibinfo {title} {Riemannian {Geometry} of {Resonant} {Optical}
  {Responses}},\ }\href {http://arxiv.org/abs/2103.01241} {\bibfield  {journal}
  {\bibinfo  {journal} {arXiv:2103.01241 [cond-mat, physics:physics]}\ }
  (\bibinfo {year} {2021})},\ \bibinfo {note} {arXiv: 2103.01241}\BibitemShut
  {NoStop}%
\bibitem [{\citenamefont {Iskin}(2018)}]{Iskin2018}%
  \BibitemOpen
  \bibfield  {author} {\bibinfo {author} {\bibfnamefont {M.}~\bibnamefont
  {Iskin}},\ }\bibfield  {title} {\bibinfo {title} {{Spin Susceptibility of
  Spin-Orbit-Coupled Fermi Superfluids}},\ }\href
  {https://doi.org/10.1103/PhysRevA.97.053613} {\bibfield  {journal} {\bibinfo
  {journal} {Phys. Rev. A}\ }\textbf {\bibinfo {volume} {97}},\ \bibinfo
  {pages} {053613} (\bibinfo {year} {2018})}\BibitemShut {NoStop}%
\bibitem [{\citenamefont {Villegas}\ and\ \citenamefont
  {Yang}(2020)}]{villegas2020anomalous}%
  \BibitemOpen
  \bibfield  {author} {\bibinfo {author} {\bibfnamefont {K.~H.~A.}\
  \bibnamefont {Villegas}}\ and\ \bibinfo {author} {\bibfnamefont
  {B.}~\bibnamefont {Yang}},\ }\bibfield  {title} {\bibinfo {title} {Anomalous
  {H}iggs oscillations mediated by {B}erry curvature and quantum metric},\
  }\href {https://arxiv.org/abs/2010.07751} {\bibfield  {journal} {\bibinfo
  {journal} {arXiv:2010.07751}\ } (\bibinfo {year} {2020})}\BibitemShut
  {NoStop}%
\bibitem [{\citenamefont {Morimoto}\ and\ \citenamefont
  {Nagaosa}()}]{morimoto_topological_2016}%
  \BibitemOpen
  \bibfield  {author} {\bibinfo {author} {\bibfnamefont {T.}~\bibnamefont
  {Morimoto}}\ and\ \bibinfo {author} {\bibfnamefont {N.}~\bibnamefont
  {Nagaosa}},\ }\bibfield  {title} {\bibinfo {title} {Topological nature of
  nonlinear optical effects in solids},\ }\href
  {https://doi.org/10.1126/sciadv.1501524} {\ \textbf {\bibinfo {volume} {2}},\
  \bibinfo {pages} {e1501524}},\ \bibinfo {note} {publisher: American
  Association for the Advancement of Science Section: Research
  Article}\BibitemShut {NoStop}%
\bibitem [{\citenamefont {Iskin}(2019)}]{Iskin19}%
  \BibitemOpen
  \bibfield  {author} {\bibinfo {author} {\bibfnamefont {M.}~\bibnamefont
  {Iskin}},\ }\bibfield  {title} {\bibinfo {title} {Geometric mass acquisition
  via a quantum metric: An effective-band-mass theorem for the helicity
  bands},\ }\href {https://doi.org/10.1103/PhysRevA.99.053603} {\bibfield
  {journal} {\bibinfo  {journal} {Phys. Rev. A}\ }\textbf {\bibinfo {volume}
  {99}},\ \bibinfo {pages} {053603} (\bibinfo {year} {2019})}\BibitemShut
  {NoStop}%
\bibitem [{\citenamefont {Iskin}(2020)}]{Iskin2020_Goldstone}%
  \BibitemOpen
  \bibfield  {author} {\bibinfo {author} {\bibfnamefont {M.}~\bibnamefont
  {Iskin}},\ }\bibfield  {title} {\bibinfo {title} {Geometric contribution to
  the goldstone mode in spin–orbit coupled fermi superfluids},\ }\href
  {https://doi.org/https://doi.org/10.1016/j.physb.2020.412260} {\bibfield
  {journal} {\bibinfo  {journal} {Physica B: Condensed Matter}\ }\textbf
  {\bibinfo {volume} {592}},\ \bibinfo {pages} {412260} (\bibinfo {year}
  {2020})}\BibitemShut {NoStop}%
\bibitem [{\citenamefont {Leykam}\ \emph {et~al.}(2018)\citenamefont {Leykam},
  \citenamefont {Andreanov},\ and\ \citenamefont {Flach}}]{Leykam:2018}%
  \BibitemOpen
  \bibfield  {author} {\bibinfo {author} {\bibfnamefont {D.}~\bibnamefont
  {Leykam}}, \bibinfo {author} {\bibfnamefont {A.}~\bibnamefont {Andreanov}},\
  and\ \bibinfo {author} {\bibfnamefont {S.}~\bibnamefont {Flach}},\ }\bibfield
   {title} {\bibinfo {title} {Artificial flat band systems: from lattice models
  to experiments},\ }\href {https://doi.org/10.1080/23746149.2018.1473052}
  {\bibfield  {journal} {\bibinfo  {journal} {Advances in Physics: X}\ }\textbf
  {\bibinfo {volume} {3}},\ \bibinfo {pages} {1473052} (\bibinfo {year}
  {2018})}\BibitemShut {NoStop}%
\bibitem [{\citenamefont {Lieb}(1989)}]{Lieb:1989}%
  \BibitemOpen
  \bibfield  {author} {\bibinfo {author} {\bibfnamefont {E.~H.}\ \bibnamefont
  {Lieb}},\ }\bibfield  {title} {\bibinfo {title} {Two theorems on the
  {H}ubbard model},\ }\href {https://doi.org/10.1103/PhysRevLett.62.1201}
  {\bibfield  {journal} {\bibinfo  {journal} {Phys. Rev. Lett.}\ }\textbf
  {\bibinfo {volume} {62}},\ \bibinfo {pages} {1201} (\bibinfo {year}
  {1989})}\BibitemShut {NoStop}%
\bibitem [{\citenamefont {Huber}\ and\ \citenamefont
  {Altman}(2010)}]{Huber2010_sawtooth}%
  \BibitemOpen
  \bibfield  {author} {\bibinfo {author} {\bibfnamefont {S.~D.}\ \bibnamefont
  {Huber}}\ and\ \bibinfo {author} {\bibfnamefont {E.}~\bibnamefont {Altman}},\
  }\bibfield  {title} {\bibinfo {title} {Bose condensation in flat bands},\
  }\href {https://doi.org/10.1103/PhysRevB.82.184502} {\bibfield  {journal}
  {\bibinfo  {journal} {Phys. Rev. B}\ }\textbf {\bibinfo {volume} {82}},\
  \bibinfo {pages} {184502} (\bibinfo {year} {2010})}\BibitemShut {NoStop}%
\bibitem [{\citenamefont {Zhang}\ and\ \citenamefont
  {Jo}(2015)}]{Zhang2015_sawtooth}%
  \BibitemOpen
  \bibfield  {author} {\bibinfo {author} {\bibfnamefont {T.}~\bibnamefont
  {Zhang}}\ and\ \bibinfo {author} {\bibfnamefont {G.-B.}\ \bibnamefont {Jo}},\
  }\bibfield  {title} {\bibinfo {title} {One-dimensional sawtooth and zigzag
  lattices for ultracold atoms},\ }\bibfield  {journal} {\bibinfo  {journal}
  {Scientific Reports}\ }\textbf {\bibinfo {volume} {5}},\ \href
  {https://doi.org/10.1038/srep16044} {10.1038/srep16044} (\bibinfo {year}
  {2015})\BibitemShut {NoStop}%
\bibitem [{\citenamefont {Oka}\ and\ \citenamefont
  {Aoki}(2009)}]{oka_photovoltaic_2009}%
  \BibitemOpen
  \bibfield  {author} {\bibinfo {author} {\bibfnamefont {T.}~\bibnamefont
  {Oka}}\ and\ \bibinfo {author} {\bibfnamefont {H.}~\bibnamefont {Aoki}},\
  }\bibfield  {title} {\bibinfo {title} {Photovoltaic {Hall} effect in
  graphene},\ }\href {https://doi.org/10.1103/PhysRevB.79.081406} {\bibfield
  {journal} {\bibinfo  {journal} {Physical Review B}\ }\textbf {\bibinfo
  {volume} {79}},\ \bibinfo {pages} {081406} (\bibinfo {year}
  {2009})}\BibitemShut {NoStop}%
\bibitem [{\citenamefont {Dehghani}\ \emph {et~al.}(2015)\citenamefont
  {Dehghani}, \citenamefont {Oka},\ and\ \citenamefont
  {Mitra}}]{PhysRevB.91.155422}%
  \BibitemOpen
  \bibfield  {author} {\bibinfo {author} {\bibfnamefont {H.}~\bibnamefont
  {Dehghani}}, \bibinfo {author} {\bibfnamefont {T.}~\bibnamefont {Oka}},\ and\
  \bibinfo {author} {\bibfnamefont {A.}~\bibnamefont {Mitra}},\ }\bibfield
  {title} {\bibinfo {title} {Out-of-equilibrium electrons and the hall
  conductance of a floquet topological insulator},\ }\href
  {https://doi.org/10.1103/PhysRevB.91.155422} {\bibfield  {journal} {\bibinfo
  {journal} {Phys. Rev. B}\ }\textbf {\bibinfo {volume} {91}},\ \bibinfo
  {pages} {155422} (\bibinfo {year} {2015})}\BibitemShut {NoStop}%
\bibitem [{\citenamefont {McIver}\ \emph {et~al.}(2020)\citenamefont {McIver},
  \citenamefont {Schulte}, \citenamefont {Stein}, \citenamefont {Matsuyama},
  \citenamefont {Jotzu}, \citenamefont {Meier},\ and\ \citenamefont
  {Cavalleri}}]{mciver_light-induced_2020}%
  \BibitemOpen
  \bibfield  {author} {\bibinfo {author} {\bibfnamefont {J.~W.}\ \bibnamefont
  {McIver}}, \bibinfo {author} {\bibfnamefont {B.}~\bibnamefont {Schulte}},
  \bibinfo {author} {\bibfnamefont {F.-U.}\ \bibnamefont {Stein}}, \bibinfo
  {author} {\bibfnamefont {T.}~\bibnamefont {Matsuyama}}, \bibinfo {author}
  {\bibfnamefont {G.}~\bibnamefont {Jotzu}}, \bibinfo {author} {\bibfnamefont
  {G.}~\bibnamefont {Meier}},\ and\ \bibinfo {author} {\bibfnamefont
  {A.}~\bibnamefont {Cavalleri}},\ }\bibfield  {title} {\bibinfo {title}
  {Light-induced anomalous {Hall} effect in graphene},\ }\href
  {https://doi.org/10.1038/s41567-019-0698-y} {\bibfield  {journal} {\bibinfo
  {journal} {Nature Physics}\ }\textbf {\bibinfo {volume} {16}},\ \bibinfo
  {pages} {38} (\bibinfo {year} {2020})}\BibitemShut {NoStop}%
\bibitem [{\citenamefont {Sentef}\ \emph {et~al.}(2015)\citenamefont {Sentef},
  \citenamefont {Claassen}, \citenamefont {Kemper}, \citenamefont {Moritz},
  \citenamefont {Oka}, \citenamefont {Freericks},\ and\ \citenamefont
  {Devereaux}}]{sentef_theory_2015}%
  \BibitemOpen
  \bibfield  {author} {\bibinfo {author} {\bibfnamefont {M.~A.}\ \bibnamefont
  {Sentef}}, \bibinfo {author} {\bibfnamefont {M.}~\bibnamefont {Claassen}},
  \bibinfo {author} {\bibfnamefont {A.~F.}\ \bibnamefont {Kemper}}, \bibinfo
  {author} {\bibfnamefont {B.}~\bibnamefont {Moritz}}, \bibinfo {author}
  {\bibfnamefont {T.}~\bibnamefont {Oka}}, \bibinfo {author} {\bibfnamefont
  {J.~K.}\ \bibnamefont {Freericks}},\ and\ \bibinfo {author} {\bibfnamefont
  {T.~P.}\ \bibnamefont {Devereaux}},\ }\bibfield  {title} {\bibinfo {title}
  {Theory of {Floquet} band formation and local pseudospin textures in
  pump-probe photoemission of graphene},\ }\href
  {https://doi.org/10.1038/ncomms8047} {\bibfield  {journal} {\bibinfo
  {journal} {Nature Communications}\ }\textbf {\bibinfo {volume} {6}},\
  \bibinfo {pages} {7047} (\bibinfo {year} {2015})}\BibitemShut {NoStop}%
\bibitem [{\citenamefont {Sato}\ \emph {et~al.}(2019)\citenamefont {Sato},
  \citenamefont {McIver}, \citenamefont {Nuske}, \citenamefont {Tang},
  \citenamefont {Jotzu}, \citenamefont {Schulte}, \citenamefont {Hübener},
  \citenamefont {De~Giovannini}, \citenamefont {Mathey}, \citenamefont
  {Sentef}, \citenamefont {Cavalleri},\ and\ \citenamefont
  {Rubio}}]{sato_microscopic_2019}%
  \BibitemOpen
  \bibfield  {author} {\bibinfo {author} {\bibfnamefont {S.~A.}\ \bibnamefont
  {Sato}}, \bibinfo {author} {\bibfnamefont {J.~W.}\ \bibnamefont {McIver}},
  \bibinfo {author} {\bibfnamefont {M.}~\bibnamefont {Nuske}}, \bibinfo
  {author} {\bibfnamefont {P.}~\bibnamefont {Tang}}, \bibinfo {author}
  {\bibfnamefont {G.}~\bibnamefont {Jotzu}}, \bibinfo {author} {\bibfnamefont
  {B.}~\bibnamefont {Schulte}}, \bibinfo {author} {\bibfnamefont
  {H.}~\bibnamefont {Hübener}}, \bibinfo {author} {\bibfnamefont
  {U.}~\bibnamefont {De~Giovannini}}, \bibinfo {author} {\bibfnamefont
  {L.}~\bibnamefont {Mathey}}, \bibinfo {author} {\bibfnamefont {M.~A.}\
  \bibnamefont {Sentef}}, \bibinfo {author} {\bibfnamefont {A.}~\bibnamefont
  {Cavalleri}},\ and\ \bibinfo {author} {\bibfnamefont {A.}~\bibnamefont
  {Rubio}},\ }\bibfield  {title} {\bibinfo {title} {Microscopic theory for the
  light-induced anomalous {Hall} effect in graphene},\ }\href
  {https://doi.org/10.1103/PhysRevB.99.214302} {\bibfield  {journal} {\bibinfo
  {journal} {Physical Review B}\ }\textbf {\bibinfo {volume} {99}},\ \bibinfo
  {pages} {214302} (\bibinfo {year} {2019})}\BibitemShut {NoStop}%
\bibitem [{\citenamefont {Nuske}\ \emph {et~al.}(2020)\citenamefont {Nuske},
  \citenamefont {Broers}, \citenamefont {Schulte}, \citenamefont {Jotzu},
  \citenamefont {Sato}, \citenamefont {Cavalleri}, \citenamefont {Rubio},
  \citenamefont {McIver},\ and\ \citenamefont {Mathey}}]{nuske_floquet_2020}%
  \BibitemOpen
  \bibfield  {author} {\bibinfo {author} {\bibfnamefont {M.}~\bibnamefont
  {Nuske}}, \bibinfo {author} {\bibfnamefont {L.}~\bibnamefont {Broers}},
  \bibinfo {author} {\bibfnamefont {B.}~\bibnamefont {Schulte}}, \bibinfo
  {author} {\bibfnamefont {G.}~\bibnamefont {Jotzu}}, \bibinfo {author}
  {\bibfnamefont {S.~A.}\ \bibnamefont {Sato}}, \bibinfo {author}
  {\bibfnamefont {A.}~\bibnamefont {Cavalleri}}, \bibinfo {author}
  {\bibfnamefont {A.}~\bibnamefont {Rubio}}, \bibinfo {author} {\bibfnamefont
  {J.~W.}\ \bibnamefont {McIver}},\ and\ \bibinfo {author} {\bibfnamefont
  {L.}~\bibnamefont {Mathey}},\ }\bibfield  {title} {\bibinfo {title} {Floquet
  dynamics in light-driven solids},\ }\href {http://arxiv.org/abs/2005.10824}
  {\bibfield  {journal} {\bibinfo  {journal} {arXiv:2005.10824 [cond-mat]}\ }
  (\bibinfo {year} {2020})}\BibitemShut {NoStop}%
\bibitem [{\citenamefont {Bukov}\ \emph {et~al.}(2015)\citenamefont {Bukov},
  \citenamefont {D'Alessio},\ and\ \citenamefont
  {Polkovnikov}}]{bukov_universal_2015}%
  \BibitemOpen
  \bibfield  {author} {\bibinfo {author} {\bibfnamefont {M.}~\bibnamefont
  {Bukov}}, \bibinfo {author} {\bibfnamefont {L.}~\bibnamefont {D'Alessio}},\
  and\ \bibinfo {author} {\bibfnamefont {A.}~\bibnamefont {Polkovnikov}},\
  }\bibfield  {title} {\bibinfo {title} {Universal high-frequency behavior of
  periodically driven systems: from dynamical stabilization to {Floquet}
  engineering},\ }\href {https://doi.org/10.1080/00018732.2015.1055918}
  {\bibfield  {journal} {\bibinfo  {journal} {Advances in Physics}\ }\textbf
  {\bibinfo {volume} {64}},\ \bibinfo {pages} {139} (\bibinfo {year}
  {2015})}\BibitemShut {NoStop}%
\bibitem [{\citenamefont {Tasaki}(1992)}]{Tasaki:1992}%
  \BibitemOpen
  \bibfield  {author} {\bibinfo {author} {\bibfnamefont {H.}~\bibnamefont
  {Tasaki}},\ }\bibfield  {title} {\bibinfo {title} {{Ferromagnetism in the
  Hubbard models with degenerate single-electron ground states}},\ }\href
  {http://link.aps.org/doi/10.1103/PhysRevLett.69.1608} {\bibfield  {journal}
  {\bibinfo  {journal} {Phys. Rev. Lett.}\ }\textbf {\bibinfo {volume} {69}},\
  \bibinfo {pages} {1608} (\bibinfo {year} {1992})}\BibitemShut {NoStop}%
\bibitem [{\citenamefont {Mielke}(1991)}]{Mielke:1991}%
  \BibitemOpen
  \bibfield  {author} {\bibinfo {author} {\bibfnamefont {A.}~\bibnamefont
  {Mielke}},\ }\bibfield  {title} {\bibinfo {title} {Ferromagnetic ground
  states for the {Hubbard} model on line graphs},\ }\href
  {https://doi.org/10.1088/0305-4470/24/2/005} {\bibfield  {journal} {\bibinfo
  {journal} {Journal of Physics A: Mathematical and General}\ }\textbf
  {\bibinfo {volume} {24}},\ \bibinfo {pages} {L73} (\bibinfo {year}
  {1991})}\BibitemShut {NoStop}%
\bibitem [{\citenamefont {Kopnin}\ \emph {et~al.}(2011)\citenamefont {Kopnin},
  \citenamefont {Heikkil\"a},\ and\ \citenamefont {Volovik}}]{kopnin:2011}%
  \BibitemOpen
  \bibfield  {author} {\bibinfo {author} {\bibfnamefont {N.~B.}\ \bibnamefont
  {Kopnin}}, \bibinfo {author} {\bibfnamefont {T.~T.}\ \bibnamefont
  {Heikkil\"a}},\ and\ \bibinfo {author} {\bibfnamefont {G.~E.}\ \bibnamefont
  {Volovik}},\ }\bibfield  {title} {\bibinfo {title} {{High-temperature surface
  superconductivity in topological flat-band systems}},\ }\href
  {https://doi.org/10.1103/PhysRevB.83.220503} {\bibfield  {journal} {\bibinfo
  {journal} {Phys. Rev. B}\ }\textbf {\bibinfo {volume} {83}},\ \bibinfo
  {pages} {220503} (\bibinfo {year} {2011})}\BibitemShut {NoStop}%
\bibitem [{\citenamefont {Brouder}\ \emph {et~al.}(2007)\citenamefont
  {Brouder}, \citenamefont {Panati}, \citenamefont {Calandra}, \citenamefont
  {Mourougane},\ and\ \citenamefont {Marzari}}]{Brouder:2007}%
  \BibitemOpen
  \bibfield  {author} {\bibinfo {author} {\bibfnamefont {C.}~\bibnamefont
  {Brouder}}, \bibinfo {author} {\bibfnamefont {G.}~\bibnamefont {Panati}},
  \bibinfo {author} {\bibfnamefont {M.}~\bibnamefont {Calandra}}, \bibinfo
  {author} {\bibfnamefont {C.}~\bibnamefont {Mourougane}},\ and\ \bibinfo
  {author} {\bibfnamefont {N.}~\bibnamefont {Marzari}},\ }\bibfield  {title}
  {\bibinfo {title} {Exponential localization of {W}annier functions in
  insulators},\ }\href {https://doi.org/10.1103/PhysRevLett.98.046402}
  {\bibfield  {journal} {\bibinfo  {journal} {Phys. Rev. Lett.}\ }\textbf
  {\bibinfo {volume} {98}},\ \bibinfo {pages} {046402} (\bibinfo {year}
  {2007})}\BibitemShut {NoStop}%
\bibitem [{\citenamefont {Tovmasyan}\ \emph {et~al.}(2016)\citenamefont
  {Tovmasyan}, \citenamefont {Peotta}, \citenamefont {T\"orm\"a},\ and\
  \citenamefont {Huber}}]{Tovmasyan:2016}%
  \BibitemOpen
  \bibfield  {author} {\bibinfo {author} {\bibfnamefont {M.}~\bibnamefont
  {Tovmasyan}}, \bibinfo {author} {\bibfnamefont {S.}~\bibnamefont {Peotta}},
  \bibinfo {author} {\bibfnamefont {P.}~\bibnamefont {T\"orm\"a}},\ and\
  \bibinfo {author} {\bibfnamefont {S.}~\bibnamefont {Huber}},\ }\bibfield
  {title} {\bibinfo {title} {{Effective Theory and Emergent $\text{SU}(2)$
  Symmetry in the Flat Bands of Attractive Hubbard Models}},\ }\href
  {https://doi.org/10.1103/PhysRevB.94.245149} {\bibfield  {journal} {\bibinfo
  {journal} {Phys. Rev. B}\ }\textbf {\bibinfo {volume} {94}},\ \bibinfo
  {pages} {245149} (\bibinfo {year} {2016})}\BibitemShut {NoStop}%
\bibitem [{\citenamefont {Zou}\ \emph {et~al.}(2018)\citenamefont {Zou},
  \citenamefont {Po}, \citenamefont {Vishwanath},\ and\ \citenamefont
  {Senthil}}]{zou_band_2018}%
  \BibitemOpen
  \bibfield  {author} {\bibinfo {author} {\bibfnamefont {L.}~\bibnamefont
  {Zou}}, \bibinfo {author} {\bibfnamefont {H.~C.}\ \bibnamefont {Po}},
  \bibinfo {author} {\bibfnamefont {A.}~\bibnamefont {Vishwanath}},\ and\
  \bibinfo {author} {\bibfnamefont {T.}~\bibnamefont {Senthil}},\ }\bibfield
  {title} {\bibinfo {title} {Band structure of twisted bilayer graphene:
  {Emergent} symmetries, commensurate approximants, and {Wannier}
  obstructions},\ }\href {https://doi.org/10.1103/PhysRevB.98.085435}
  {\bibfield  {journal} {\bibinfo  {journal} {Physical Review B}\ }\textbf
  {\bibinfo {volume} {98}},\ \bibinfo {pages} {085435} (\bibinfo {year}
  {2018})}\BibitemShut {NoStop}%
\bibitem [{\citenamefont {Po}\ \emph {et~al.}(2018{\natexlab{b}})\citenamefont
  {Po}, \citenamefont {Zou}, \citenamefont {Vishwanath},\ and\ \citenamefont
  {Senthil}}]{po_origin_2018}%
  \BibitemOpen
  \bibfield  {author} {\bibinfo {author} {\bibfnamefont {H.~C.}\ \bibnamefont
  {Po}}, \bibinfo {author} {\bibfnamefont {L.}~\bibnamefont {Zou}}, \bibinfo
  {author} {\bibfnamefont {A.}~\bibnamefont {Vishwanath}},\ and\ \bibinfo
  {author} {\bibfnamefont {T.}~\bibnamefont {Senthil}},\ }\bibfield  {title}
  {\bibinfo {title} {Origin of {Mott} {Insulating} {Behavior} and
  {Superconductivity} in {Twisted} {Bilayer} {Graphene}},\ }\href
  {https://doi.org/10.1103/PhysRevX.8.031089} {\bibfield  {journal} {\bibinfo
  {journal} {Physical Review X}\ }\textbf {\bibinfo {volume} {8}},\ \bibinfo
  {pages} {031089} (\bibinfo {year} {2018}{\natexlab{b}})}\BibitemShut
  {NoStop}%
\bibitem [{\citenamefont {Ahn}\ \emph {et~al.}(2019)\citenamefont {Ahn},
  \citenamefont {Park},\ and\ \citenamefont {Yang}}]{ahn_failure_2019}%
  \BibitemOpen
  \bibfield  {author} {\bibinfo {author} {\bibfnamefont {J.}~\bibnamefont
  {Ahn}}, \bibinfo {author} {\bibfnamefont {S.}~\bibnamefont {Park}},\ and\
  \bibinfo {author} {\bibfnamefont {B.-J.}\ \bibnamefont {Yang}},\ }\bibfield
  {title} {\bibinfo {title} {Failure of {Nielsen}-{Ninomiya} {Theorem} and
  {Fragile} {Topology} in {Two}-{Dimensional} {Systems} with {Space}-{Time}
  {Inversion} {Symmetry}: {Application} to {Twisted} {Bilayer} {Graphene} at
  {Magic} {Angle}},\ }\href {https://doi.org/10.1103/PhysRevX.9.021013}
  {\bibfield  {journal} {\bibinfo  {journal} {Physical Review X}\ }\textbf
  {\bibinfo {volume} {9}},\ \bibinfo {pages} {021013} (\bibinfo {year}
  {2019})}\BibitemShut {NoStop}%
\bibitem [{\citenamefont {Su}\ and\ \citenamefont
  {Lin}(2018{\natexlab{b}})}]{su:2018}%
  \BibitemOpen
  \bibfield  {author} {\bibinfo {author} {\bibfnamefont {Y.}~\bibnamefont
  {Su}}\ and\ \bibinfo {author} {\bibfnamefont {S.-Z.}\ \bibnamefont {Lin}},\
  }\bibfield  {title} {\bibinfo {title} {Pairing symmetry and spontaneous
  vortex-antivortex lattice in superconducting twisted-bilayer graphene:
  {B}ogoliubov-de {G}ennes approach},\ }\href
  {https://doi.org/10.1103/PhysRevB.98.195101} {\bibfield  {journal} {\bibinfo
  {journal} {Phys. Rev. B}\ }\textbf {\bibinfo {volume} {98}},\ \bibinfo
  {pages} {195101} (\bibinfo {year} {2018}{\natexlab{b}})}\BibitemShut
  {NoStop}%
\bibitem [{\citenamefont {Gonzalez-Arraga}\ \emph {et~al.}(2017)\citenamefont
  {Gonzalez-Arraga}, \citenamefont {Lado}, \citenamefont {Guinea},\ and\
  \citenamefont {San-Jose}}]{gonzalez-arraga:2017}%
  \BibitemOpen
  \bibfield  {author} {\bibinfo {author} {\bibfnamefont {L.~A.}\ \bibnamefont
  {Gonzalez-Arraga}}, \bibinfo {author} {\bibfnamefont {J.~L.}\ \bibnamefont
  {Lado}}, \bibinfo {author} {\bibfnamefont {F.}~\bibnamefont {Guinea}},\ and\
  \bibinfo {author} {\bibfnamefont {P.}~\bibnamefont {San-Jose}},\ }\bibfield
  {title} {\bibinfo {title} {Electrically controllable magnetism in twisted
  bilayer graphene},\ }\href {https://doi.org/10.1103/PhysRevLett.119.107201}
  {\bibfield  {journal} {\bibinfo  {journal} {Phys. Rev. Lett.}\ }\textbf
  {\bibinfo {volume} {119}},\ \bibinfo {pages} {107201} (\bibinfo {year}
  {2017})}\BibitemShut {NoStop}%
\end{thebibliography}%

\clearpage
\appendix

\section{Quadratic intra-band coupling}
\label{sec:calcQuadraticIntra}
Here we present the calculation for the quadratic intra-band coupling
\begin{equation}
    L_{n, \mu \nu}^{AA}(\fk) = \bra{n(\fk)} \left( \dmu \dnu H_0 \right) \ket{n(\fk)}.
\end{equation}
From now, we leave out the explicit $k$-dependence of the quantities for brevity.
From the Schr\"odinger equation we obtain
\begin{equation}
    \begin{aligned}
    &\bra{n} \left( \dmu \dnu H \right) \ket{n} + \bra{n} \left( \dmu H \right) \ket{\dnu n} + \bra{n} \left( \dnu H \right) \ket{\dmu n}
    \\
    & \hspace{5mm} = \dmu \dnu \varepsilon_n 
    + \left( \dmu \varepsilon_n \right) \langle n \ket{\dnu n} + \left( \dnu \varepsilon_n \right) \langle n \ket{\dmu n}. 
    \end{aligned}
    \label{eq:QuadraticIntraIntermediateResult1}
\end{equation}
We next simplify the second and third term on the LHS, as exemplified here for the second term:
\begin{equation}
    \begin{aligned}
    & \bra{n} \left( \dmu H \right) \ket{\dnu n}\\
    & = \sum_{n'} \bra{n} \left( \dmu H \right) \ket{n'} \langle n' \ket{\dnu n}\\
    & = \sum_{n' \neq n} \underbrace{\bra{n} \left( \dmu H \right) \ket{n'}}_{\overset{\text{Sec.~\ref{sec:LinInter}}}{=} (\varepsilon_n - \varepsilon_n') \langle  \dmu n  \ket{n'} } \langle n' \ket{\dnu n}\\
    & + \underbrace{\bra{n} \left( \dmu H \right) \ket{n}}_{\overset{\text{Sec.~\ref{sec:LinIntra}}}{=} \dmu \varepsilon_n} \langle n \ket{\dnu n}\\
    & = 2 \sum_{n' \neq n} (\varepsilon_n {-} \varepsilon_{n'}) \langle \dmu n \ket{n'} \langle n' \ket{\dnu n} + \left( \dmu \varepsilon_n \right) \langle n \ket{\dnu n}.
    \end{aligned}
    \label{eq:QuadraticIntraIntermediateResult2}
\end{equation}
Inserting this and the analogous term, with $\mu$ and $\nu$ exchanged, back into Eq.~(\ref{eq:QuadraticIntraIntermediateResult1}) 
we arrive at the final result
\begin{equation}
    \begin{aligned}
    &L_{n, \mu \nu}^{AA}(\fk) = \dmu \dnu \varepsilon_n(\fk) {-} \sum_{n \neq n'} (\varepsilon_n(\fk) {-} \varepsilon_{n'}(\fk)) \times\\
    & \times \left(\langle \dmu n(\fk) \ket{n'(\fk)} \langle n'(\fk) \ket{\dnu n(\fk)} + \hc \right),
    \label{eq:QuadraticIntraAppendix}
    \end{aligned}
\end{equation}
where we have restored the explicit $\fk$ dependence.
This is the result quoted in Sec.~\ref{sec:QuadraticIntra} Eq.~(\ref{eq:QuadraticIntra}).
The above calculation is analogous to that of Ref.~\onlinecite{Iskin19}.

\section{Light-matter coupling in Dirac Hamiltonian} \label{Sec:LMC_DIRAC}
The Dirac Hamiltonian to describe a single Dirac point can generally written as
\begin{equation}
   H_\text{D} = v_F (\sigma_x k_x + \sigma_y k_y),
   \label{Eq:H_DIRAC}
\end{equation}
where $v_F$ denotes the band velocity, $\fk = (k_x,k_y)$ the two-dimensional lattice momentum, and $\sigma_{x,y}$ the corresponding Pauli matrices. The orthonormal eigenvectors are
\begin{equation}
    \bm{v_1} = \frac{1}{\sqrt{2}}(-\frac{k_x+\I k_y}{|\fk|},1)^T, \quad \bm{v_2} = \frac{1}{\sqrt{2}}(\frac{k_x+\I k_y}{|\fk|},1)^T,
    \label{Eq:DIRAC_EV}
\end{equation}
with corresponding eigenvalues
\begin{equation}
    \varepsilon_1 = -v_F|\fk| , \qquad \varepsilon_2 = v_F|\fk|.
\end{equation}
The momentum derivatives of the Dirac Hamiltonian are
\begin{equation}
    \partial_x H_\text{D} = v_F \sigma_x , \qquad \partial_y H_\text{D} = v_F \sigma_y.
\end{equation}
In the eigenbasis, Eq.~(\ref{Eq:DIRAC_EV}), these take the form
\begin{equation}
    \begin{aligned}
   S \partial_x H_\text{D} S^{-1} &= \frac{v_F}{|\fk|} (-\sigma_z k_x - \sigma_y k_y), \\
   S \partial_y H_\text{D} S^{-1} &= \frac{v_F}{|\fk|} (-\sigma_z k_y + \sigma_y k_x).
    \end{aligned}
\end{equation}
The second momentum derivatives vanish since the Dirac Hamiltonian only contains terms that are linear in momentum.

\section{Floquet convergence}
\label{sec:Floquet convergence}
\begin{figure*}
    \centering
    \includegraphics[scale=0.45]{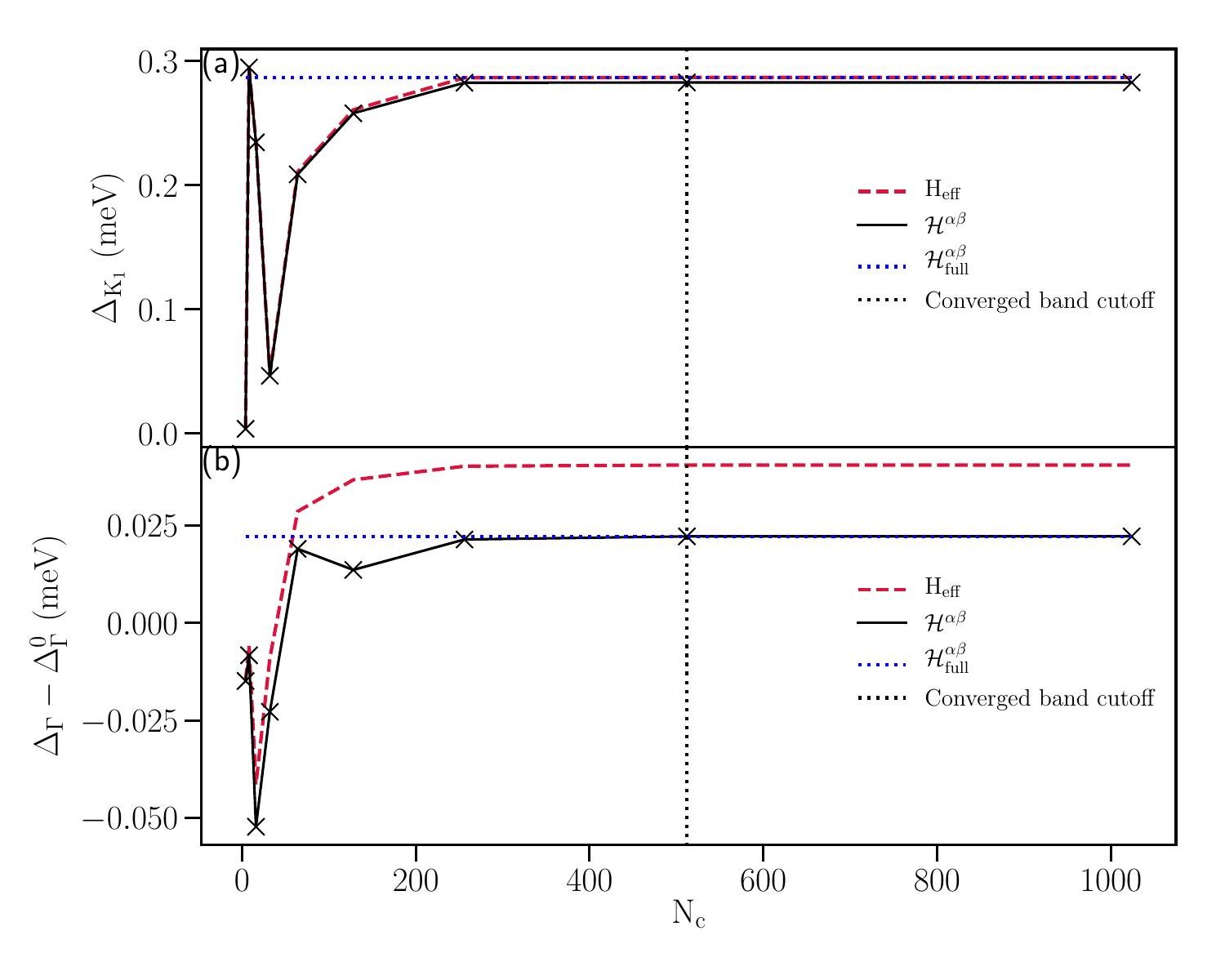}
    \caption{Convergence of Floquet effects as function of band cutoff $N_c$ for a driving frequency $\Omega=\REV{3}$ eV and a driving amplitude $A_0=\REV{0.01}$ $\text{\AA}^{-1}$. Black solid lines indicate the values computed with the truncated Floquet Hamiltonian $\mathcal{H}^{\alpha\beta}$. \REV{The blue dashed line indicates the reference values extracted from the full Floquet Hamiltonian without band truncation and with full exponential LMC. Red dashed lines indicate the values computed via the effective Floquet Hamiltonian $H_\text{eff}$ in the HFA. For both quantities, the Floquet gap at $K_1$ (a) and the bandwidth renormalization at $\Gamma$ (b), we find good convergence for a band cutoff $N_c=512$. We use this cutoff for all numerical Floquet calculations presented in Sec.~\ref{Sec.:C}. The truncated Floquet Hamiltonian (black) and the full Floquet Hamiltonian (blue) show consistent results at $N_c$, rendering field effects beyond second order insignificant in the chosen driving regime. While the high frequency approximation works well at the K-point, it overestimates the bandwidth renormalization at the $\Gamma$ point. However, it should be noted that the HFA is predominantly introduced for means of a profound understanding of the underlying physical mechanism, not to provide quantitatively exact results.}} 
    \label{fig:FLOQ_GAP_CON}
\end{figure*}

In Fig.~\ref{fig:FLOQ_GAP_CON} we show the convergence of the Floquet gap at $K_1$ and the electronic bandwidth renormalization of the low-energy band manifold at $\Gamma$ as a function of the number $\text{max}(\alpha)$ of considered photon sectors. For the laser parameters used in this paper, we find good convergence when we retain only contributions up to first order $\alpha=\pm 1$.  
\begin{figure*}
    \centering
    \includegraphics[scale=0.45]{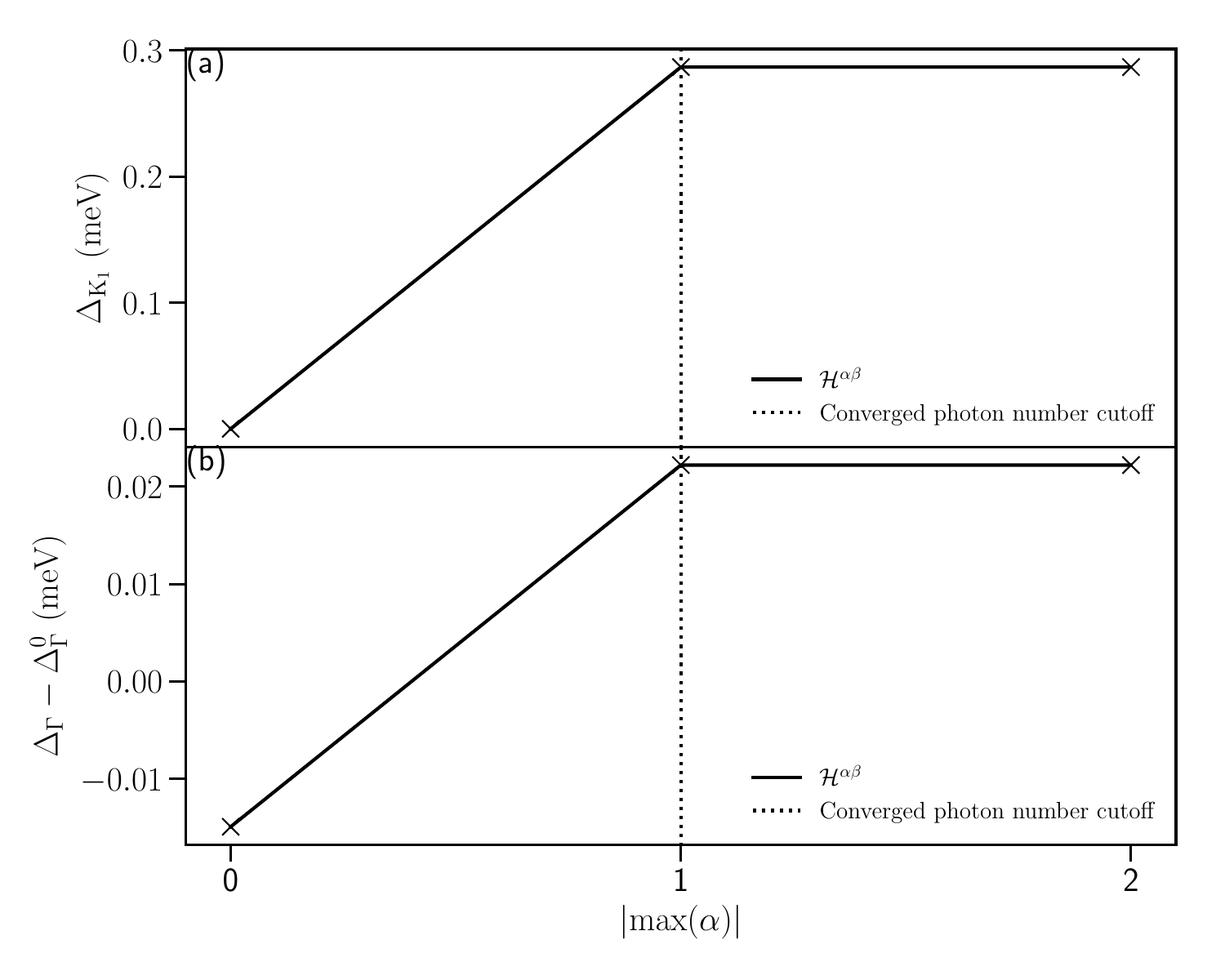}
    \caption{Convergence of Floquet effects as function of considered photon sectors for a driving frequency $\Omega=\REV{3}$ eV and a driving amplitude $A_0=\REV{0.01}$ $\text{\AA}^{-1}$. Black solid lines indicate the values computed with the truncated Floquet Hamiltonian $\mathcal{H}^{\alpha\beta}$. For both quantities, the Floquet gap at $K_1$ (a) and the bandwidth renormalization at $\Gamma$ (b), we find good convergence at linear photon order $|\text{max}(\alpha)|=1$, \REV{which we use throughout this work.}}
    \label{fig:FLOQ_GAP_CON_M}
\end{figure*}
In Fig.~\ref{fig:FLOQ_GAP_CON} we plot the convergence of the Floquet gap at $K_1$ and the electronic bandwidth renormalization of the low-energy band manifold at $\Gamma$ as a function of the band cutoff $N_c$. We find good convergence for a band cutoff of $N_c=512$ bands.  
\end{document}